\documentclass[acmsmall, nonacm, screen]{acmart} 
\PassOptionsToPackage{table}{xcolor}
\usepackage{tabularx}
\AtBeginDocument{%
  \providecommand\BibTeX{{%
    \normalfont B\kern-0.5em{\scshape i\kern-0.25em b}\kern-0.8em\TeX}}}

\setcopyright{acmlicensed}
\copyrightyear{2025}
\acmYear{2025}
\acmDOI{XXXXXXX.XXXXXXX}

\acmJournal{TOCHI}
\acmVolume{37}
\acmNumber{4}
\acmArticle{111}
\acmMonth{8}

\usepackage{threeparttable}
\usepackage{tabularx}
\usepackage{booktabs}
\usepackage{multirow}
\usepackage{makecell}
\usepackage{colortbl}
\usepackage{booktabs}
\usepackage{array}
\usepackage{pgf}
\usepackage{collcell}
\usepackage{hhline}
\usepackage{longtable}
\usepackage{multirow}
\usepackage{amsmath}
\usepackage{adjustbox}
\usepackage{wrapfig}
\usepackage{float}
\usepackage{enumitem}
\usepackage{url}
\usepackage{hyperref}
\usepackage{ragged2e}
\usepackage{tabularray}
\usepackage{setspace}
\usepackage{graphicx}
\usepackage{subcaption}

\definecolor{high}{HTML}{628c77} 
\definecolor{mid}{HTML}{b9f2d5}  
\definecolor{low}{HTML}{FFFFFF}  

\definecolor{high1}{HTML}{7262ac} 
\definecolor{mid1}{HTML}{FFFFFF}  
\definecolor{low1}{HTML}{2e974e}  

\newcommand{\gradientcelld}[8]{
\xdef\lowvalx{#2}%
\xdef\midvalx{#3}%
\xdef\maxvalx{#4}%
\xdef\lowcolx{#5}%
\xdef\midcolx{#6}%
\xdef\highcolx{#7}%
\xdef\opacityx{#8}%
\ifdimcomp{#1pt}{>}{\maxvalx pt}{\cellcolor{\highcolx!100.0!\midcolx!\opacityx}#1}{
\ifdimcomp{#1pt}{<}{\midvalx pt}{%
\ifdimcomp{#1pt}{<}{\lowvalx pt}{\cellcolor{\midcolx!0.0!\lowcolx!\opacityx}#1}{
     \pgfmathparse{int(round(100*(#1/(\midvalx-\lowvalx))-(\lowvalx*(100/(\midvalx-\lowvalx)))))}%
    \xdef\tempa{\pgfmathresult}%
    \cellcolor{\midcolx!\tempa!\lowcolx!\opacityx}#1%
}}{
     \pgfmathparse{int(round(100*(#1/(\maxvalx-\midvalx))-(\midvalx*(100/(\maxvalx-\midvalx)))))}
    \xdef\tempb{\pgfmathresult}%
    \cellcolor{\highcolx!\tempb!\midcolx!\opacityx}#1%
}}
}

\newcommand{\gradientcell}[1]{
    \gradientcelld{#1}{-0.04}{0.36}{0.76}{low}{mid}{high}{60}
    }

\newcommand{\g}[1]{
    \gradientcelld{#1}{-0.18}{0}{0.13}{low1}{mid1}{high1}{60}
    }

\newcommand{\gradientcellboldd}[8]{
\xdef\lowvalx{#2}%
\xdef\midvalx{#3}%
\xdef\maxvalx{#4}%
\xdef\lowcolx{#5}%
\xdef\midcolx{#6}%
\xdef\highcolx{#7}%
\xdef\opacityx{#8}%
\ifdimcomp{#1pt}{>}{\maxvalx pt}{\cellcolor{\highcolx!100.0!\midcolx!\opacityx}\textbf{#1}}{
\ifdimcomp{#1pt}{<}{\midvalx pt}{%
\ifdimcomp{#1pt}{<}{\lowvalx pt}{\cellcolor{\midcolx!0.0!\lowcolx!\opacityx}\textbf{#1}}{
     \pgfmathparse{int(round(100*(#1/(\midvalx-\lowvalx))-(\lowvalx*(100/(\midvalx-\lowvalx)))))}%
    \xdef\tempa{\pgfmathresult}%
    \cellcolor{\midcolx!\tempa!\lowcolx!\opacityx}\textbf{#1}%
}}{
     \pgfmathparse{int(round(100*(#1/(\maxvalx-\midvalx))-(\midvalx*(100/(\maxvalx-\midvalx)))))}
    \xdef\tempb{\pgfmathresult}%
    \cellcolor{\highcolx!\tempb!\midcolx!\opacityx}\textbf{#1}%
}}
}

\newcommand{\gradientcellbold}[1]{
    \gradientcellboldd{#1}{-0.04}{0.36}{0.76}{low}{mid}{high}{60}
}

\usepackage{graphicx}
\usepackage{color}
\definecolor{darkred}{HTML}{7e0f12}
\definecolor{darkgreen}{rgb}{0.0, 0.5, 0.0}
\definecolor{purple}{HTML}{7262ac}
\begin{document}

\title{Exploring the Human-LLM Synergy in Advancing Theory-driven Qualitative Analysis}


\author{Han Meng}
\email{han.meng@u.nus.edu}
\orcid{0009-0003-2318-3639}
\author{Yitian Yang}
\email{e1498132@u.nus.edu}
\orcid{0009-0000-7530-2116}
\affiliation{
  \institution{Department of Computer Science, National University of Singapore}
  \streetaddress{21 Lower Kent Ridge Road}
  \city{Singapore}
  \country{Singapore}
  \postcode{119077}
}
\author{Wayne Fu}
\email{wayne_fu@acm.org}
\orcid{0009-0007-4947-0045}
\affiliation{
  \institution{Amazon Research, Amazon.com Inc}
  \country{USA}
}
\author{Jungup Lee}
\email{swklj@nus.edu.sg}
\orcid{0000-0002-8243-0543}
\affiliation{
  \institution{Department of Social Work, National University of Singapore}
  \streetaddress{21 Lower Kent Ridge Road}
  \city{Singapore}
  \country{Singapore}
  \postcode{119077}
}
\author{Yunan Li}
\email{liyunan@u.nus.edu}
\orcid{0009-0004-6162-4172}
\author{Yi-Chieh Lee}
\email{yclee@nus.edu.sg}
\orcid{0000-0002-5484-6066}
\affiliation{
  \institution{Department of Computer Science, National University of Singapore}
  \streetaddress{21 Lower Kent Ridge Road}
  \city{Singapore}
  \country{Singapore}
  \postcode{119077}
}

\renewcommand{\shortauthors}{et al.}

\begin{abstract}
Qualitative coding is a demanding yet crucial research method in the field of Human-Computer Interaction (HCI).
While recent studies have shown the capability of large language models (LLMs) to perform qualitative coding within theoretical frameworks, their potential for collaborative human-LLM discovery and generation of new insights beyond initial theory remains underexplored. 
To bridge this gap, we proposed \texttt{CHALET}, a novel approach that harnesses the power of human-LLM partnership to advance theory-driven qualitative analysis by facilitating iterative coding, disagreement analysis, and conceptualization of qualitative data. 
We demonstrated \texttt{CHALET}'s utility by applying it to the qualitative analysis of conversations related to mental-illness stigma, using the attribution model as the theoretical framework. 
Results highlighted the unique contribution of human-LLM collaboration in uncovering latent themes of stigma across the cognitive, emotional, and behavioral dimensions. 
We discuss the methodological implications of the human-LLM collaborative approach to theory-based qualitative analysis for the HCI community and beyond.
\end{abstract}


\maketitle

\section{Introduction}

\begin{figure}
    \centering
    \includegraphics[width=0.8\textwidth]{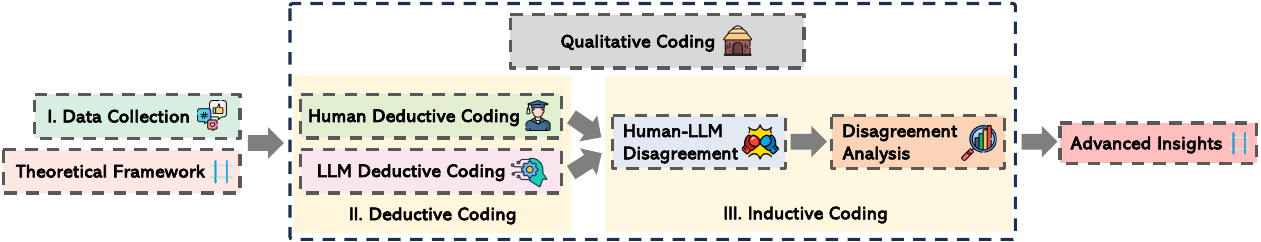}
    \caption{Overview of the \includegraphics[scale=0.004]{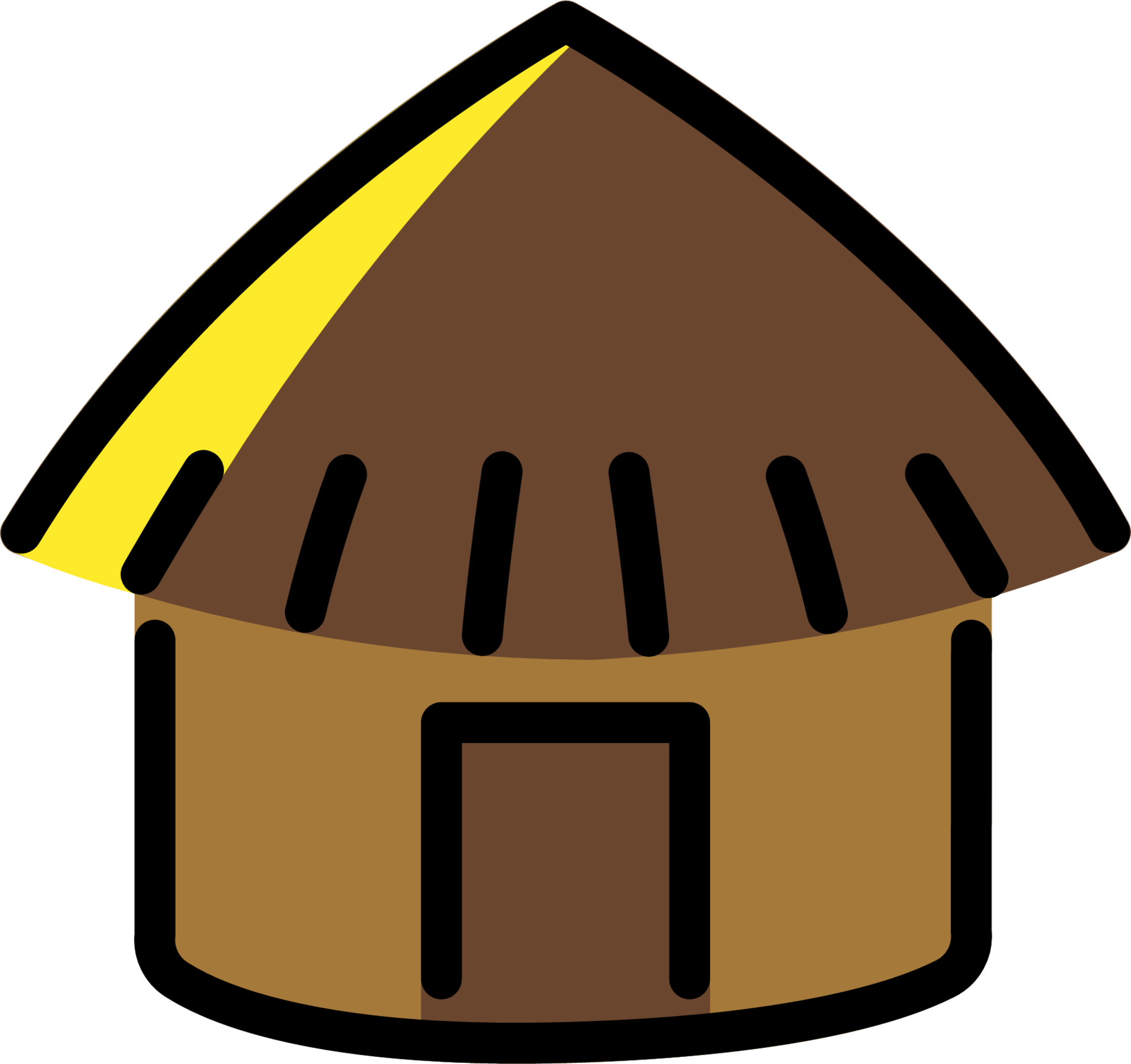} \texttt{CHALET} approach. In this work, we propose this approach to advance theory-driven qualitative analysis.}
    \Description{This figure provides ..}
    \label{fig:intro}
\end{figure}

Qualitative coding \cite{qualitative_method_adams_2008}, while commonly used in HCI research \cite{qualitative_hci_beneteau_2020, qualitative_quadri_2024, qualitative_hci_varanasi_2021, inductive_coding_practice_jardine_2024} to interpret human behavior \cite{new_attribution_ong_2020} as part of the analysis to guide technology development \cite{new_tech_extension_ameen_2022}, is an arduous process \cite{deductive_labelling_xiao_2023, qualitative_proscons_miles_1979}. 
The advent of LLMs, such as GPT-4\footnote{\url{https://chatgpt.com/}} and Llama \cite{llama2_touvron_2023}, offers the potential to support the qualitative-coding process by performing text analysis, consistently processing language, and simulating diverse analytical perspectives \cite{llm_for_text_analysis_meng_2025}.
In addition, leveraging the consistency of LLMs may help overcome some of the challenges of qualitative coding, such as replication difficulties \cite{human_coding_bias_leeson_2019} and fatigue-induced human errors \cite{bias_qualitative_collier_1996}.
Given these potential benefits, LLMs are increasingly used to facilitate deductive approaches of qualitative coding \cite{deductive_labelling_xiao_2023, llm_deductive_hou_2024, llm_deductive_kirsten} in which they can learn and apply theoretical frameworks\footnote{For simplicity, the term `theoretical framework' is used here in a broad sense. For finer distinctions between theoretical frameworks, theories, and hypotheses, see Muthukrishna \& Henrich's analyses \cite{theory_definition_muthukrishna_2019}.} and the associated human-generated codebooks to code qualitative data (texts). 

Although there has been a growing body of literature that examines the capability of LLMs in performing theory-driven qualitative coding \textit{within} existing coding schemes by simulating human coders \cite{llm_deductive_hou_2024, llm_deductive_kirsten, coding_deductive_llm_tai_2024}, their potential to function as a tool to collaborate with human coders to identify data patterns and extend theoretical constructs \textit{beyond} the initial coding frameworks remains largely unexplored. 
In fact, existing research primarily focuses on replicating human coding with LLMs, quantifying LLM-human coder \textit{agreement} \cite{deductive_labelling_xiao_2023, coding_deductive_llm_tai_2024} to demonstrate LLMs' potential as \textit{human coder replacements}. 
Yet this emphasis overshadows the value of having LLMs as \textit{collaborators}. 
Specifically, we propose and examine the novel process by which analyzing disagreements between LLMs and human coders could support collaborative, human-LLM inductive analyses that refine and enrich the codebook and its use. 
This represents a fundamental shift in perspective, changing the view of disagreements from problems to be resolved to opportunities for theoretical discovery.

We name the proposed approach \includegraphics[scale=0.004]{1.png} \texttt{CHALET} (\textbf{C}ollaborative \textbf{H}uman-LLM \textbf{A}na\textbf{L}ysis for \textbf{E}mpowering \textbf{T}heory-driven qualitative coding). 
Figure \ref{fig:intro} illustrates the major steps in \texttt{CHALET}, in which LLMs play a critical role: 1) LLM deductive coding, 2) identification of human-LLM coding disagreements, and 3) human-LLM collaborative inductive coding. 
We assessed our approach with a case study of the attribution model of mental-illness stigma \cite{attribution_model_corrigan_2003}. 
This theory, which links causes attributed to behaviors with the stigma of mental illness, connects perceptions of responsibility to stigma-driven emotions and actions \cite{attribution_model_corrigan_2003}. 
Emerging concepts such as microaggression \cite{microaggression_stigma_gibson_2023} and mental-illness invalidation \cite{microaggression_scale_gonzales_2015} necessitate a deeper qualitative analysis of the attribution model to reflect contemporary social-stigma perspectives.

Our work contributes to the HCI community in several ways. 
Firstly, we introduced \texttt{CHALET}, a systematic approach for theory-driven human-LLM collaborative qualitative analysis. 
In our case study, we showed how HCI researchers can leverage human-LLM \textit{collaboration} to derive new theoretical constructs that neither perspective could achieve independently.
Second, we reframed \textbf{human-LLM coding disagreement} from misalignments to reconcile into an opportunity for theoretical discovery. 
Our disagreement analysis reveals how differences between human and LLM interpretations can expose hidden ambiguity in existing coding schemes and generate fresh theoretical insights.
Finally, based on our observations, we reflected on the practical implications of future technology-enhanced qualitative research methods and the epistemological considerations, benefits, and challenges of adopting LLMs in HCI research. 
We explored how agency is distributed between humans and LLMs in collaborative settings and discussed broader implications for how computational HCI approaches can productively engage with interpretive inquiry.

\section{Related Work}

\subsection{The Importance of Qualitative Analysis in HCI}

Qualitative analysis is an essential research methodology in various fields, with qualitative coding being one of its main techniques. 
Qualitative coding involves categorizing and labeling data to identify a broad spectrum of patterns, themes, and hypotheses. 
Deductive coding \cite{deductive_method_fereday_2006} applies pre-defined codes derived from existing theories (etic approach), while inductive coding (emic approach) \cite{new_attribution_hanafiah_2015} generates codes from the data itself; hybrid deductive-inductive coding \cite{traditional_fereday_2006} combines both approaches, starting with pre-defined codes and allowing new codes to emerge from the data concurrently. 
Qualitative analysis plays a crucial role in understanding intricate human cognition, behavior, and other psychological constructs, allowing researchers to gain deep insights into the thoughts, emotions, and motivations that drive human actions \cite{qualitative_proscons_miles_1979}. 
This approach is valuable in the field of HCI, where understanding the intricacies of digital interaction and user behavior is essential for designing interfaces and systems \cite{hci_qualitative_coding_method_lazar_2017}. 

However, coding, categorizing, and making meaning from qualitative data is inherently challenging, given its heavy reliance on human understanding and interpretation of conversations, texts, or behaviors to map onto often complex social and psychological constructs \cite{positionality_holmes_2020}. 
The recent proliferation of generative AI technologies \cite{ai_qc_feuston_2021, ai_qc_jiang_2021, ai_qc_muller_2016, ai_qc_rietz_2021, ai_qc_yan_2014}, natural language processing (NLP) \cite{computer_deductive_nelson_2021, nlp_qc_marathe_2018}, and LLMs \cite{llm_inductive_jakub_2023} are beginning to show promise in supporting these research activities by helping researchers consistently read and process data and rapidly generate diverse analytical viewpoints \cite{llm_intelligent_sartori_2023}.

\subsection{LLM-assisted Text Classification}

Building on these capabilities, LLMs have begun to gain traction in various research domains, including education, sociology, psychology, and other social sciences \cite{chatgpt_edu_prather_2023, llm_psycho_demszky_2023}.
Recently, a trend in research has been to explore the potential of LLMs as textual data annotators \cite{llm_multilingual_labelling_steve_2024} and classifiers \cite{classify_llm_sun_2023}. 
Studies comparing LLMs with human annotators have found that in many tasks \cite{llm_label_method_gilardi_2023, labeling_llm_aldeen_2023}, LLMs can reach human expert-level performance \cite{labeling_llm_kaikaus_2023} even in low-resource scenarios \cite{labeling_llm_rouzegar_2024}, outperforming traditional pre-trained models while requiring less data \cite{labeling_llm_nasution_2024}.

Specifically, they have outperformed crowdsourced annotations from platforms like Amazon Mechanical Turk in objective text-classification tasks like relevance determination and topic detection \cite{llm_label_method_he_2024}, and surpassed fine-tuned models in automatic genre identification \cite{label_llm_kuzman_2023} and zero-shot misinformation labeling \cite{llm_label_misinfo_hoes_2023}.
Their capabilities extend to various types of text, helping alleviate human annotation burden \cite{labeling_llm_gray_2023} across political content \cite{label_llm_politic_twi_heseltine_2024}, legal documents \cite{labeling_llm_savelka_2023, labeling_llm_prasad_2024, labeling_llm_azad_2023}, health-related posts (e.g., COVID-19 vaccination-related data) \cite{llm_label_tekumalla_2023, labeling_llm_guo_2024}, and financial news \cite{labeling_llm_uymaz_2023}.
More impressively, in social-annotation tasks that require interpreting social norms, cultural context, language subtleties, and innate human elements (e.g., emotional tones and behavioral indicators), LLMs continue to show promise \cite{labeling_llm_dubourg_2024}.
They have performed well in areas challenging even for humans, such as detecting implicit hate speech \cite{label_llm_huang_2023}, stance detection \cite{labeling_llm_liyanage_2024}, online toxicity \cite{labeling_llm_ravi_2023}, and emotional valence \cite{labeling_llm_maceda_2023}. 
Previous studies have also demonstrated that 
LLMs can classify psychosocial content, including personality traits \cite{chatbot_various_task_amin_2023}, sentiment \cite{labeling_llm_fu_2024}, suicide tendencies \cite{chatbot_various_task_amin_2023}, mental disorders \cite{labeling_llm_nowacki_2024}, social determinants \cite{labeling_llm_ramachandran_2023}, and propaganda \cite{labeling_llm_hasanain_2024}, without specialized training.

Current research has made significant progress in using LLMs for text classification. 
However, these efforts take a relatively exploratory approach. 
Theoretically grounded qualitative analysis, on the other hand, requires a more rigorous definition and mapping of social and psychological constructs to linguistic features and labels. 
These mappings can be difficult even for human coders, who often require considerable training before they can consistently perform the tasks. 
Existing studies that prioritize conceptualization-agnostic classifications risk underutilizing the data's innate theoretical value and complex intrinsic patterns.
Furthermore, emphasizing classification performance and end-state \textit{results} while neglecting the interpretive \textit{process} may limit insights into how, when, and why these theoretical constructs develop \cite{theory_boyd_2021}, potentially diminishing the overall theoretical contribution of the research.
To remedy this situation, \textit{qualitative coding} \cite{hci_qualitative_coding_method_lazar_2017} is well-poised to compensate for the lack of theoretical groundwork and in-depth analysis. 

\subsection{Qualitative Coding with Technology Support}

\subsubsection{LLM-supported Deductive Coding}

When performing theory-driven deductive coding, human coders typically adhere to a holistic codebook to ensure that the coding is aligned with the pre-defined characteristics or constructed concepts being applied \cite{coding_deductive_llm_tai_2024}. 
Manual methods of deductive coding, such as iterating between refining the codebook and applying codes to the data, present scalability challenges when working with large datasets, where maintaining coding consistency becomes increasingly difficult as coder fatigue and careless errors may accumulate \cite{manual_code_mckibben_2022}.
Accordingly, technological advances have ushered in a variety of methods to support deductive coding and content analysis \cite{ai_qc_feuston_2021, ai_qc_jiang_2021, computer_deductive_baden_2020}, including rule/dictionary-based approaches \cite{computer_deductive_nelson_2021, ai_qc_crowston_2010}, machine learning \cite{ai_qc_muller_2016, ai_qc_rietz_2020}, and NLP \cite{ai_qc_guetterman_2018}.
For instance, supervised learning approaches, such as SVM, have been utilized to analyze leadership behaviors \cite{ai_qc_liew_2014}.

The advent of LLMs has further promised efficiency and consistency by allowing the analysis of a larger, more diverse array of datasets and the detection of language patterns.
Several studies \cite{deductive_labelling_xiao_2023, prompt_practice_dunivin_2024, llm_deductive_kirsten, coding_deductive_llm_tai_2024} have reported fair to substantial agreement between human experts and LLMs when using expert-developed codebooks. 
The LLM-assisted approach has been applied to multiple theoretical frameworks \cite{qualitative_research_llm_bano_2023, llm_deductive_bijker_2024}, such as Schwartz's theory of human values \cite{human_value_theory_schwartz_2012} and the Theoretical Domains Framework \cite{tdf_cane_2012}. 
Specifically, LLMs have demonstrated their versatility in various analytical contexts, handling children's speech \cite{deductive_labelling_xiao_2023}, historical media representations \cite{prompt_practice_dunivin_2024}, business communication \cite{llm_deductive_omizo_2024}, and technology usage patterns \cite{llm_deductive_kirsten}. 
Applications in the field of education have been particularly fruitful, with researchers applying LLMs to classify discourse in collaborative learning environments \cite{llm_deductive_siiman_2023}, assess literary analysis skills \cite{llm_deductive_zhang_2024}, and code students' knowledge-building processes \cite{llm_deductive_hou_2024}.
Moreover, Hou et al. \cite{llm_deductive_hou_2024} found that prompt-engineering techniques can further improve agreement between human expert-coded and LLM-coded data.

While these studies have evaluated LLMs' proficiency to code in alignment with humans, less emphasis has been placed on the factors that lead to coder disagreement. 
Traditionally, analyzing the reasons for these disagreements is crucial \cite{disagreement_zade_2018} to generate theoretical insights by facilitating collaborative data reinterpretation and allowing human coders to reconsider the assigned codes. 
Only a few studies \cite{ncoder_zambrano_2023, deductive_labelling_chew_2023} have recognized the value of understanding and utilizing such disagreement to refine the codebook accordingly. 
Results from these studies not only suggested that analysis of disagreement could clarify code definitions but also hinted that LLMs have the potential to work in tandem with human researchers \cite{creativity_llm_mirowski_2023}. 
This kind of human-LLM collaboration can be beneficial in several ways, such as by helping human researchers discover, define, and conceptualize new constructs. 
To further explore this understudied area, we propose an approach that leverages the systematic analyses of human-LLM disagreements to derive new theoretical insights through collaborative qualitative coding.

\subsubsection{LLM-supported Inductive Coding}

Inductive coding is a bottom-up method by which researchers interpret raw text-based data to identify patterns and construct ideas, independent of pre-existing conceptualizations or hypotheses \cite{codebook_method_boyatzis_1998}. 
Initially, the inductive coding process involved manually finding observations in the data, deriving patterns, forming hypotheses, and finally conceptualizing the phenomenon \cite{traditional_coding_procedure_richards_2017}. 
Subsequently, several computer-assisted methods have emerged to facilitate inductive coding for researchers \cite{computer_find_codes_nelson_2020, human_coding_bias_leeson_2019, ai_inductive_gebreegziabher_2023}, including platforms such as Atlas.ti \cite{atlas_sas_2016}, MaxQDA\footnote{\url{https://www.maxqda.com}}, nVivo\footnote{\url{https://lumivero.com/products/nvivo/nvivo-product-tour/}}, and nCoder \cite{ncoder_zambrano_2023}. 
In addition, various AI/NLP-based techniques \cite{inductive_labelling_hingle_2023, ai_qc_jiang_2021, ai_inductive_hong_2022} have been developed to aid in this process.
For example, unsupervised machine learning methods, such as topic modeling \cite{ai_qc_baumer_2017, ai_qc_eads_2021, tradition_computer_inductive_lennon_2021, ai_inductive_bakharia_2016, ai_inductive_hong_2022} and clustering algorithms (e.g., $k$-means \cite{computer_deductive_nelson_2021} and HDBSCAN \cite{traditional_nlp_inductive_parfenova_2024}), have been applied to help code social concepts like inequality in news coverage.

The emergence of LLMs has further shown promise as a viable approach alongside conventional AI/NLP approaches \cite{inductive_labelling_gamieldien_2023, llm_inductive_sinha_2024, llm_inductive_yan_2024, llm_inductive_morgan_2023}. 
When evaluated against human-led coding on identical datasets, LLMs demonstrated comparable performance in inferring most of the key themes \cite{llm_inductive_mathis_2024, llm_inductive_barany_2024, llm_inductive_perkins_2024, inductive_labelling_gamieldien_2023}.
LLMs can support most steps of such an inductive process, from generating initial codes \cite{content_analysis_coding_toolgpt_gao_2023, llm_inductive_zenimoto_2024, llm_inductive_lam_2024} and searching for themes \cite{llm_inductive_jakub_2023, llm_inductive_lee_2024, llm_inductive_fuller_2024} to reviewing and defining those themes \cite{llm_inductive_paoli_2023, inductive_labelling_dai_2023, llm_inductive_khan_2024}.
Specifically, in the initial coding phase, LLMs excel at generating descriptive codes \cite{llm_inductive_bryda_2024} and offering on-demand code suggestions \cite{content_analysis_coding_tool_gao_2023}. 
For theme identification, they can cluster responses \cite{llm_inductive_islam_2024, llm_inductive_singh_2024, llm_inductive_turobov_2024} and extract essential commonalities from codes \cite{llm_inductive_lee_2024}. 
During the review and definition stages, LLMs help refine codes \cite{llm_inductive_lopez_2024} and critique codebooks \cite{llm_inductive_nguyen_2024, llm_inductive_khan_2024}. 
For instance, Deiner et al. \cite{llm_inductiv_deiner_2024} showcased the LLM's ability to code health-related corpora inductively.

While studies show that LLMs can help suggest codes and themes from qualitative data, such as interview transcripts, their utility may be limited with large datasets. 
Previous approaches involved feeding large volumes of data directly into LLMs, without a structured methodology, which might be constrained by LLMs' input length limitations \cite{llm_inductive_paoli_2023}. 
In such situations, human researchers may be unable to make full use of LLMs or may encounter higher technical barriers when working with these models.
In addition, previous studies have primarily focused on distinguishing relatively overt topics \cite{llm_inductive_morgan_2023}, but identifying inner attributions, tacit knowledge, and socio-psychological constructs requires a deeper level of social awareness, which presents a more challenging undertaking that has not been extensively explored.
Thus, our paper explores new ways of human-LLM collaboration to help address both the data-quantity burden and the increased analytical complexity of interpreting such intricate social issues.

\subsubsection{Hybrid Deductive/Inductive Coding Approach}

The hybrid approach \cite{hybrid_proudfoot_2023, hybrid_xu_2020, traditional_enrich_hamad_2016, hybrid_bowe_2020, hybrid_swain_2018} integrates deductive coding using a priori templates with inductive coding that allows new themes to emerge from the data. 
This method begins with the development of a codebook based on theoretical frameworks, which is then applied to the text to identify meaningful units. 
As coding progresses, coders assign inductive codes to data units describing new themes observed in the data, expanding or separating from pre-determined codes. 
Through iterative analysis, codes are linked to discover patterns, with themes clustered and assigned succinct phrases to capture underlying meanings, ultimately leading to a conceptualization grounded in both theory and raw data \cite{traditional_fereday_2006, theory_driven_bonikowski_2022}.
This method is an indispensable and integral part of theory-driven qualitative analysis. 

Although LLMs have been used tentatively in deductive and inductive coding, their use in the hybrid coding approach remains largely underexplored. 
To date, no research has incorporated LLMs or other AI techniques into the process of hybrid qualitative analysis. 
This gap highlights the need for further exploration to address this unknown, as it may lead to overlooking the full scope of human-LLM collaboration in facilitating \textit{theory-driven} concept induction and extension.
Our research aims to address this gap by introducing a partnered human-LLM approach that provides LLM assistance in a hybrid manner, merging deductive and inductive coding. This contribution enhances technology-supported qualitative analysis.

\begin{figure}
    \centering
    \includegraphics[width=\textwidth]{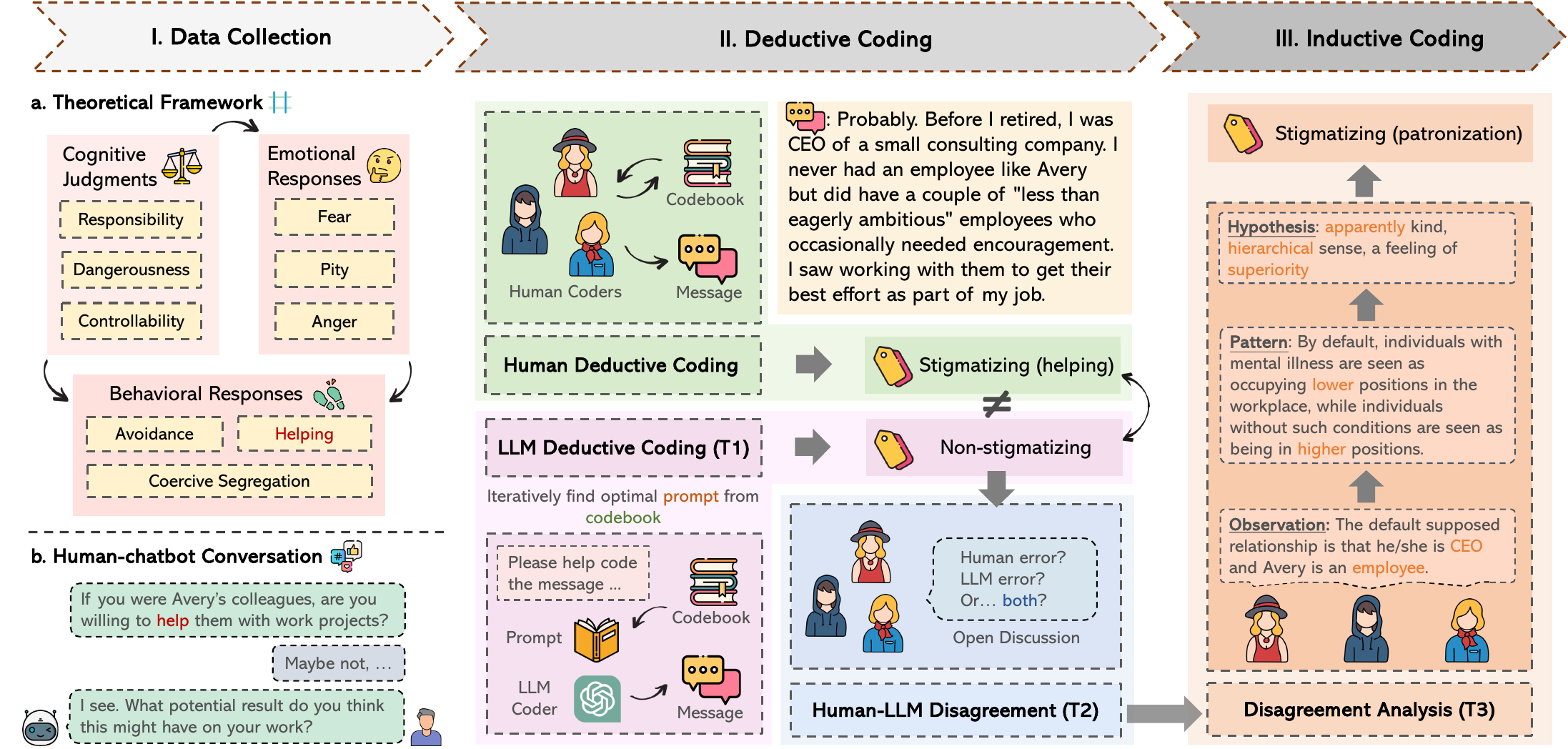}
    \caption{The \includegraphics[scale=0.004]{1.png} \texttt{CHALET} systematic approach: a case study on mental-illness stigma. \textbf{I. Data Collection}: \textbf{(a) Theoretical Framework}. The attribution model \cite{attribution_model_corrigan_2003} is selected as the theoretical underpinning. This model covers concepts of \textit{responsibility}, \textit{dangerousness}, \textit{controllability} (i.e., cognitive judgments), \textit{fear}, \textit{pity}, \textit{anger} (i.e., emotional responses), \textit{avoidance}, \textit{helping}, and \textit{coercive segregation} (i.e., behavioral responses). \textbf{(b) Human-chatbot Conversation}. Qualitative data is collected interactively from participants through chatbot-posed questions that align with the theoretical model. For example, participants are queried about their willingness to assist people with mental illness (corresponding to the \textit{helping} code). \textbf{II. Human-LLM Synergistic Deductive Coding}. Human coders use the attribution model to develop a codebook and code the response as \textit{Stigmatizing}. This codebook is then learned by LLMs (T1), which, in contrast, code the same response as \textit{Non-stigmatizing}. \textbf{III. Collaborative Inductive Coding}. Disagreements between human coders and LLMs (T2) inform further qualitative analysis, revealing a perceived CEO-employee relationship and an underlying assumption of a hierarchical social structure. This is hypothesized to indicate a sense of superiority, resulting in the final coding of \textit{Stigmatizing (patronization)} (T3).}
    \Description{This figure illustrates..}
    \label{fig:method}
\end{figure}

\section{\texttt{CHALET}: Collaborative Human-LLM AnaLysis for Empowering Theory-driven Qualitative Coding}

\includegraphics[scale=0.004]{1.png} \texttt{CHALET} is a hybrid qualitative coding approach that leverages the co-action between humans and LLMs. 
Figure \ref{fig:method} shows its major components. 
Specifically, \texttt{CHALET} has three major steps, and within these steps, there are three key tasks (T1, T2, and T3) that highlight the uniqueness of human-LLM collaboration. 
Below is an overview of these steps and their associated key tasks:

\begin{enumerate}
    \item \textit{Data collection based on theoretical framework.} 
    In the first step, a theoretical framework is selected by researchers to guide the analysis. Qualitative data is then collected under certain conditions designed to align with this framework. 
    \item \textit{Human-LLM synergistic deductive coding.} 
    In the second step, humans and LLMs perform synergistic deductive coding of the data. A codebook is usually developed by human coders to analyze the data qualitatively. Then, the LLMs learn from the human coders using various aspects of the codebook, such as code definitions and/or keywords, as well as the human-coded data. In this step, there are two key tasks {\bf (T1, T2)}:
    \begin{itemize}
    \item \textbf{T1:} Instruct LLMs to consistently apply the human-established codebook in deductive coding by identifying a prompt design that generates output closely aligned with human coders, thus avoiding unnecessary, biased discordance analysis.
    \item \textbf{T2:} Identify and categorize key mismatches between coding done by humans and LLMs, determining whether each represents a coding error or signals a potential new code.
    \end{itemize}
    \item \textit{Collaborative inductive coding.} 
    In the third step, disagreements between human coders and LLMs are analyzed and conceptualized. 
    The conceptualization, through human-LLM collaborative coding, may lead to the emergence of new codes and (sub-)themes. 
    In this step, one key task {\bf (T3)} is:
    \begin{itemize}
    \item \textbf{T3:} Reconcile human-LLM coding disagreements through human-LLM collaborative coding to develop novel theoretical constructs and generate qualitative insights.
\end{itemize}
\end{enumerate}

To illustrate our approach and highlight LLMs' roles, we present in Figure \ref{fig:method} (and throughout the paper) a \textit{case study} of qualitative-data analysis grounded in the attribution model -- a theoretical model of mental-illness stigma \cite{attribution_model_corrigan_2003}.
Human coders and LLMs perform deductive data coding to determine if a participant's message is considered \textit{stigmatizing} and, if so, how it can be coded into one of these categories based on the attribution model (T1). 
Codes from human coders and LLMs are compared, and disagreements will be identified (T2). 
These disagreements are used for collaborative inductive coding (T3), in which humans and LLMs work together to derive new stigma-related themes/codes and enhance the qualitative-analysis process.
We will describe each step and task in detail next.

\subsection{Data Collection}

As foundational models, LLMs require domain-specific data to make accurate inferences \cite{fewshot_brown_2020}. 
This characteristic makes targeted data collection essential for developing agentic LLMs that can support effective human-AI collaboration.
Our first step is to collect rich qualitative data from participants using methods such as interviews, focus groups, and human-chatbot interactions \cite{collect_data_asensio-cuesta_2021_obesity}, as this crucial step influences and shapes the subsequent analysis and results. 
A key principle of our approach is that each question designed to elicit responses from participants should be carefully curated to correspond to a specific construct or concept within the chosen theoretical framework, facilitating theory-driven qualitative analysis that generates structured, theoretically relevant data.

\subsection{Human-LLM Synergistic Deductive Coding}

After data collection, we conduct human deductive coding to develop a codebook. 
We subsequently devise a method to transform the human-generated codebook into a set of instructions that could be provided to the LLM, with the goal of improving the LLM's ability to perform deductive coding more accurately and in closer alignment with human coders.
In particular, this step entails both human and LLM coders coding the same set of messages, paving the way for in-depth comparison, further analysis of their coding results, and collaboration between human and LLM coders.

\subsubsection{Human Deductive Coding}

Initially, human coders engage in theory-informed  \textit{deductive coding} \cite{coding_manual_saldana_2016} of the collected data using the theoretical framework. 
The process begins by identifying key concepts from the theoretical framework and developing a preliminary codebook to maximize \textit{inter-rater reliability}\footnote{While we use inter-rater reliability to verify that our codebook can be applied consistently in a deductive manner, we recognize that optimizing for higher inter-rater reliability is not always appropriate for qualitative research in HCI, particularly in inductive or interpretivist work \cite{irr_mcdonald_2019}.} \cite{interrater_agreement_clarke_2023} and coding coherence. 
The codebook typically includes code names, definitions, qualifiers, exclusions \cite{deductive_method_fereday_2006}, and examples \cite{deductive_coding_method_neuendorf_2017}, serving as a guide throughout the coding process \cite{codebook_method_boyatzis_1998}. 
Human coders then code subsets of data and set checkpoints, allowing the codebook to evolve iteratively until satisfactory inter-rater reliability is achieved \cite{hci_qualitative_coding_method_lazar_2017}.
Following prior practices \cite{deductive_coding_example_warner_2019, deductive_coding_example_chopra_2021}, at least two coders participate in both the initial codebook development and its iterative refinement. 
The coders resolve disagreements at each checkpoint through open discussion with the main researcher. 
Whenever they need domain expertise to resolve any disagreement, they may consult a domain specialist.
After a sufficient amount of data has been coded, a stratified sampling re-check mechanism is used to enhance coding consistency and maintain the reliability of the coding process. 

\subsubsection{LLM Deductive Coding (T1)}

To instruct the LLM to code, we need to guide LLMs such that they are capable of dissecting the texts and constructs in ways that conform to the theoretical framework. 
To this end, the codebook, curated and finalized during the human deductive coding process, serves as a valuable resource for this guidance. 
Unlike previous work that either focuses on general text classification \cite{llm_label_method_gilardi_2023, labeling_llm_aldeen_2023} without theoretical grounding or shows that LLMs can perform deductive coding using relatively simple, ad-hoc prompts \cite{deductive_labelling_xiao_2023, prompt_practice_dunivin_2024, llm_deductive_kirsten, coding_deductive_llm_tai_2024}, we conduct a guided exploration of prompt design based on the human-produced codebook. 
With this focus, understanding the effects of different prompts provided to LLMs \cite{deductive_labelling_xiao_2023} on discrepancies with humans is crucial to our approach.
The main reason is that we want to find prompts that elicit deductive coding outputs from LLMs that most closely match those produced by well-trained human coders. 
Given that we do not know, a priori, which prompts are best for LLMs to perform deductive coding, our exploration of prompt design can provide useful information to contextualize the performance of LLMs in such complex socio-cognitive tasks. 
Results will also better align LLMs with those of humans before we identify their disagreements in the next step. 

To this end, we propose to first systematically investigate the effect of incorporating various codebook components, such as code definitions, keywords, coding rules, and examples, into the LLM instructions. 
Next, we explore whether providing additional components beyond the codebook enhances human-LLM coding agreement. 
This exploration guides us to evaluate combinations of codebook-derived components and supplementary details to assemble prompts that yield better LLM coding results. 
We also fine-tune the model configurations, textual structures, output formats (with or without short justifications for assigned codes), and output strategies (single output vs. majority voting from multiple outputs) to improve human-LLM agreement and enhance the interpretability of the results.

Once the inter-rater reliability between human and LLM coding has reached a reasonably high level, we proceed to code the remaining parts of the dataset not coded by humans using the LLM. 
This is followed by a sampling of these newly LLM-coded, previously human-uncoded data to verify that the consistency between human and LLM coding is maintained \cite{bias_labelling_llm_ashwin_2023} in order to create a validated dataset. 
This LLM-based procedure not only facilitates human-LLM disagreement identification in our approach but can generally increase our confidence that the LLMs are performing at a level comparable to trained human coders. 
In situations in which it is important to reduce human effort in large-scale qualitative analyses, this step can be done iteratively to validate how well the model is capable of performing (most of) the coding, potentially alleviating the workload of manual coding methods.

\subsubsection{Human-LLM Disagreement Analysis (T2)}

Disagreement analysis \cite{disagreement_zade_2018} is essential for our methodology. 
Instead of merely aiming to minimize the human-LLM disagreements \cite{ncoder_zambrano_2023, deductive_labelling_chew_2023}, our approach seeks to \textit{benefit} from them. 
Specifically, we show how these discrepancies can be used to identify areas in which theoretical constructs require improvement, thereby enriching the coding schemes.

Upon completion of both human and LLM deductive coding, each message in a subset of the data is associated with a human-assigned code, along with multiple LLM-generated codes produced using different prompts. 
We then identify a subset of messages for the disagreement analysis. 
While there are different ways to determine the selection criterion, for the purpose of demonstrating the general validity of our approach, we decided to adopt a relatively conservative criterion -- we only selected messages in which the LLM-generated codes from \textit{all} prompt variants differ from the human-assigned code for the same message. 
In other words, any message with at least one LLM code matching the human code was excluded from the disagreement discussion. 
Generally, a more conservative criterion will lead to a smaller subset of messages (i.e., less effort), while relaxing the criterion may lead to more messages for further analysis (i.e., more effort but potentially more improvement). 
We provided a detailed discussion of disagreement-selection criteria and their implications in Section \ref{sec:dis_discussion}.

Subsequently, all human coders independently review these \textit{human-LLM disagreement} and engage in open discussion to understand in what way they can provide information for further qualitative analysis. All coders are required to decide whether each human-LLM disagreement belongs to one of the following categories: 

\begin{itemize}
    \item \textbf{Human Coding Error}: During human deductive coding, coders made incorrect decisions.
    \item \textbf{LLM Coding Error}: Alternatively, the disagreement may stem from LLM limitations in interpreting context or nuance, leading to difficulties in coding certain messages, particularly those with complex or implicit patterns.
    \item \textbf{New Code}: It is also plausible that both humans and LLMs have erred; the human-LLM disagreement signals a novel concept that has not been accounted for in the current coding scheme.
\end{itemize}

If, after open discussion, the coders identify potential new codes, they will initiate a systematic induction to generate and conceptualize new concepts therein and to form umbrella themes through collaborative analysis with the LLMs.

\subsection{Collaborative Inductive Coding (T3)}
\label{sec:ind_method}

We reconcile LLM-human coding disagreements through inductive coding, where human coders and LLMs work in tandem to develop new codes and capture the content of disputed messages. 
Building upon manual hybrid coding \cite{hybrid_proudfoot_2023, hybrid_xu_2020, traditional_enrich_hamad_2016, hybrid_bowe_2020, hybrid_swain_2018}, this step facilitates the conceptualization of qualitative data by leveraging human expertise in socio-psychological understanding and LLMs' capabilities to rapidly generate code suggestions from messages and synthesize theme proposals from codes \cite{llm_inductive_paoli_2023}.
We involve three human coders to facilitate triangulation and follow a two-cycle coding procedure \cite{coding_manual_saldana_2016}: first for descriptive capture, followed by theoretical formation.

To align with previous practices \cite{inductive_coding_practice_jardine_2024, inductive_coding_practice_xiao_2023}, three coders first independently review the messages involved in the human-LLM discrepancies. 
They jot down notable observations and recurring patterns, listing each interesting disagreement in a spreadsheet, along with relevant excerpts and their initial interpretations. 
These individual spreadsheets are then merged for collective review. 
In open discussions, coders share their observations about the commonalities among the disagreement messages, progressively evolving their understanding. 
When a new pattern is identified that goes beyond the original coding scheme, coders work together, aided by LLM suggestions, to assign a preliminary new code name to capture this emerging concept. 
This iterative process continues until all messages with human-LLM coding discrepancies have been reviewed. 
Subsequently, coders revisit and revise these new codes, collapsing the initial codes into fewer, more analytic constructs. 
Further discussion and affinity diagramming \cite{affinity_diagramming_harboe_2015} are conducted to synthesize these concepts, organize them into sub-themes, and develop overarching themes to elucidate LLM-human discordance. 
These newly discovered (sub-)themes and their associated codes offer fresh insights that enrich our understanding of the social construct under study.

In the next section, we will present a case study that illustrates how we applied this method. 
We aim to show that analyzing disagreements between humans and LLMs could offer distinct advantages.
To preview our results, we found that LLMs, trained on a broad spectrum of global knowledge and characterized by their semantic-driven, rule-abiding nature, can help human coders think outside the box, identify controversial gray areas that may be overlooked or variably classified due to subjectivity, and consistently apply coding rules, potentially introducing new analytical views into the theory-driven qualitative analysis process.

\section{Case Study: Advancing Theory-driven Qualitative Analysis of Mental-illness Stigma}
\label{sec:casestudy}

\subsection{Data Collection}
\label{sec:datacollection}

To demonstrate how to collect theory-informed qualitative data for \texttt{CHALET} in practice, we decided to use human-chatbot conversations, as AI chatbots have proven effective for qualitative-data collection \cite{collect_data_alkoudmani_2023_health, collect_data_asensio-cuesta_2021_obesity, collect_data_casas_2018_food, collect_data_erazo_2020_covid} by promoting self-disclosure \cite{disclosure_lee_2022} through techniques like gamification and humanization \cite{game_pas_2020, humanization_rhim_2022} that transform survey questions into interactive communications, offering a middle ground between scalability and depth.
Here, however, we do not claim that \texttt{CHALET} is inherently constrained by the specific choices made in our case study, including data source and collection method, domain, theoretical framework, participant demographics, or language; rather, it should potentially extend to other data types and fields where data collection is guided by social theories, though such extensions would require future empirical validation.
Data collection for the case study ran between November 2023 and February 2024.

\subsubsection{Human-chatbot Conversation Design} 
\label{sec:conversation_cs}

\begin{figure}
    \centering
    \includegraphics[width=0.8\textwidth]{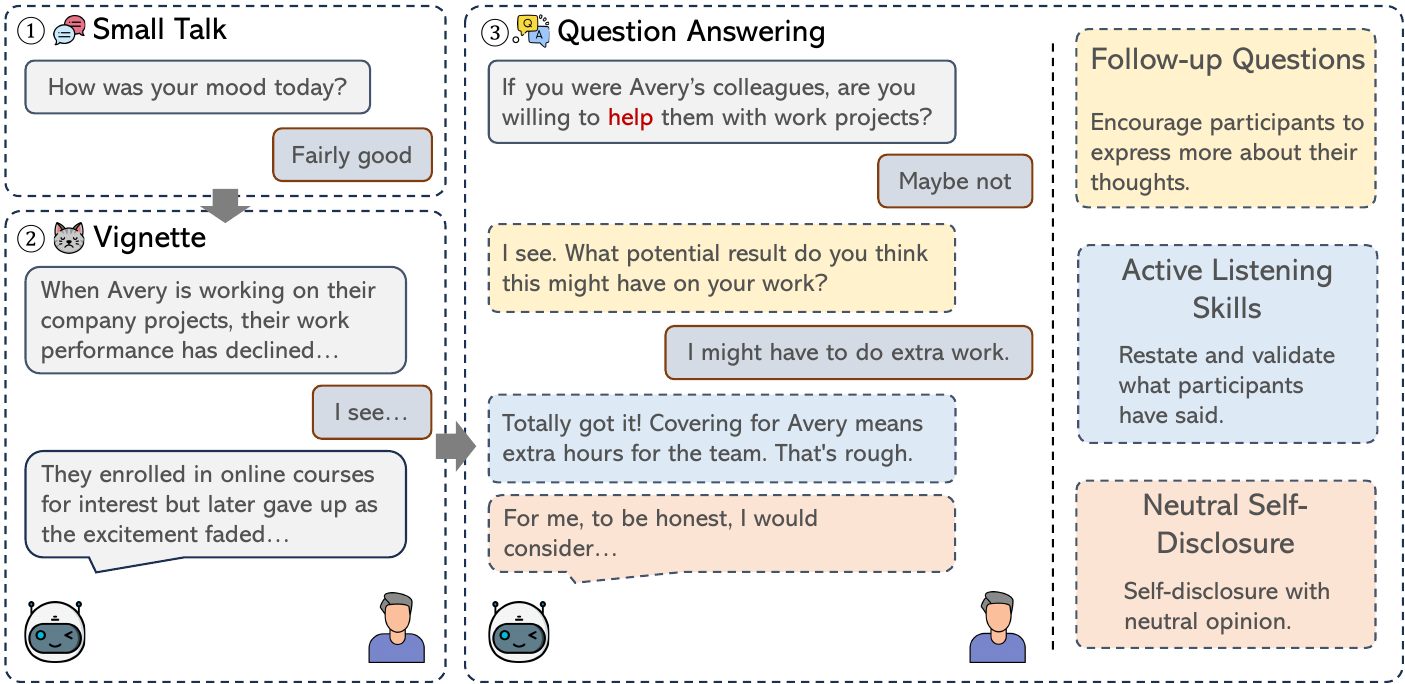}
    \caption{Conversation flow design. The conversation comprises three main components: \textbf{small talk}, \textbf{vignette delivery}, and \textbf{question answering}. In the question-answering phase, \textbf{follow-up questions} are used to gather additional information \cite{follow_up_q_han_2021}; \textbf{active-listening skills} are implemented to engage participants \cite{active_chatbot_xiao_2020}; \textbf{neutral self-disclosure} is applied to facilitate mutual disclosure \cite{disclosure_lee_2022} while maintaining neutrality.}
    \Description{This figure outlines..}
    \label{fig:ca}
\end{figure}

We designed our chatbot, \textit{Nova}, with a gender-neutral name and avatar.
The conversation began with a \textit{small-talk} session to build rapport. 
Next, Nova delivered a \textit{vignette} about Avery's experience with mental illness, specifically major depressive disorder, over several messages, intermittently prompting participants for brief responses. 
After the vignette, Nova posed \textit{open-ended questions} inviting participants to share their own feelings and opinions. 
These core questions, which assessed stigmatizing attitudes toward mental illness, were interspersed with a strategically placed \textit{mid-session break} to re-engage participants with lighthearted dialogue about hobbies and to alleviate potential emotional distress.
Three strategies were designed into the chatbot to facilitate participants' self-disclosure: \textit{follow-up questions}, \textit{active-listening skills}, and \textit{neutral self-disclosure} (see Figure \ref{fig:ca} and \textit{Supplementary Materials} for details).

\paragraph{\textcolor{darkred}{\textbf{Vignette Design.}}} 
\textit{Vignettes} \cite{vignette_link_2004}, which are brief, evocative stories, are effective tools for eliciting opinions and attitudes toward people with mental illness in chatbot-mediated social contact \cite{chatbot_reduce_kim_2020, chatbot_reduce_sebastian_2017, chatbot_aq27_practice_lee_2023}, as they provide detailed stimuli that may resonate with participants' experiences \cite{vignette_marwaha_2002}, while avoiding excessive emotionally charged situations \cite{vignette_gourlay_2014}. 
We designed a vignette based on the DSM-5 \cite{dsm5_apa_2013}, which focused on \textit{Avery}, a person experiencing depressive symptoms.
The vignette described Avery's symptoms without technical or medical jargon and did not explicitly specify Avery's gender or age. 
It illustrated how the symptoms negatively affected Avery's life, including work, relationships, family, and social interactions with strangers. 
We avoided including more extreme symptoms such as self-injurious behavior or suicidal ideation for ethical considerations. 

\paragraph{\textcolor{darkred}{\textbf{Conversation-question Design.}}}
The chatbot questions presented in Table \ref{tab:question} were adapted from the original attribution model \cite{attribution_model_corrigan_2003} and the Attribution Questionnaire-27 (AQ-27) \cite{aq27_corrigan_2012, aq_supplement_armstrong_2003}, which was developed based on the attribution model and operationalized concepts into measurable items (i.e., \textit{blame}, \textit{fear}, \textit{pity}, \textit{anger}, \textit{helping}, \textit{avoidance}, \textit{coercion}, \textit{segregation}, and \textit{dangerousness}). 
Following the practice of Lee et al. \cite{chatbot_aq27_practice_lee_2023}, we combined coercion-segregation and fear-dangerousness pairs to create a more concise conversation flow and designed seven chatbot questions corresponding to the attributions of \textit{responsibility}, \textit{fear}, \textit{pity}, \textit{anger}, \textit{helping}, \textit{social distance} (i.e., \textit{avoidance}), and \textit{coercive segregation}. 
All materials that emanated from the chatbot were reviewed and refined by a specialist (one of the co-authors with expertise in mental health) and a professional psychiatrist. 
To foster self-disclosure and reduce social desirability bias, we embedded questions in vivid, relatable scenarios using relatively indirect phrasing \cite{sd_indirect_q_fisher_1993}.
We also formulated one to two follow-up questions for each question, taking advantage of LLMs' strong eloquence and flexibility, to probe for underlying reasons, potential outcomes, or specific triggering scenarios.

\begin{table}[tbp]
\small
\caption{The chatbot question scripts.}
\label{tab:question}
\begin{tabularx}{\columnwidth}{p{0.74\columnwidth}X}
\toprule
\multicolumn{1}{m{0.74\columnwidth}}{\textbf{Question Scripts}} & \multicolumn{1}{m{0.24\columnwidth}}{\textbf{Attribution Types}}\\

\midrule
As mentioned in the story, Avery is facing difficulties in their relationships with colleagues and their work performance. Do you think Avery's current situation is primarily \textbf{a result of their actions}? Please share your thoughts. & Responsibility \\
\specialrule{0em}{2pt}{2pt}
Let's say you were one of Avery's friends at the party or other social event. Would you feel \textbf{anger} toward them because of their actions, like losing their temper and yelling at someone? & Anger \\ \specialrule{0em}{2pt}{2pt}
I'd like to know -- would you feel \textbf{concern} and \textbf{sympathy} for Avery's mental state? & Pity \\
\specialrule{0em}{2pt}{2pt}
If you were to travel or engage in activities with Avery, would you have any feelings of \textbf{fear} or \textbf{threat} regarding the potential consequences of your interaction? & Fear \\ \specialrule{0em}{2pt}{2pt}
I wonder if you were one of Avery's colleagues or friends, then would you be inclined to \textbf{help} them with tasks such as work projects? & Helping \\
\specialrule{0em}{2pt}{2pt}
Aside from being a colleague or a friend, I'm also curious, if you're Avery's neighbor, would you think it's better to \textbf{separate} them from the neighborhood and have them \textbf{hospitalized} for treatment? & Coercive Segregation \\
\specialrule{0em}{2pt}{2pt}
If you were in the process of selecting a \textbf{tenant} for your home, would you feel comfortable entrusting it to someone like Avery? & Social Distance \\
\bottomrule
\end{tabularx}
\end{table}

\begin{table}[tbp]
  \centering
  \small
  \caption{Participant Characteristics. \textbf{Mental illness experience} refers to whether participants had close friends or family members affected by mental illness.}
    \begin{tabular}{llc}
    \toprule
    \multicolumn{2}{c}{} & \multicolumn{1}{p{11em}}{\textbf{ALL (N=1,002) n (\%)}} \\
    \midrule
    \multirow{3}[2]{*}{\textbf{Mental Illness Experience}} & Yes   & 526 (52.5) \\
          & No    & 299 (29.84) \\
          & Maybe & 177 (17.66) \\
    \midrule
    \multirow{6}[2]{*}{\textbf{Age}} & 21-24 & 70 (6.99) \\
          & 25-34 & 234 (23.35) \\
          & 35-44 & 211 (21.06) \\
          & 45-54 & 155 (15.47) \\
          & 55-64 & 131 (13.07) \\
          & 65+   & 201 (20.06) \\
    \midrule
    \multirow{4}[2]{*}{\textbf{Gender}} & Male  & 540 (53.89) \\
          & Female & 460 (45.91) \\
          & Prefer not to say & 1 (0.1) \\
          & Prefer to self-describe: Trans Man & 1 (0.1) \\
    \midrule
    \multirow{6}[2]{*}{\textbf{Ethnicity}} & White & 695 (69.36) \\
          & Black or African American & 210 (20.96) \\
          & Asian & 47 (4.69) \\
          & Mixed & 26 (2.59) \\
          & American Indian or Alaska Native & 6 (0.6) \\
          & Other (Hispanic, Chicano, etc.) & 18 (1.8) \\
    \midrule
    \multirow{9}[2]{*}{\textbf{Education}} & Less than primary & 2 (0.2) \\
          & Primary & 24 (2.4) \\
          & Some secondary & 14 (1.4) \\
          & Secondary & 139 (13.87) \\
          & Vocational or similar & 104 (10.38) \\
          & Some college but no degree & 190 (18.96) \\
          & University -- Bachelor's degree & 328 (32.73) \\
        & \multicolumn{1}{p{17.915em}}{Graduate or professional degree (MA, MS, MBA, PhD, law degree, medical degree, etc.)} & \multirow{2}{*}{197 (19.66)} \\
          & Prefer not to say & 4 (0.4) \\
    \bottomrule
    \end{tabular}%
  \label{tab:participant}%
\end{table}%

\subsubsection{Participants and Procedure}

We amassed participants from diverse backgrounds in Western countries, mostly the United States and the United Kingdom, through two online platforms, Prolific and Qualtrics.
During recruitment, we clearly outlined the study's general scope and duration, and participants' right to withdraw at any point. 
Our inclusion criteria regulated that all participants must 1) be at least 21 years old, 2) speak English as their first language, 3) be willing to read material related to mental illness, and 4) have no immediate or urgent mental-health concerns.
The last criterion is incorporated because of the potential risk that vignettes about mental illness could cause distress to individuals grappling with such issues \cite{ethic_mental_illness_roberts_2002}. 
We recruited 1,002 participants, none of whom reported having an ongoing mental illness. 
We used IP addresses to filter out duplicate participants. 
The participants were relatively balanced in terms of gender (53.89\% male, 45.91\% female) and featured a broad age distribution, with the largest age groups being 25-34 (23.35\%) and 35-44 (21.06\%). 
Approximately half (52.5\%) had friends or family affected by mental illness. 
Detailed participant characteristics are presented in Table \ref{tab:participant}.

This one-time study took about 30 minutes to complete, and participants were paid approximately US\$6.30 for their time. 
We first obtained participants' consent and informed them that their data would remain confidential. 
Participants were forewarned that the study would involve a scenario related to mental illness and were given the option to withdraw if they felt uncomfortable. 
Next, participants were asked basic demographic questions. 
Participants then engaged in the human-chatbot conversation for approximately 20 minutes. 
A post-study debriefing explained the research objective of assessing stigma attitudes and informed participants about common types of mental-illness stigma. 

\subsubsection{Quality of Collected Data}
\label{sec:data_quality}

In total, we collected 7,014 messages, with each participant providing seven messages corresponding to the seven attributions under study. 
We measured the duration of the human-chatbot conversations and the word counts of the participants' contributions to them, with the former averaging $t=17.63$ minutes ($SD = 8.18$). 
The mean word counts per participant message, broken down by attribution, were 43.82 for \textit{responsibility} ($SD = 14.68$), 43.40 for \textit{social distance} ($SD = 15.49$), 41.50 for \textit{helping} ($SD = 14.20$), 40.86 for \textit{anger} ($SD = 13.45$), 40.18 for \textit{coercive segregation} ($SD = 13.70$), 39.74 for \textit{fear} ($SD = 14.14$), and 39.30 for \textit{pity} ($SD = 13.98$).
Our human coding of 4,200 messages from the first 600 participants showed that only 47 (1.12\%) were excluded due to being excessively brief (<5 words), incomplete, or off-topic. 
Word-frequency analysis further demonstrated that the collected messages were generally on-task, with high-frequency words closely aligned with their respective attributions -- for instance, ``\textit{hospitalization}" and ``\textit{treatment}" dominated messages about \textit{coercive segregation}, ``\textit{threatened}" and ``\textit{frightened}" were prominent in \textit{fear}-related messages, and ``\textit{support}" appeared frequently in \textit{helping}-related messages.

\subsection{Human-LLM Synergistic Deductive Coding}

\subsubsection{Human Deductive Coding} 
\label{sec:humandeductive}

The attribution model \cite{attribution_model_corrigan_2003} informed the human deductive-coding process of our case study. 
To clarify the terminology used in this paper: \textit{codes} are labels that assign symbolic meaning to each message. 
\textit{Attributions} are concepts derived from the attribution model \cite{attribution_model_corrigan_2003}, including \textit{responsibility}, \textit{social distance}, \textit{fear}, \textit{anger}, \textit{coercive segregation}, \textit{helping}, and \textit{pity}. 
Each code is generated to represent one of these attributions or as \textit{non-stigmatizing} if it does not match any of the stigma-related attributions. 
Three coders were recruited from the research team (two female and one male), two of whom had computer science backgrounds and one of whom specialized in the social sciences. 
Each had previous experience with qualitative coding.
Notably, we set the number and backgrounds of the human coders, the intervals at which agreement is checked, and the coding scheme as examples rather than fixed prerequisites of the \texttt{CHALET} framework; instead, we document these choices to illustrate how \texttt{CHALET} allows human coders to develop a coherent understanding of their data and theoretical constructs and produce a well-written codebook to instruct the LLM.


\paragraph{\textcolor{darkred}{\textbf{Procedure.}}}
We started with \textit{pilot coding}, followed by \textit{formal coding} with ongoing inter-rater reliability checks \cite{coding_manual_saldana_2016}. 
First, as a \textit{pilot} phase, two coders coded 70 messages (10 for each attribution) based on the attribution model \cite{attribution_model_corrigan_2003}, discussed the coded messages and disagreements, and created an initial codebook.


The coders then proceeded to \textit{formal coding}, in which a third coder was trained by the initial two coders. 
They randomly selected 700 messages from 100 participants, which two coders independently coded through frequent checkpoints: after coding 10 participants (Cohen's $\kappa =0.55$), then another 10 participants (Cohen's $\kappa =0.53$), followed by four sets of 20 participants each (Cohen's $\kappa =0.66$, 0.79, 0.76, and 0.72, respectively).
At each checkpoint, the coders discussed coding decisions and rationales to resolve discrepancies, added empirically derived coding rules \& specifications, revised definitions, and assigned keywords for each code.
The versioned codebook evolution is available in \textit{Supplementary Materials}.


Once achieving satisfactory inter-rater reliability (Cohen's $\kappa > 0.6$ \cite{hci_qualitative_coding_method_lazar_2017}), the coders stabilized the codebook and set up \textit{longer} checkpoints for minor refinements: coding of 50 participants (Cohen's $\kappa=0.74$), then another 50 participants (Cohen's $\kappa=0.66$), followed by 100 participants (Cohen's $\kappa=0.69$), and two sets of 150 participants each (Cohen's $\kappa=0.69$ and 0.66, respectively). 
After completing coding of 4,200 messages from 600 participants, coders verified coding consistency through stratified sampling, recoding 50 messages from the first 700 coded messages, 40 from the next 1,400 messages, and 10 from the final 2,100 messages. 
Upon completion, the coders consulted with the specialist to resolve any remaining disagreements, outputting both the human-generated codes and a developed codebook (see \textit{Supplementary Materials}).

\paragraph{\textcolor{darkred}{\textbf{Results.}}}
The derived coding scheme comprised eight codes: seven stigma-related codes (i.e., \textit{responsibility}, \textit{social distance}, \textit{fear}, \textit{anger}, \textit{coercive segregation}, \textit{helping}, and \textit{pity}) and one \textit{non-stigmatizing} code. 

\begin{figure}
\centering
\includegraphics[width=0.7\textwidth]{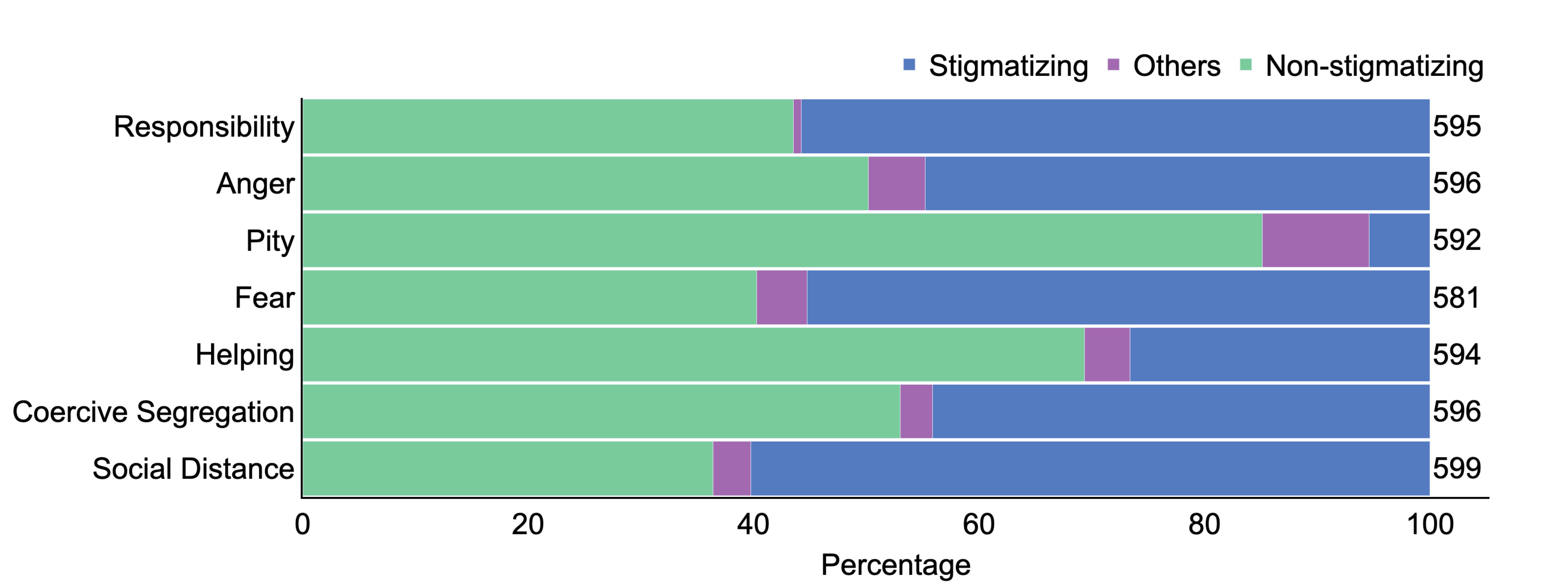}
\caption{Distribution of human-assigned codes across seven attributions, where each message was coded as \textit{Stigmatizing}, \textit{Non-stigmatizing}, or \textit{Stigmatizing (others)}. Codes were assigned through iterative deductive coding with inter-rater reliability checks (Cohen's $\kappa$ > 0.6), following the procedures described in Section \ref{sec:humandeductive}. See \textit{Supplementary Materials} for the complete codebook.}
\label{fig:deductive_info}
\end{figure}

Figure \ref{fig:deductive_info} shows the distribution of codes generated by human coders\footnote{For each message, human coders considered the specific attribution question (e.g., \textit{responsibility}) in Table \ref{tab:question} that elicited the message and then categorized it into one of three codes: 1) \textit{Stigmatizing} for that specific attribution (e.g., \textit{Stigmatizing (responsibility)}), 2) \textit{Non-stigmatizing}, or 3) \textit{Stigmatizing (others)} if the message showed stigma but related to other attributions. We consider this three-way classification potentially easier than having human coders select from all eight possible codes at once.}. 
Of the total 4,153 human-coded messages, 46.01\% ($n=$1,911) contain stigmatizing attitudes.
Among the stigmatizing codes, \textit{responsibility} was the most prevalent, accounting for 9.53\% ($n=395$) of all codes. 
\textit{Social distance} followed closely with 9.15\% ($n=379$), and \textit{fear} made up 8.86\% ($n=367$) of the total codes. 
\textit{Anger} and \textit{coercive segregation} constituted 7.19\% ($n=298$) and 6.57\% ($n=272$) of the codes, respectively. 
The least common stigmatizing codes were \textit{helping} with 3.81\% ($n=158$) and \textit{pity} with 1.01\% ($n=42$). 
Table \ref{table:ind} presents example messages for each \textit{stigmatizing} code. 
For \textit{non-stigmatizing} messages, one example came from P383, who strongly opposed coercive segregation and found hospitalization suggestions offensive:

\begin{quote}
   ``\textit{No, I think that is a deeply offensive and appalling suggestion. I think that is a strange question to ask. The real question is what right would anyone have to take her Avery away or hospitalise her against her will? She is harming no one.}" (P383)
\end{quote}

We calculated inter-rater reliability (Cohen's $\kappa$) for each attribution to verify our codebook's clarity.
Five attributions showed relatively high inter-rater reliability (Cohen's $\kappa \approx 0.70$): \textit{responsibility} (0.71), \textit{anger} (0.75), \textit{fear} (0.69), \textit{coercive segregation} (0.71), and \textit{social distance} (0.70).
\textit{Helping} (0.56) and \textit{pity} (0.36) showed lower inter-rater reliability, possibly due to their limited occurrence in the collected data -- only 158 out of 594 instances for \textit{Stigmatizing (helping)} and 88 out of 592 for \textit{Stigmatizing (pity)} -- preventing human coders from fully developing reflexive understanding between stigmatizing and non-stigmatizing attitudes.
The overall agreement was satisfactory (\textbf{Cohen's $\kappa = 0.71$}) \cite{hci_qualitative_coding_method_lazar_2017}, and recheck coding yielded Cohen's $\kappa$ of 0.69.
Throughout the human deductive coding, we output a list of code definitions, keywords, coding rules \& specifications, and example messages for LLMs to learn from.

\subsubsection{LLM Deductive Coding (T1)}
\label{sec:casestudy_llmdeduc}

To instantiate \texttt{CHALET}'s T1, in our case study, we explored and evaluated 23 prompt variants to instruct LLMs to apply the human-established codebook, maximize LLM-human agreement (as measured by comparing LLM-generated codes to existing human-assigned codes on 4,200 messages, represented by Cohen's $\kappa$), and reduce disagreement potentially caused by less-tuned LLM interpretations.

\paragraph{\textcolor{darkred}{\textbf{T1 Procedure.}}}
We systematically assess and operate two key elements of prompt design: \textit{components inside the codebook} and \textit{components not covered by the codebook}.

\textbf{Which components of the codebook substantially impact LLM-human coding agreement?} 
The human-written codebook consists of several components for each code: a \textit{code name}, several \textit{keywords}, \textit{coding rules \& specifications}, and one \textit{example}. 
We investigated whether these codebook components, along with other materials previously provided to human coders (i.e., the vignette as background knowledge, questions from Table \ref{tab:question}, and participants' messages), improved LLM-human agreement or caused potential information overload that reduced it.
Thus, we systematically steered the presence or absence of each component in the prompts and compared their effect on human-LLM coder agreement (Cohen's $\kappa$).

Furthermore, we compared LLM-human agreement under two distinct prompt scenarios: ``target-code" and ``all-code" prompts \cite{prompt_practice_dunivin_2024}. 
In the target-code scenario, the LLM was presented with seven separate tasks, one for each stigma attribution. 
For each task, the LLM was provided with codebook components for only the target-stigma code and the \textit{Non-stigmatizing} code and was asked to classify each message into one of three codes: the target-stigma code, \textit{Non-stigmatizing}, and \textit{Stigmatizing (others)} for all other-stigma codes. 
In the all-code scenario, the LLM completed a single task in which it was given codebook components for all seven stigma codes and the \textit{Non-stigmatizing} code and was required to classify the messages into one of these eight categories. 
The reason for this comparison is to test whether familiarity with the full codebook, especially the coding rules and examples of non-target codes, improves the LLM's performance in matching human coding decisions on the target code.
For example, when coding one stigma attribution (e.g., \textit{responsibility}), the LLM might only need to know how the codebook defines stigmatizing and non-stigmatizing cases of that attribution, with responses that fit neither definition automatically coded as \textit{Stigmatizing (others)}.

\textbf{Beyond the codebook, what additional components might influence LLM-human agreement?} 
To avoid limiting our exploration to codebook components alone, and inspired by research showing the benefits of in-context learning in language processing tasks \cite{icl_dong_2022}, we first investigated the effect of adding more human-coded examples to the prompt. 
These examples were carefully selected by human coders from existing human-coded messages. 
In addition, research has shown that the order of examples in prompts can affect LLM performance \cite{prompt_sensitivity_lu_2022}. 
Hence, we studied the impact of example order by adjusting the sequence of examples for three codes (i.e., \textit{Stigmatizing}, \textit{Non-stigmatizing}, and \textit{Stigmatizing (others)}) under the condition of having only target-code codebook components, as exhaustive permutation testing would be computationally expensive for all eight codes in the all-code scenario.
Finally, given the general effectiveness of the \textit{chain-of-thought} prompting strategy \cite{cot_wei_2022}, which breaks down complex reasoning tasks into sequential smaller steps, we explored whether incorporating explicit reasoning (e.g., ``\textit{The view that those with mental illness should change their perspective and consider others suggests self-help ability and personal responsibility.}'') would improve LLM's consistency with human coding.

We used \texttt{gpt-4-1106-preview} for all analyses. 
Our prompt design and model parameters encoded insights from prior practices \cite{prompt_practice_dunivin_2024, prompt_practice_he_2024, prompt_practice_hou_2024, prompt_practice_reiss_2023, prompt_practice_tornberg_2024, prompt_guidance_ziems_2024}, alongside our own empirical explorations (detailed in \textit{Supplementary Materials}), which included temperature setting tests (1, 0.5, and 0), LLM reasoning output for coding decisions, and majority voting with multiple LLM outputs (5 outputs).
We selected the following configuration: a temperature setting of 0, a role-playing approach in which the LLM was instructed to assume the role of ``\textit{a competent coder for mental-illness stigma}," a textual structure with \textit{instructions} (e.g., codebook components) preceding \textit{content} (e.g., the messages to be coded), and requesting the LLM to provide reasoning for its generated codes.
We opted to forgo adopting majority voting. 
Figure \ref{fig:prompt_structure} in Appendix \ref{app:prompt} shows our prompt structure.

This process improves LLMs' coding capabilities, enables them to assist in deductive coding, and allows subsequent disagreement analysis to more likely reflect genuine perspectival differences rather than instructional inadequacies.
Using the prompt with the highest LLM-human agreement, we had LLMs code previously human-uncoded messages to reduce human effort.
Given that prior research has shown that LLMs may introduce biases into qualitative coding, such as over-predicting certain codes due to training-data preconceptions \cite{bias_labelling_llm_ashwin_2023}, we randomly selected 25 messages (6.25\% of the human-uncoded messages) from each code for human review.

\paragraph{\textcolor{darkred}{\textbf{T1 Results.}}}
Although the specific findings about which prompt variants yield higher agreement are particular to this case study, our results can provide practical guidance on how \texttt{CHALET} as a methodological framework can help researchers effectively utilize LLMs in deductive coding, such as improving the consistency between human and LLM coders and understanding the general patterns and tendencies of LLMs when they deviate from human coders.

\begin{figure}[htbp]
\centering
\begin{subfigure}{0.499\textwidth}
    \includegraphics[width=\linewidth]{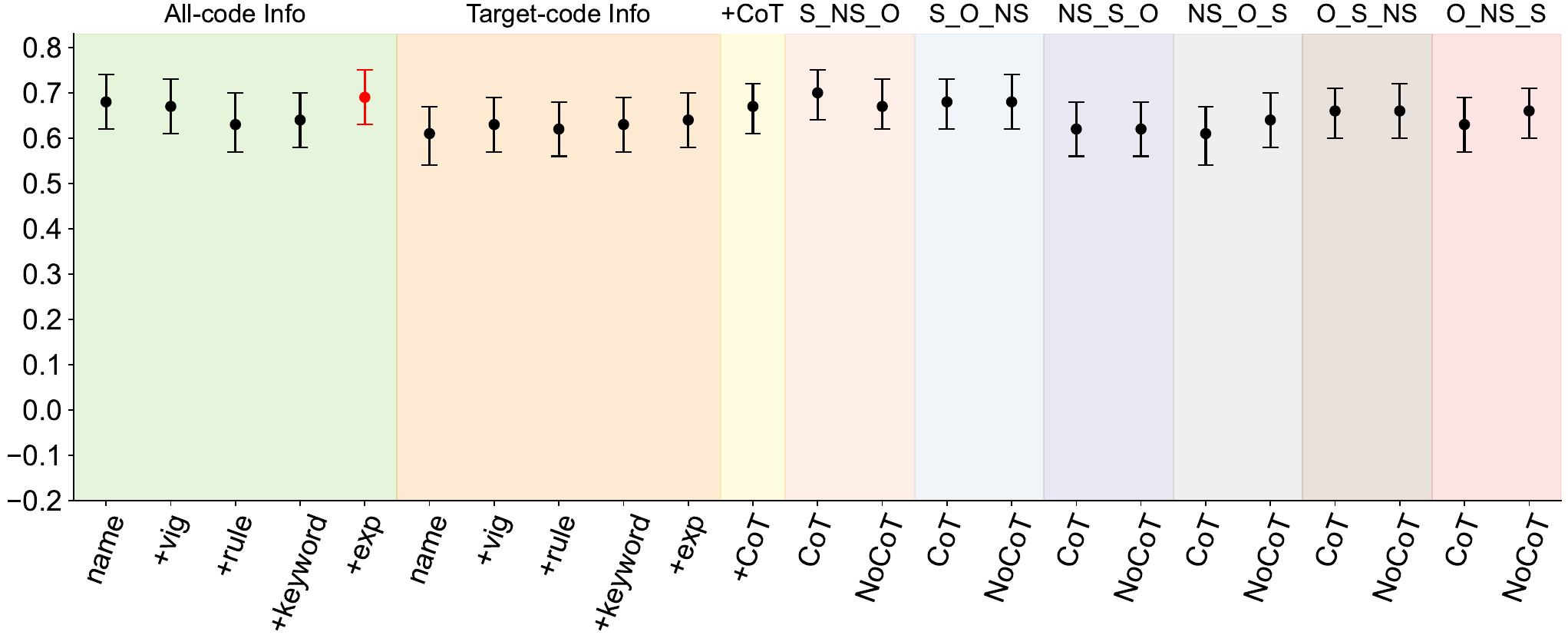}
    \caption{Responsibility}
\end{subfigure}\hfill
\begin{subfigure}{0.499\textwidth}
    \includegraphics[width=\linewidth]{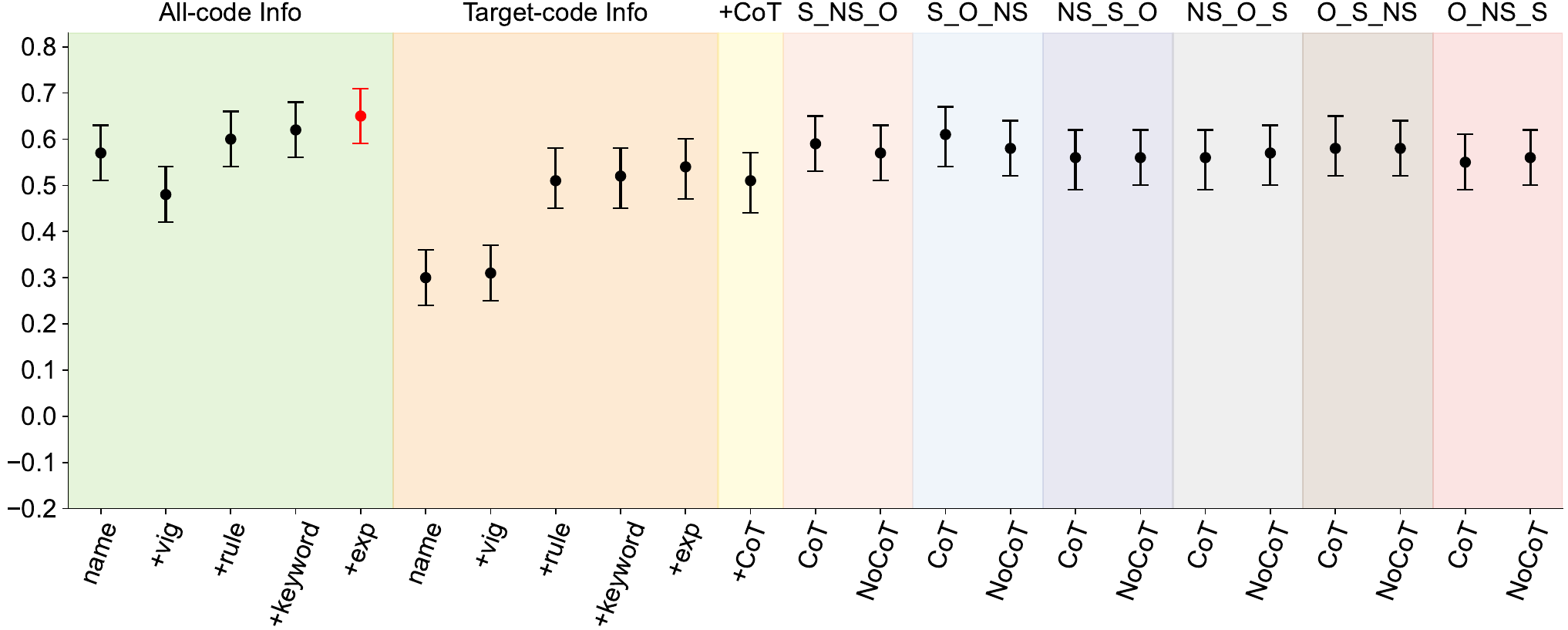}
    \caption{Anger}
\end{subfigure}

\vspace{1pt} 

\begin{subfigure}{0.499\textwidth}
    \includegraphics[width=\linewidth]{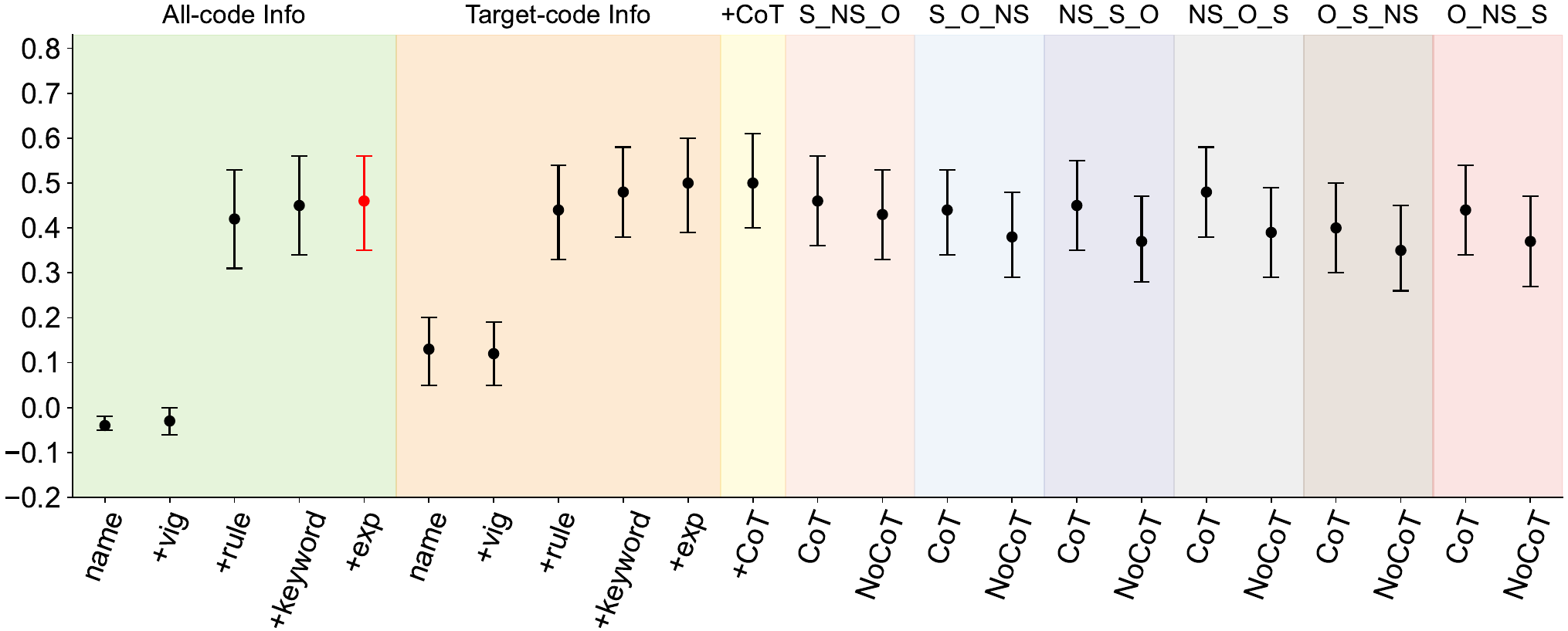}
    \caption{Pity}
\end{subfigure}\hfill
\begin{subfigure}{0.499\textwidth}
    \includegraphics[width=\linewidth]{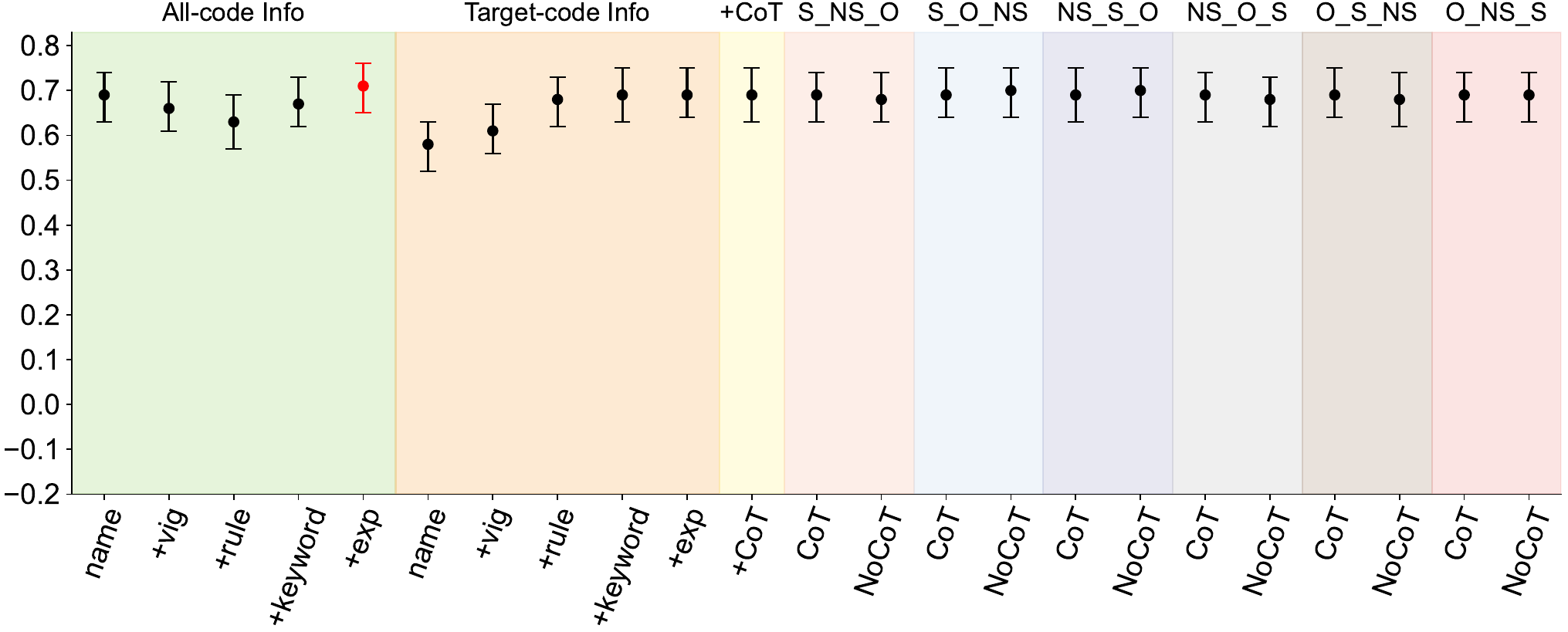}
    \caption{Fear}
\end{subfigure}

\vspace{1pt}

\begin{subfigure}{0.499\textwidth}
    \includegraphics[width=\linewidth]{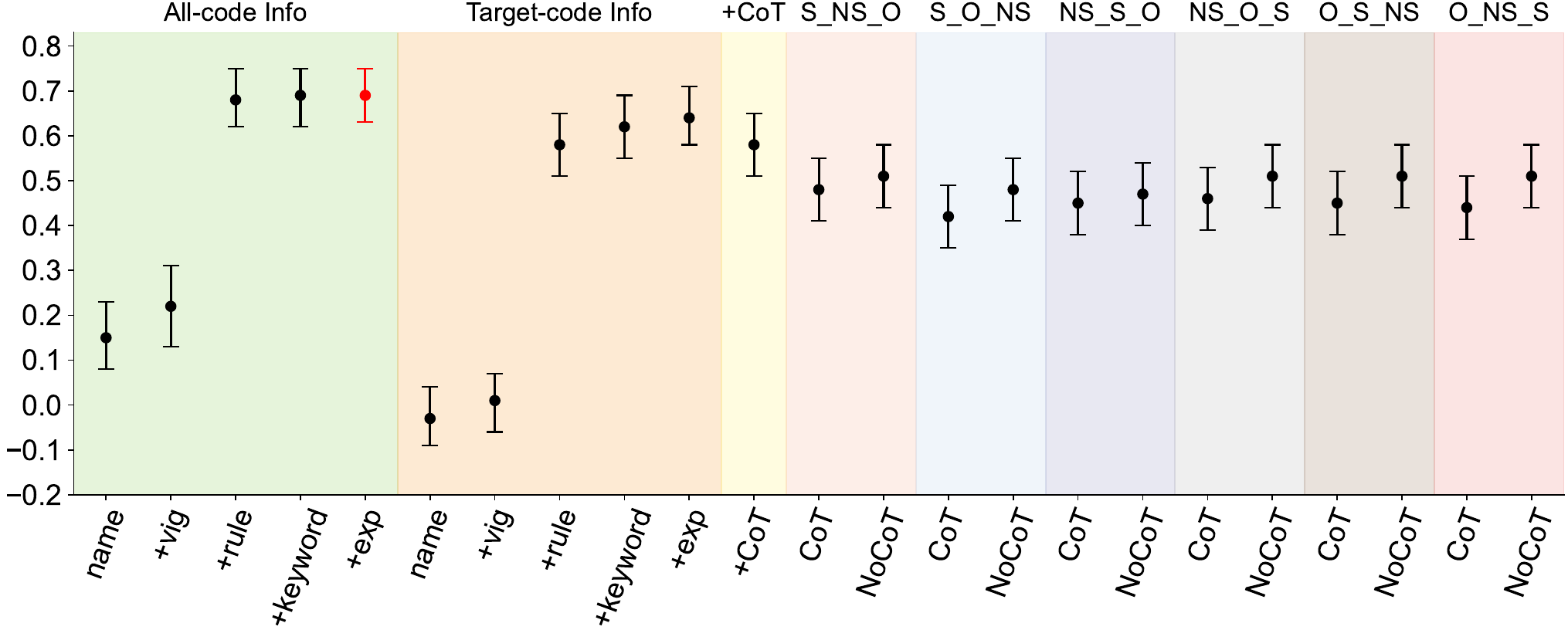}
    \caption{Helping}
\end{subfigure}\hfill
\begin{subfigure}{0.499\textwidth}
    \includegraphics[width=\linewidth]{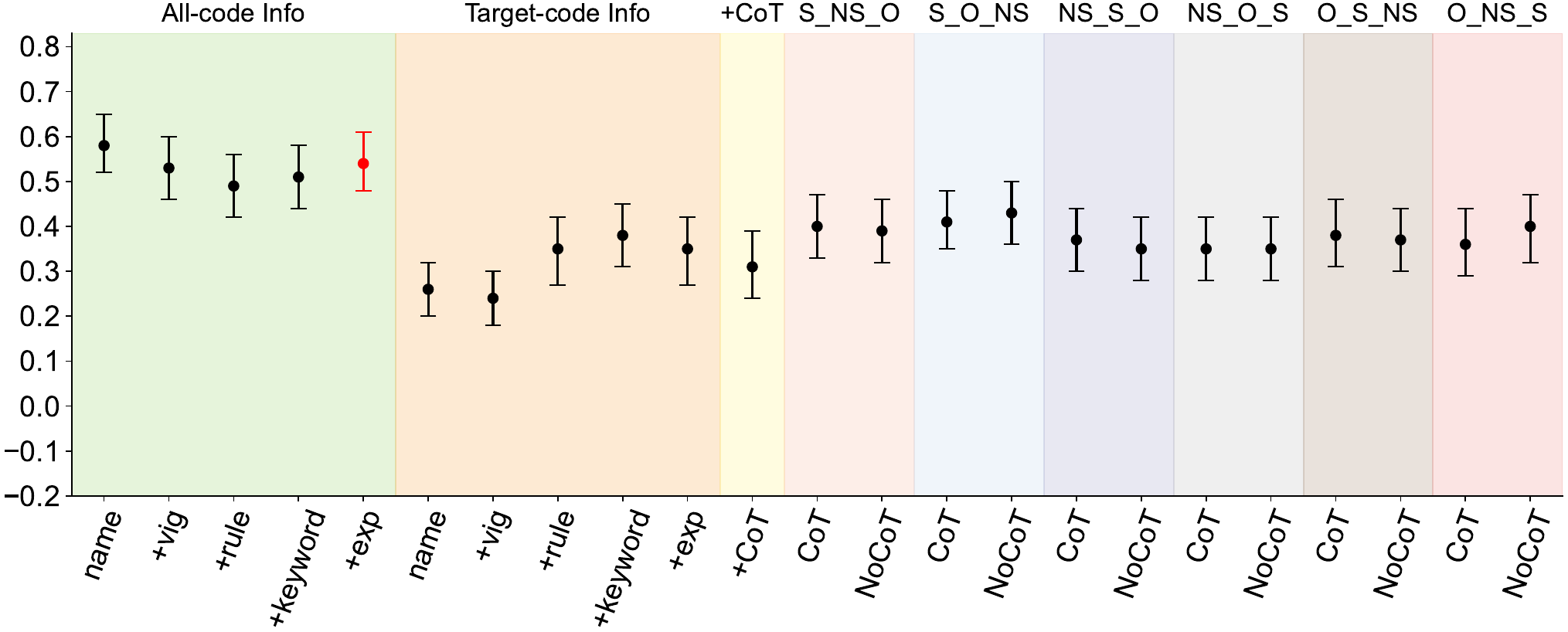}
    \caption{Coercive Segregation}
\end{subfigure}

\vspace{1pt}

\begin{subfigure}{0.499\textwidth}
    \includegraphics[width=\linewidth]{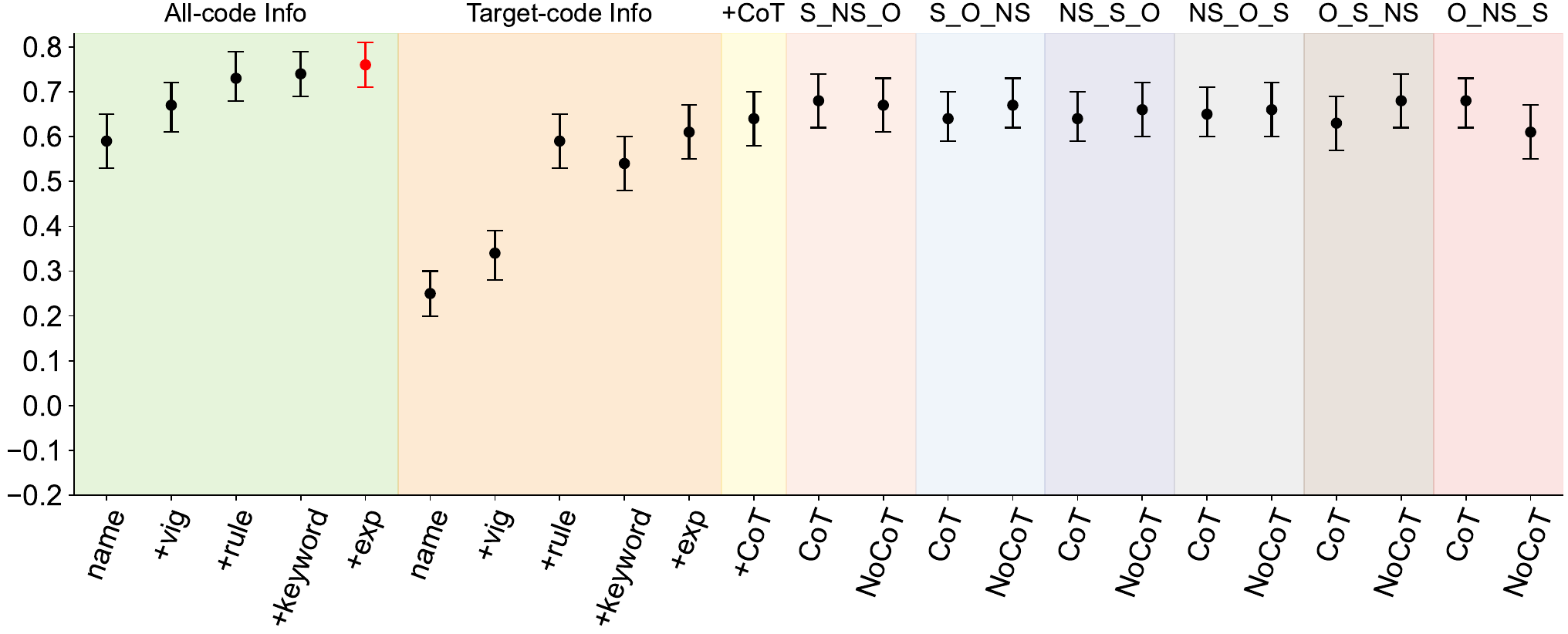}
    \caption{Social Distance}
\end{subfigure}\hfill
\begin{subfigure}{0.499\textwidth}
    \includegraphics[width=\linewidth]{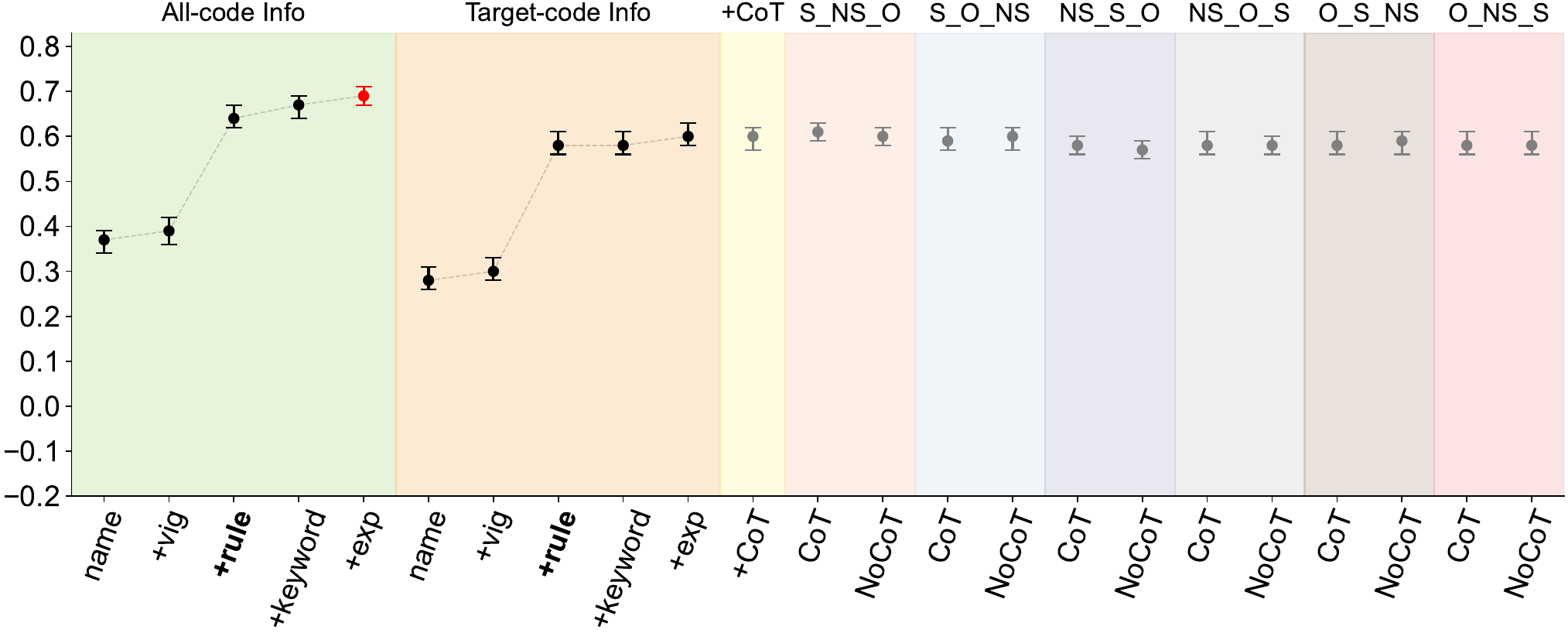}
    \caption{\textbf{Total}}
\end{subfigure}

\caption{Cohen's $\kappa$ between LLM-generated and human-assigned codes under different prompts. The $x$-axis shows 23 prompts with different codebook-component combinations. We assign aliases $L1$-$L23$ to these prompts from left to right for brevity. (a)-(g) represents the LLM-human coding consistency for each attribution using prompts with varying codebook components. (h) displays the aggregated Cohen's $\kappa$ across all attributions. \textbf{All-code info} refers to prompts that provide the same amount of codebook components for all codes, while \textbf{target-code info} indicates that the prompt only provides codebook components for the corresponding attribution, with other attributions coded collectively as \textit{Stigmatizing (others)}. \textit{name}: code name only; \textit{+vig}: added vignette; \textit{+rule}: added rules; \textit{+keyword}: added keywords; \textit{+exp}: added example. \textbf{CoT}: chain-of-thought; \textbf{NoCoT}: no chain-of-thought. \textbf{S}: \textit{Stigmatizing}; \textbf{NS}: \textit{Non-stigmatizing}; \textbf{O}: \textit{Stigmatizing (others)}. \textbf{S\_NS\_O} denotes the order of the three examples as \textit{Stigmatizing}, \textit{Non-stigmatizing}, and \textit{Stigmatizing (others)}. The other notations similarly indicate other permutations. Error bars indicate 95\% confidence intervals for Cohen's $\kappa$ estimates across varying prompt conditions. The result highlighted in \textbf{\textcolor{red}{red}} represents the one with components that fully match the human-written codebook, albeit in a different format. Prompt configurations were systematically varied to test codebook component effects on LLM-human agreement; detailed prompt templates and design rationale are provided in \textit{Supplementary Materials}.}
\label{fig:llmdeductive}
\end{figure}

\textbf{Components inside the codebook: Inclusion of the full codebook improved human-LLM coding consistency, with different components showing varying effects.} 
We present the results in Figure \ref{fig:llmdeductive} and Appendix \ref{app:full}'s Table \ref{tab:exp1}.
First, by comparing the provision of only \textit{target-code} components with including components from \textit{all codes}, we found that the latter approach improves the consistency between LLM and human coding -- for instance, Cohen's $\kappa$ rose from 0.60 to 0.75 when providing all codebook components for all codes.
This suggested that components from other codes actually help LLM learn holistically rather than act as noise, enabling it to differentiate multiple attributions at a level comparable to human coders.

Second, we incrementally added the presence of different codebook components in the prompts and observed changes in LLM-human coding consistency. 
We conducted this analysis using target-code prompts, with similar trends observed for all-code prompts.
Overall, adding vignettes, keywords, and examples yielded small improvements to consistency, with Cohen's $\kappa$ increasing by 0.02 (0.28 to 0.30), 0.01 (0.58 to 0.59), and 0.01 (0.59 to 0.60), respectively. 
The inclusion of coding rules substantially increased consistency, driving Cohen's $\kappa$ up by 0.28 (0.30 to 0.58) (see Table \ref{tab:exp1}). 
Compared to keywords and examples, rules helped define the intension and extension of each code in a more explicit, structured, and logical manner, which presumably enabled LLM to reason about text semantics in a way similar to how human coders use them to make consistent judgments. 

To illustrate this, P388 stated, ``\textit{I will feel a bit angry because I want all guests to have a good time, but I try to make people understand Avery's behavior. They need help and should be empathized with.}" 
The human coders unanimously coded this message as \textit{Stigmatizing (anger)}. 
Without rules, the LLM coded it as \textit{Non-stigmatizing}, focusing on the empathetic aspects of the message. 
Yet, when provided with coding rules, the LLM correctly coded it as \textit{Stigmatizing (anger)}, recognizing that expressions of anger, even if mild, take precedence when coding stigma. 
Here, the coding rules helped the LLM distinguish between intension (i.e., feelings of anger) and extension (i.e., attempts at understanding) in the participant's message and prioritized the former, much like human coders do.


Specifically, for each attribution, the vignettes increased consistency slightly, e.g., with Cohen's $\kappa$ increased by 0.02 for \textit{responsibility} (0.61 to 0.63) and by 0.03 for \textit{fear} (0.58 to 0.61). 
In the same vein, keywords and examples yielded small gains in consistency of 0.01 to 0.04 -- exemplified by \textit{anger}, where Cohen's $\kappa$ improved from 0.51 to 0.52 with keywords and further increased to 0.54 with an example. 
Rules proved to be most impactful, particularly for codes with low initial agreement, such as \textit{helping} (0.01 to 0.58) and \textit{pity} (0.12 to 0.44). 
For \textit{helping}, the LLM, for example, correctly coded P365's statement ``\textit{I probably would help her but only if I knew she was seeking some type of medical help}" as stigmatizing, following the coding rule that conditional assistance based on treatment-seeking behavior reflects stigma.

Interestingly, in the case of \textit{pity}, for instance, the LLM interpreted P437's statement ``\textit{I feel emotional and compassionate toward her}" as subtly stigmatizing without coding rules, reasoning that viewing Avery's situation as deserving of sympathy, though well-intentioned, could constitute an unintentional form of stigma. 
This LLM insight, despite lowering consistency, 1) stimulated a unique analytical perspective for examining \textit{pity}, 2) intrigued us to deliberate how expressions of pity might inadvertently convey condescension, superiority, and patronization, and 3) potentially mitigated our one-sidedness by challenging our preconceived notion of pity as purely positive, a view also supported by the theory \cite{attribution_model_corrigan_2003}, thus sparking us to conceptualize a new form of implicit stigma in the later disagreement analysis.

\textbf{Components not covered by the codebook: The inclusion of additional components beyond the codebook minimally affected LLM-human coding consistency.} 
Figure \ref{fig:llmdeductive} and Appendix \ref{app:full}'s Table \ref{tab:exp2} compare the LLM-human consistency between prompts using only the codebook and those using additional components.
In general, adding human-reasoning processes, more examples, and varied example orders showed minimal effect on LLM-human consistency, with the overall Cohen's $\kappa$ difference ranging from -0.01 to 0.03 across all settings. 
This might be attributed to several factors, such as the applicability and specificity of the supplementary context, its compatibility with the existing codebook, or the LLM's performance ceiling based on the codebook alone. 

For attributions such as \textit{anger}, \textit{fear}, \textit{coercive segregation}, and \textit{social distance}, more examples slightly bolstered consistency, with Cohen's $\kappa$ increasing by up to 0.13 as shown in Table \ref{tab:exp2}, likely because LLM could better code messages with strong similarities to these examples.
Conversely, for \textit{helping}, adding more examples led to lower coding consistency, as Cohen's $\kappa$ dropped by 0.09 to 0.18, potentially due to input overload complicating the model's task. 
For \textit{pity} and \textit{responsibility}, Cohen's $\kappa$ fluctuated inconsistently (-0.10 to 0.05 and -0.03 to 0.06 respectively) when examples were added in different orders or reasoning text was included. 

The effect of the example order also varied. 
Specifically, the \textit{Stigmatizing}$\_$\textit{Non-stigmatizing}$\_$\textit{Stigmatizing (others)} ($S\_NS\_O$) order without CoT increased Cohen's $\kappa$ by 0.09 for \textit{anger} and decreased it by 0.09 for \textit{helping}; similarly, the $NS\_S\_O$ order with CoT showed changes in Cohen's $\kappa$ with improvements for \textit{anger} (0.08) and \textit{social distance} (0.03), but a decline for \textit{responsibility} (-0.02).
Overall, while our explorations provided a preliminary illustration of how different prompt designs can affect LLM-human consistency in deductive coding, future larger-scale studies are needed to carefully investigate why these variations affect each code differently, and how LLMs should be optimally configured for these coding tasks.

Beyond Cohen's $\kappa$, we explored how example order affects the frequency distribution of LLM-generated codes.
We quantified how often each code (i.e., \textit{Stigmatizing}, \textit{Non-stigmatizing}, and \textit{Stigmatizing (others)}) was assigned by the LLM when placed in different positions within the target-code prompts.
The results reveal that codes positioned at the \textit{beginning} appeared \textit{more} frequently in LLM outputs.
Across all codes, presenting the \textit{Stigmatizing} code first resulted in an average of 1,418.75 \textit{Stigmatizing} codes, compared to 1,310.25 when presented second, and 1,254.25 when presented last. 
A chi-square test further confirmed that these differences in code frequencies when \textit{Stigmatizing}, \textit{Non-stigmatizing}, or \textit{Stigmatizing (others)} was presented first in the prompt were statistically significant ($\chi^2$(4, $N$ = 4,153) = 85.04, $p<$ .001) and not due to chance.
We discussed the implications of the observed \textit{primacy effect} when using LLMs for qualitative coding in \textit{Supplementary Materials}.

By consolidating all the aforementioned findings, we identified the highest agreement (Cohen's $\kappa$) between humans and LLMs for each attribution: \textit{responsibility} (0.69), \textit{anger} (0.65), \textit{pity} (0.50), \textit{fear} (0.71), \textit{helping} (0.69), \textit{coercive segregation} (0.58), and \textit{social distance} (0.76). 
Upon applying the prompts that achieved these agreements to code previously human-uncoded messages, our validation through human review of the randomly selected messages yielded high human-LLM agreement (Cohen's $\kappa$ = 0.87), substantiating the validity of our LLM coding.

\textbf{Summary of T1 Results.} 
Following our prompt-design exploration, we achieved a high overall human-LLM agreement (\textbf{Cohen's $\kappa=0.69$}), comparable to the human-human agreement (\textbf{Cohen's $\kappa=0.71$}). 
Looking at specific attributions, we found higher human-LLM agreement than human-human agreement for \textit{social distance} (0.76 vs. 0.70), \textit{helping} (0.69 vs. 0.56), and \textit{pity} (0.50 vs. 0.36), comparable agreement for \textit{fear} (0.71 vs. 0.69) and \textit{responsibility} (0.69 vs. 0.71), and lower agreement for \textit{anger} (0.65 vs. 0.75) and \textit{coercive segregation} (0.58 vs. 0.71).
As Figure \ref{fig:llmdeductive} shows, the results presumably point to the fact that: 1) LLMs process different components of the codebook in a similar way to human coders, with varied consistency for different codes; 2) providing holistic context (e.g., all-code components) generally improves LLM-human coding agreement, except for codes with limited samples where it may introduce noise; and 3) the generally high human-LLM consistency and the LLM's human-like codebook-learning capability suggest that discrepancies between human coders and LLMs have meaningful implications for subsequent disagreement analyses, as they are more likely to shed light on improving the coding scheme rather than merely reflecting LLM limitations. 

\subsubsection{Human-LLM Disagreement Analysis (T2)}
\label{sec:disagree_cs}

We shifted our focus to investigating the sources, underlying causes, and implications of human-LLM disagreements to demonstrate one possible way of implementing \texttt{CHALET}'s T2 in practice, though we recognize that the specific causes of discrepancies may vary across different case studies and domains.

\paragraph{\textcolor{darkred}{\textbf{T2 Procedure.}}} 
After identifying messages where all 23 LLM-generated codes (produced by the 23 prompts shown on the $x$-axes of Figure \ref{fig:llmdeductive}) consistently differed from the human codes, we conducted a disagreement analysis using Chinh et al.'s open-discussion method \cite{code_disagreement_solve_chinh_2019}.
Three coders first independently audited these disagreements in separate spreadsheets by 1) recalling and reflecting their previous coding (Section \ref{sec:humandeductive}) rationale, 2) revisiting each message's content, 3) reviewing all 23 LLM-generated codes and their justification, and 4) categorizing each message as stemming from \textit{human coding error}, \textit{LLM coding error}, or potentially indicating a \textit{new code}.

The coders then merged their individual analyses into a shared document for collaborative review. 
During open discussions, they prioritized analyzing messages where their independent categorizations differed. 
For messages where no coder had indicated potential new codes, they resolved disagreements by sharing their interpretations of the message content and discussing how it aligned with existing codes.
When at least one coder suggested a potential new code, they discussed why these messages might not fit into existing codes, examined to what extent they differed from clearly coded messages, re-evaluated and re-defined the boundaries of current codes, and explored potential semantic, syntactic, and linguistic patterns in these ambiguous messages.
If, after discussion, the coders believe that the message may indicate a new concept, it is retained for further exploration in the next step. 

\paragraph{\textcolor{darkred}{\textbf{T2 Results.}}}
Of the total 4,143 messages, 273 (6.50\%) had human codes that disagreed with all 23 LLM-generated codes. 
Specifically, among messages responding to questions about different attributions, responses to \textit{coercive segregation} had the highest number of disagreements (88 messages, 32.23\% of all disagreements), followed by responses to \textit{anger} (51, 18.68\%) and \textit{fear} (47, 17.22\%). 
Lower rates of disagreement appeared in responses to \textit{responsibility} (32, 11.72\%), \textit{social distance} (24, 8.79\%), \textit{pity} (18, 6.59\%), and \textit{helping} (13, 4.76\%).

\textbf{Human-LLM Disagreement Categorization.} 
Among these disagreements, the coders identified 51 (18.68\%) messages indicating human coding errors. 
For example, in the following case, humans coded the message as \textit{Stigmatizing (anger)} while the LLM coded it as \textit{Non-stigmatizing}.
Upon review, the coders agreed with the LLM's interpretation, noting that they appeared to have hastily coded the message as expressing anger due to the presence of the word ``\textit{perhaps,}'' a type of oversight that can occur when processing large volumes of data, where coders may react to specific terms without fully considering their context:

\begin{quote}
    ``\textit{Perhaps but if I knew they had mental health issues I would be understanding, and I would not get angry at them in response as that solves nothing.}'' (P3)
\end{quote}

The analysis also revealed 41 (15.02\%) messages with LLM coding errors. 
In the example below, the LLM coded the message as \textit{Non-stigmatizing}, while the human coders coded it as \textit{Stigmatizing (responsibility)}. 
The coders could not agree with the LLM's interpretation, as phrases like ``\textit{should ask for help}" imply personal responsibility and agency, which constitute a form of stigmatization:

\begin{quote}
    ``\textit{I believe they should ask for help, and not assume things. Not at all.}'' (P386)
\end{quote}

Most significantly, the coders found 181 (66.30\%) messages that suggested the need for new codes. 
The following example illustrates the complexity of these messages:

\begin{quote}
    ``{\it I would feel embarrassment more than anger. But I would stand up for them and explain Avery's condition to the others so they would have some sympathy. I would also like to chat with Avery after they cool down and explain how their emotions might dampen the mood of a social gathering.}'' (P242)
\end{quote}

Initially, the human coders coded this message as \textit{Stigmatizing (anger)}, due to the implicit aversion expressed. 
The LLM's interpretations varied when different prompts were used: eleven variants led it to identify the message as \textit{Stigmatizing (pity)}, viewing the embarrassment as a lack of sympathy, while twelve others saw it as \textit{Non-stigmatizing}, emphasizing the willingness to help and the expression of pity.
By observing and discussing other examples like this one, the coders hypothesized a potential pattern in the prompt variations: prompts that produced \textit{Stigmatizing (pity)} codes often (8 out of 11) included examples of the \textit{Stigmatizing (others)} code and provided more components (e.g., full codebook details, additional examples, and chain-of-thought text). 
In contrast, when prompts generated \textit{Non-stigmatizing} codes, they tended (6 out of 12) to contain fewer components (e.g., only code names, vignettes, and keywords).

Although there was no clear consensus between the human and LLM codes, during discussions, the coders acknowledged the imprecision of their initial judgment of \textit{Stigmatizing (anger)} upon deeper reflection, noting that the expressed embarrassment and perceived aversion differed from other \textit{Stigmatizing (anger)} messages. 
The coders also rejected the notion that embarrassment equates to a simple lack of pity, as embarrassment often stems from social discomfort rather than an absence of sympathy, and dismissed \textit{Non-stigmatizing} as an inaccurate code.
The coders concluded that the message conveyed a stigmatizing attitude not captured by the current coding scheme, and thus marked it as a \textit{new code}, prompting further analysis and theoretical development. 

\textbf{Language Patterns in Human-LLM Disagreements.}
Interestingly, the coders found that in 160 (58.60\%) of the 273 disagreement messages, the LLM consistently coded messages as \textit{Non-stigmatizing} while humans coded them as \textit{Stigmatizing}, with the reverse scenario occurring in only 14 (5.12\%) messages. 
During the open discussion, the coders discovered that messages that LLMs coded as \textit{Non-stigmatizing} but humans identified as \textit{Stigmatizing} shared distinct language patterns, as detailed in Table \ref{table:lf}. 
Given that 33 of these 160 messages (20.63\%) were marked as LLM coding errors, identifying these patterns revealed crucial areas where LLMs fall short, particularly in interpreting implicit, indirect, and tacit expressions of stigma.

\begin{table}[tbp]
\small
\renewcommand{\arraystretch}{1.22}
\caption{Language patterns in messages coded as \textit{Stigmatizing} by humans but \textit{Non-stigmatizing} by LLMs across all 23 prompt configurations, with one contrasting case where both agreed on \textit{Non-stigmatizing} to illustrate the distinction. These patterns were identified through open discussion among human coders who reviewed 160 messages (out of 273 in which all LLM codes differed from human codes). See Section \ref{sec:disagree_cs} for the selection criteria and analysis details.}
\label{table:lf}
\begin{tabular}{>{\raggedright\arraybackslash}p{0.01\linewidth} 
                  >{\raggedright\arraybackslash}p{0.01\linewidth} 
                  >{\raggedright\arraybackslash}p{0.53\linewidth}
                  >{\raggedright\arraybackslash}p{0.17\linewidth}}
\toprule
\multicolumn{2}{p{0.20\linewidth}}{\textbf{Language Patterns}} & \textbf{Quotes} & \textbf{Human Codes} \\
\midrule

\multicolumn{2}{p{0.20\linewidth}}{\begin{minipage}[t]{\linewidth} \setlength{\baselineskip}{10pt}\textbf{`Someone Else Might Think'}\end{minipage}} & \begin{minipage}[t]{\linewidth}
    \setlength{\baselineskip}{10pt}{\textcolor{darkred}{Neighbors may find it hard} to understand Avery’s outbursts and strange behavior if they do not know them very well. I think you jumped a step. Surely they could talk to someone regularly before hospitalization becomes necessary. (P189)}\end{minipage} & Paternalism \\
\multicolumn{2}{p{0.20\linewidth}}{\begin{minipage}[t]{\linewidth} \setlength{\baselineskip}{10pt}\textbf{Over-conjecture \& Misconception}\end{minipage}} & \\
      & \multicolumn{1}{p{0.13\linewidth}}{\textit{Over-conjecture}} & \begin{minipage}[t]{\linewidth}
      \setlength{\baselineskip}{10pt}{It would depend on whether he was impacting the neighborhood. I think he \textcolor{darkred}{would mostly keep to himself}. (P47)} \end{minipage} & \begin{minipage}[t]{\linewidth} \setlength{\baselineskip}{10pt} {Disconnection and Functional Incapability} \end{minipage} \\
      & \multicolumn{1}{p{0.13\linewidth}}{\textit{Misconception}} & \begin{minipage}[t]{\linewidth}
      \setlength{\baselineskip}{10pt}{If they are not threatening to me or others then I am comfortable. She \textcolor{darkred}{isolates herself} from others so she would not feel a need to interact with them. (P144)} \end{minipage} & \begin{minipage}[t]{\linewidth} \setlength{\baselineskip}{10pt} {Disconnection and Functional Incapability} \end{minipage}\\
\multicolumn{2}{p{0.20\linewidth}}{\textbf{Need \& Suggestion}} & \\
      & \multicolumn{1}{p{0.13\linewidth}}{\textit{Need}} & \begin{minipage}[t]{\linewidth}
      \setlength{\baselineskip}{10pt}{I am not sure about that. It is very complex. Avery has a lot going on. She \textcolor{darkred}{definitely needs to} talk to someone. No weakness there. (P550)} \end{minipage} & Paternalism \\
      & \multicolumn{1}{p{0.13\linewidth}}{\textit{Suggestion}} & \begin{minipage}[t]{\linewidth}
      \setlength{\baselineskip}{10pt}{As long as they are not hurting anyone, I would \textcolor{darkred}{suggest} that they seek treatment to help them feel better. I doubt if they need to be hospitalized. Their behavior is upsetting to them and to others, but they have not actually tried to physically hurt anyone. (P460)} \end{minipage} & Non-stigmatizing\\
\multicolumn{2}{p{0.20\linewidth}}{\begin{minipage}[t]{\linewidth} \setlength{\baselineskip}{10pt}\textbf{Individual-level \& Stereotypical-level}\vspace{0.6mm} \end{minipage}} & \\
      & \multicolumn{1}{p{0.13\linewidth}}{\textit{Individual-level}} & \begin{minipage}[t]{\linewidth}
      \setlength{\baselineskip}{10pt}{So far other members of the house are not at risk. Also, isolating \textcolor{darkred}{him} might worsen his situation. He could start having suicidal thoughts thereby hurting himself. (P282)} \end{minipage} & Self-neglect \\
      & \multicolumn{1}{p{0.13\linewidth}}{\begin{minipage}[t]{\linewidth} \setlength{\baselineskip}{10pt}\textit{Stereotypical-level}\end{minipage}} & \begin{minipage}[t]{\linewidth}
      \setlength{\baselineskip}{10pt}{Certainly, \textcolor{darkred}{people with depression} tend to face such difficulties and challenges. They find it quite hard to concentrate and see things through. They easily get distracted. And if not carefully managed, it progresses and gets worse and may lead to suicidal thoughts creeping into their subconscious. (P187)} \end{minipage} & Self-neglect\\
\bottomrule
\end{tabular}
\end{table}

Specifically, LLMs may have overlooked how language patterns like \textit{someone else might think} can represent personal views, a nuance that human coders recognized through their social awareness and interpersonal experience. 
This psycholinguistic nuance, known as ``\textit{distancing language,}" \cite{distancing_language_nook_2017} involves expressing potentially stigmatizing views through third-party perspectives to disguise personal opinions in socio-cultural contexts. 
In addition, LLMs struggled to differentiate between the description of \textit{potential outcomes} and \textit{current states} -- failing to recognize that over-conjecturing about potential behaviors itself constitutes implicit stigma and microaggression.
LLMs also tended to miss the tonal difference between \textit{need} and \textit{suggestion}, where ``\textit{need}" carries undertones of paternalistic attitudes and power asymmetry, while ``\textit{suggestion}" reflects a more equitable relational stance. 
Further, it appeared difficult for the LLMs to distinguish whether stigma was directed at the vignette characters only or at people with depression as a whole, a critical distinction that the human coders used to separate \textit{individual-level} prejudice from \textit{stereotypical} assumptions.


\subsection{Collaborative Inductive Coding (T3)}
\label{sec:induct_res}

To tackle T3, we took the final step of resolving disagreements and uncovering fresh insights absent in the original coding scheme through inductive analysis of human-LLM discrepancies. 
This part of the case study not only shows how to execute T3 of \texttt{CHALET} concretely, but also empirically propels the deciphering of mental-illness stigma.

\paragraph{\textcolor{darkred}{\textbf{T3 Procedure.}}} 
We performed inductive coding on the 181 messages identified as potentially containing new codes during our disagreement analysis. 
Inspired by prior research \cite{llm_inductive_paoli_2023}, we adopted a human-LLM collaborative approach in which human coders first generated initial codes, LLMs then refined them and proposed theme mergers, and human coders iteratively curated the final coding scheme.
The \texttt{gpt-4-1106-preview} model was used for all analyses.

Concretely, three coders first familiarized themselves with all the messages and then independently pre-coded them by taking notes on recurring data patterns. 
Each coder symbolically assigned summative, salient, essence-capturing, and/or evocative codes to each message and created analytic memos -- sites for reflexive thinking and self-dialogue about the data \cite{coding_manual_saldana_2016} -- in separate spreadsheets. 
These codes and notes included both direct excerpts from participant messages (e.g., ``\textit{will lose job}") and their analytical insights (e.g., ``\textit{humble}", ``\textit{hierarchy}").
Coders then merged their spreadsheets and openly discussed to exchange, explicate, and evolve their interpretations. 
They used shared digital whiteboards as affinity diagrams (see \textit{Supplementary Materials}) to arrange the proposed codes and analytic memos.

Next, the coders presented these initial human-generated codes and corresponding messages to the LLM and asked it to revise code names, generate articulate, theory-grounded terminology, and provide preliminary descriptions for each code, drawing on its broad psychological knowledge \cite{llm_psycho_demszky_2023}.
The human coders then discussed and assessed the LLM's outputs and decided whether to adopt them outright, modify them, or prompt the LLM for additional revisions. 
The LLM's suggestions (e.g., ``\textit{inaction}") were also added to the affinity diagram.

The coders subsequently aggregated and theorized the codes by identifying commonalities among them.
They prompted the LLM to propose preliminary groupings of codes into sub-themes. 
Three coders, along with the mental-health specialist on the research team, then organized and re-grouped the post-it notes on the affinity diagram according to LLM's grouping suggestions and offered critiques as needed. 
Critiques may include requests to separate or merge certain codes and/or address overly broad, generic, and logically inconsistent groupings.
The coders iteratively directed the LLM to revise and re-generate code-grouping proposals and conceptualized the emerging sub-themes based on their own interpretations. 
Once the coders reached consensus on the hierarchical coding scheme, they synthesized their analytic memos and the LLM's outputs to extract the essence of each sub-theme and formulate overarching definitions. 
Following the attribution model \cite{attribution_model_corrigan_2003}, they categorized these sub-themes into three themes -- \textit{cognitive judgments}, \textit{emotional responses}, and \textit{behavioral responses}.
This cyclical process synthesized human analysis, LLM input, and collaborative concept development.


As part of our methodological choices, we compared it to an LLM-only method to evaluate whether autonomous LLM coding is sufficient and to understand the extent to which the collaborative approach added value.
For the autonomous LLM approach, three prompts\footnote{1) ``\textit{We are conducting inductive coding on mental-illness stigma. Identify up to 181 of the most relevant codes in the text, provide a name for each code in no more than 3 words, 4 lines of meaningful and dense description of the code, and a quote from the participant for each code no longer than 7 lines,}" 2) ``\textit{Determine which items in the following list of topics are completely unique,}" and 3) ``\textit{Determine how all the codes can be grouped into themes. Group only the code numbers and provide a name and description for each theme}" \cite{llm_inductive_paoli_2023}.} were used. 
First, the LLM identified initial codes with supporting excerpts from participant messages and generated descriptions. 
Second, it reduced duplicate codes but preserved distinct excerpts. 
Third, we had the LLM group codes into themes with descriptions at higher temperature ($\geq$ 0.5) for greater creativity, and then refine theme names based on the codes, descriptions, and excerpts \cite{llm_inductive_paoli_2023}.
These two approaches were evaluated by means of expert-review sessions with the primary researcher and the mental-health specialist.

\paragraph{\textcolor{darkred}{\textbf{T3 Results.}}} 
Through collaborative inductive coding, the coders discovered 12 new sub-themes and 8 codes beyond the original attribution model \cite{attribution_model_corrigan_2003}. 
This expanded coding scheme, detailed in Table \ref{table:ind}, deepens our understanding of mental-illness stigma by taking into account more implicit stigmatizing mentality and other less easily detectable microaggressions, such as microassault, microinsult, and microinvalidation \cite{microaggression_stigma_barber_2020, microaggression_stigma_gibson_2023, microaggression_stigma_gonzales_2015, microaggression_scale_ertem_2022, microaggression_racial_sue_2007}, and by revealing how tacit, unintentional, and socially normalized forms of stigmatization are expressed through apparently neutral or even positive language.

To illustrate how our human-LLM inductive coding enabled the identification of these implicit forms of stigma, we present an example of how the coders derived and conceptualized a new sub-theme from P219's message:

\begin{quote}
``\textit{I might not be frightened or being threatened regarding the potential implications of your interaction. Everything might be fine if you talk sensibly and do not cross your limits.}" (P219)
\end{quote}

Initially, human coders coded this message as \textit{Stigmatizing (fear)}, while the LLM coded it as \textit{Non-stigmatizing}. 
After the discussion, the coders flagged potentially problematic phrases like ``\textit{talk sensibly}" and ``\textit{do not cross your limits}," which were added to the affinity diagram. 
These phrases implied heightened vigilance -- a pattern the coders initially coded as ``\textit{treat them differently}". 
When presented to the LLM, it suggested the code name ``\textit{special care}'', a term the coders adopted as it better captured the duality of this behavioral intention -- while seemingly expressing consideration, this type of ``care" actually reinforced an us-versus-them divide, since such extraordinary precautions would not normally be present in ordinary social interactions.
The coders then grouped similar expressions across messages, such as ``\textit{be aware}," ``\textit{do not trigger}," and ``\textit{everything will be fine as long as you maintain composure}." 
While the LLM proposed ``\textit{preemptive self-protection}" as a sub-theme, consultation with the mental-health specialist led to the term ``\textit{differential support}" (sub-theme 3.3 in Table \ref{table:ind}) to capture this out-group discrimination. 
This iterative coding process was applied to all 181 messages, yielding the expanded coding scheme.

{
\small
\renewcommand{\arraystretch}{1}
\setlength{\tabcolsep}{4pt}
\setlength{\parskip}{-2pt}
\begin{longtable}{>{\raggedright\arraybackslash}p{0.01\linewidth} 
                  >{\raggedright\arraybackslash}p{0.01\linewidth} 
                  >{\raggedright\arraybackslash}p{0\linewidth} 
                  >{\raggedright\arraybackslash}p{0\linewidth} 
                  >{\raggedright\arraybackslash}p{0.68\linewidth}}
\caption{The expanded coding scheme for mental-illness stigma. The \textcolor{darkgreen}{green} sub-themes represent those in the original attribution model \cite{attribution_model_corrigan_2003}. Other sub-themes were generated through collaborative inductive coding.} \label{table:ind} \\
\toprule
\multicolumn{4}{p{0.25\linewidth}}{\textbf{Themes (Sub-themes, Codes)}} & \textbf{Quotes} \\
\midrule
\endfirsthead

\multicolumn{5}{c}%
{{}} \\
\toprule
\multicolumn{4}{p{0.25\linewidth}}{\textbf{Themes (Sub-themes, Codes)}} & \textbf{Quotes} \\
\midrule
\endhead

\midrule
\multicolumn{5}{r}{{}} \\
\endfoot

\endlastfoot

\multicolumn{5}{p{\linewidth}}{\textbf{Theme 1: Cognitive Judgment}} \\
      & \multicolumn{3}{p{0.2\linewidth}}{\textcolor{darkgreen}{Responsibility}} & \begin{minipage}[t]{\linewidth} \setlength{\baselineskip}{10pt}
      \textcolor{darkgreen}{I really cannot say as I do not have the full details about Avery's life, but if their current situation is triggered by maybe a habit of theirs, then they need to step up and do better. (P45)}
      \end{minipage} \\
      & \multicolumn{3}{p{0.2\linewidth}}{\textit{Sub-theme 1.1: Self-neglect}} & \\[12pt]
      &  & \multicolumn{2}{p{0.2\linewidth}}{\begin{minipage}[t]{\linewidth} \setlength{\baselineskip}{10pt} Code 1.1.1: Think they will suicide and do self-harm \end{minipage}} & \begin{minipage}[t]{\linewidth}
      \setlength{\baselineskip}{10pt}
      Certainly, people with depression tend to face such difficulties and challenges. They find it quite hard to concentrate and see things through. They easily get distracted. And if not carefully managed, it progresses and gets worse and may lead to suicidal thoughts creeping into their subconscious. (P187) 
      \end{minipage} \\[42pt]    
      &  & \multicolumn{2}{p{0.2\linewidth}}{\begin{minipage}[t]{\linewidth} \setlength{\baselineskip}{10pt} Code 1.1.2: Think they cannot take care of themselves very well\end{minipage}} & \begin{minipage}[t]{\linewidth}
      \setlength{\baselineskip}{10pt}
      Not scared of Avery, just concerned for their well-being. I might be concerned about whether or not they would show up and if they would be prepared, dressed appropriately, and such. I might also take on responsibility for booking things if I were worried they would not get that kind of task done. (P333) 
      \end{minipage} \\
      & \multicolumn{3}{p{0.22\linewidth}}{\begin{minipage}[t]{\linewidth} \setlength{\baselineskip}{10pt} \textit{Sub-theme 1.2: Vulnerability}\end{minipage}} & \\[12pt]
      &  & \multicolumn{2}{p{0.2\linewidth}}{\begin{minipage}[t]{\linewidth} \setlength{\baselineskip}{10pt} Code 1.2.1: Think they will be harmed by others\end{minipage}} & \begin{minipage}[t]{\linewidth}
      \setlength{\baselineskip}{10pt} I would feel worried for them regarding their safety. Others may retaliate to the anger Avery displayed. (P394) \end{minipage} \\[21pt]
      &  & \multicolumn{2}{p{0.2\linewidth}}{\begin{minipage}[t]{\linewidth} \setlength{\baselineskip}{10pt} Code 1.2.2: Think they will bring trauma\end{minipage}} & \begin{minipage}[t]{\linewidth}
      \setlength{\baselineskip}{10pt} Not frightened or threatened but maybe triggered. Triggering my own depression because of being around someone who is in a completely “down” state of feeling. (P600) \end{minipage} \\[21pt]
      & \multicolumn{3}{p{0.2\linewidth}}{\textit{Sub-theme 1.3: Crime-prone}} & \begin{minipage}[t]{\linewidth}
      \setlength{\baselineskip}{10pt} Certainly not. It is not a humane way of dealing with someone with a mental illness like depression specifically. If they are still of sound mind and can make decisions for themselves they have the ability to turn their lives around. However, if their depression was making them act violently toward others or influenced them to partake in criminal behavior then yes I agree they should be hospitalized, only in an extreme case. (P144) \end{minipage} \\
      & \multicolumn{3}{p{0.22\linewidth}}{\begin{minipage}[t]{\linewidth} \setlength{\baselineskip}{10pt} \textit{Sub-theme 1.4: Minimization}\end{minipage}} & \\[12pt]
      &  & \multicolumn{2}{p{0.2\linewidth}}{\begin{minipage}[t]{\linewidth} \setlength{\baselineskip}{10pt} Code 1.4.1: Think they will deny that they have a mental illness\end{minipage}} & \begin{minipage}[t]{\linewidth}
      \setlength{\baselineskip}{10pt} I would let them know I will be there to assist if needed. Help may be declined. They might not think they have a problem. (P270)  \end{minipage} \\[21pt]
      &  & \multicolumn{2}{p{0.2\linewidth}}{\begin{minipage}[t]{\linewidth} \setlength{\baselineskip}{10pt} Code 1.4.2: Think they have no problem and can recover themselves\end{minipage}} & \begin{minipage}[t]{\linewidth}
      \setlength{\baselineskip}{10pt} I probably would, but I would want to stress that they should feel free to talk to me about any issues. Avery seems like a good person. They enjoy learning. They currently have problems, but I feel those can be overcome. (P637) \end{minipage} \\[32pt]
      & \multicolumn{3}{p{0.22\linewidth}}{\begin{minipage}[t]{\linewidth} \setlength{\baselineskip}{10pt} \textit{Sub-theme 1.5: Disconnection and Functional Incapability}\end{minipage}} & \\[21pt]
      &  & \multicolumn{2}{p{0.2\linewidth}}{\begin{minipage}[t]{\linewidth} \setlength{\baselineskip}{10pt} Code 1.5.1: Think they are not well-integrated into society\end{minipage}} & \begin{minipage}[t]{\linewidth}
      \setlength{\baselineskip}{10pt} While feeling concerned for the well-being and the potential impact on the neighborhood, I might also be uncertain about the appropriateness of separating Avery and advocating for hospitalization without a clear understanding of their specific needs. (P435) \end{minipage} \\[32pt]
      &  & \multicolumn{2}{p{0.2\linewidth}}{\begin{minipage}[t]{\linewidth} \setlength{\baselineskip}{10pt} Code 1.5.2: Think they are incapable of functioning normally\end{minipage}} & \begin{minipage}[t]{\linewidth}
      \setlength{\baselineskip}{10pt} Assuming there are no issues with paying bills, I see no reason why not. Depression or depressive episodes on their own do not bother me to the point where I would deny the person to be a tenant, but if the situation leads to an inability to fulfill payments, then that would be an issue. (P420) \end{minipage}\\
\multicolumn{5}{p{\linewidth}}{\textbf{Theme 2: Emotional Response}} \\
      & \multicolumn{3}{p{0.2\linewidth}}{\textcolor{darkgreen}{Anger}} & \begin{minipage}[t]{\linewidth}
      \setlength{\baselineskip}{10pt} \textcolor{darkgreen}{To a certain extent, I will feel a bit angry because I want all my guests to have a good time, but I try to make people understand Avery's behavior. They need help and should be empathized with. (P64)} \end{minipage}\\[21pt]
      & \multicolumn{3}{p{0.2\linewidth}}{\textcolor{darkgreen}{Pity}} & \begin{minipage}[t]{\linewidth}
      \setlength{\baselineskip}{10pt} \textcolor{darkgreen}{Unlikely. I have my own concerns and Avery may not need my input. I do not want to get involved. (P148)} \end{minipage}\\[12pt]
      & \multicolumn{3}{p{0.2\linewidth}}{\textcolor{darkgreen}{Fear}} & \begin{minipage}[t]{\linewidth}
      \setlength{\baselineskip}{10pt} \textcolor{darkgreen}{Yes definitely. I will need to have more information on what triggers her episodes. When she is in a crowded place or if she sees joyful people. (P154)} \end{minipage}\\[12pt]
      & \multicolumn{3}{p{0.2\linewidth}}{\textit{Sub-theme 2.1: Apathy}} & \begin{minipage}[t]{\linewidth}
      \setlength{\baselineskip}{10pt} It is not about separation or not, but what the doctors deem necessary. Nothing, we all have our own issues to deal with. (P165) \end{minipage}\\[12pt]
      & \multicolumn{3}{p{0.2\linewidth}}{\textit{Sub-theme 2.2: Frustration}} & \begin{minipage}[t]{\linewidth}
      \setlength{\baselineskip}{10pt} I would not if they were open about their struggles. However, if they never shared then I might feel hurt. Openness is key. If they are open to sharing then I would understand the struggle and not read into their behavior too much because I would know they are having a bad day with chemical imbalance. (P533) \end{minipage}\\[42pt]
      & \multicolumn{3}{p{0.22\linewidth}}{\begin{minipage}[t]{\linewidth} \setlength{\baselineskip}{10pt}  \textit{Sub-theme 2.3: Embarrassment}\end{minipage}} & \begin{minipage}[t]{\linewidth}
      \setlength{\baselineskip}{10pt} I would feel embarrassment more than anger. But I would stand up for them and explain to the others of Avery’s condition so they would have some sympathy. I would also like to have a chat with Avery after they cool down and explain how their emotions might cause a damper on the mood of a social gathering. (P242) \end{minipage}\\[42pt]
      & \multicolumn{3}{p{0.2\linewidth}}{\begin{minipage}[t]{\linewidth} \setlength{\baselineskip}{10pt} \textit{Sub-theme 2.4: Compassion Fatigue}\end{minipage}} & \begin{minipage}[t]{\linewidth}
      \setlength{\baselineskip}{10pt} I will be neutral in that situation. I believe there is a limit to how much you can accommodate anger and pain from a loved one. (P434) \end{minipage}\\[12pt]
\multicolumn{5}{p{\linewidth}}{\textbf{Theme 3: Behavioral Response}} \\
      & \multicolumn{3}{p{0.2\linewidth}}{\textcolor{darkgreen}{Social Distance}} & \begin{minipage}[t]{\linewidth}
      \setlength{\baselineskip}{10pt} \textcolor{darkgreen}{Honestly, I do not think I would. I would be concerned that they would not take care of the place, or they would lose their job and not be able to pay. (P260)} \end{minipage}\\[21pt]
      & \multicolumn{3}{p{0.2\linewidth}}{\textcolor{darkgreen}{Helping}} & \begin{minipage}[t]{\linewidth}
      \setlength{\baselineskip}{10pt} \textcolor{darkgreen}{No. I would feel like it would have a negative impact. Lack of assistance on her part. (P67)} \end{minipage}\\[12pt]
      & \multicolumn{3}{p{0.2\linewidth}}{\textcolor{darkgreen}{Coercive Segregation}} & \begin{minipage}[t]{\linewidth}
      \setlength{\baselineskip}{10pt} \textcolor{darkgreen}{Yes. She will get worse. Her mental state might spill out into arguments with innocent people. (P136)} \end{minipage}\\[12pt]
      & \multicolumn{3}{p{0.2\linewidth}}{\begin{minipage}[t]{\linewidth} \setlength{\baselineskip}{10pt} \textit{Sub-theme 3.1: Differential Support}\end{minipage}} & \begin{minipage}[t]{\linewidth}
      \setlength{\baselineskip}{10pt} Not frightened, but I would be aware of her sensitive position always, in order I could choose my words more wisely. When you are depressed you already feel so bad about things, that anything can make it feel even worse. That is the care you need to have. (P374) \end{minipage}\\[32pt]
      & \multicolumn{3}{p{0.2\linewidth}}{\textit{Sub-theme 3.2: Paternalism}} & \begin{minipage}[t]{\linewidth}
      \setlength{\baselineskip}{10pt} No, I feel that Avery is feeling withdrawn due to their depression. Avery needs someone to “take the reins” and listen to their needs. Avery needs friendship to support mental health and mental health referrals. (P239) \end{minipage}\\[21pt]
      & \multicolumn{3}{p{0.22\linewidth}}{\begin{minipage}[t]{\linewidth} \setlength{\baselineskip}{10pt} \textit{Sub-theme 3.3: Condescension}\end{minipage}} & \begin{minipage}[t]{\linewidth}
      \setlength{\baselineskip}{10pt} No, that seems too far. I am not a doctor and would not know if that is best for them. They do not seem to be physically violent or self-harming and they need to be taught how to deal with everyday life rather than being removed from it. (P130) \end{minipage}\\
\bottomrule
\end{longtable}
}

\textbf{Reliability Validation.} 
To validate our findings and assess the extent to which our expanded coding scheme could improve human-LLM agreement, we applied the same LLM deductive-coding approach (Section \ref{sec:casestudy_llmdeduc}) to the 273 discrepant messages identified in Section \ref{sec:disagree_cs}.
We retained the prompt structure from Figure \ref{fig:prompt_structure}, using an all-code prompt that included code names, the vignette, coding rules \& specifications, keywords, and one example for each of the 19 sub-themes (including both the original seven attributions and twelve newly uncovered sub-themes) from Table \ref{table:ind}, plus a \textit{Non-stigmatizing} code. 
This coding process achieved a Cohen's $\kappa$ of 0.23.
Notably, analyzing these persistent discrepancies informed the coders to further clarify the definitions of the newly discovered codes.
After clarification, Cohen's $\kappa$ improved to 0.26. 
Although the increase in consistency is small, we should point out that it is a relatively substantial advance in the human-LLM agreement compared to when we used the original attribution model \cite{attribution_model_corrigan_2003}, which failed to produce any matches with human coding.
Yet, it's worth noting that 111 out of 273 messages (40.66\%) were still recognized as LLM coding errors by human coders, with 79 (71.17\%) of these retaining their previous inaccurate codes. 
This shows the potential limitation of applying LLMs in this qualitative-coding task. 
When we then applied the expanded coding scheme to all 4,200 human-coded messages, Cohen's $\kappa$ also improved from 0.69 to 0.72.

We also conducted human coding on messages from the remaining uncoded 402 participants (out of 1,002 total) to further evaluate how well our findings generalized to new data.
We randomly sampled 100 messages and had three human coders analyze them using both the original attribution model \cite{attribution_model_corrigan_2003} and our expanded coding scheme.
The comparison revealed an improvement in human-human agreement: using the attribution model, coders failed to reach a consensus on 23 out of 100 messages (Cohen's $\kappa$ = 0.67), while using the extended coding scheme, they disagreed on only 9 messages (Cohen's $\kappa$ = 0.87). 
This confirms that our expanded coding scheme could presumably be applicable to previously unseen data and increase coding consistency.

\textbf{Comparison with Autonomous LLM Coding.} 
The results of the autonomous LLM coding are reported in Appendix \ref{sec:app_llm_ind}.
Three coders and the mental-health specialist evaluated both approaches by rating their generated codebooks on a five-point Likert scale (1 = lowest, 5 = highest) using measures of ease of use, code clarity, mutual exclusivity, and exhaustiveness.
These criteria were chosen based on previous work on codebook quality \cite{coding_manual_saldana_2016} and LLM-assisted codebook development \cite{llm_inductive_barany_2024}. 
The collaborative approach received higher ratings on all measures (code clarity: 4.75 vs. 2.50, mutual exclusivity: 4.00 vs. 1.75, ease of use: 4.00 vs. 3.25, exhaustiveness: 4.00 vs. 3.50). 
We empirically found that the autonomous LLM approach produced five relatively descriptive, superficial codes without explanatory interpretation -- e.g., ``\textit{Traveling Concerns}" (codes 7.1 and 11.1) captured visible actions but missed the underlying \textit{fear}, while ``\textit{Tenant Acceptance}" (codes 4.1 and 10.2) described housing decisions without recognizing the participant's \textit{distancing} tendency. 
Themes like ``\textit{Mental Health Focus}" (theme 3) were overly generic with vague descriptions of ``\textit{topics specifically related to mental health issues}".
The codebook in Table \ref{tab:llm_ind} showed potential redundancy, exemplified by ``\textit{Community and Social Support}" (theme 9) and ``\textit{Support Systems}" (theme 2) describing overlapping support-related concepts. 
Quantitatively, of the 26 codes generated, 16 were duplicates (61.54\%), such as the code ``\textit{Anger Management}" appearing four times in different themes. 
These findings suggest that \texttt{CHALET}'s human participation and co-creation remain crucial when inductively coding psychological constructs like mental-illness stigma.

\section{Discussion}

Our proposed systematic approach, \includegraphics[scale=0.004]{1.png} \texttt{CHALET}, provides a roadmap for collaborative human-LLM theory-driven qualitative analysis. 
Through three interlocking tasks -- deductive coding, disagreement analysis, and inductive coding -- we showed that the approach can direct human coders' attention to aspects of the codebook needing revision, provide alternative perspectives that enrich understanding of core theoretical constructs underlying the coding schemes, and facilitate the development of new themes through human-LLM collaboration.
We hope these observations can provide the driving forces to broaden the use of LLM in qualitative coding, in the sense that LLM is more than just an automated tool that \textit{replaces} human coders; rather, it can \textit{collaborate} with human coders to enhance the overall coding process.

In the first task, human coders performed \textbf{deductive coding} to provide examples and interpretations to train LLMs to acquire qualitative-coding skills (\textbf{T1}). 
We demonstrated that by systematically exploring prompt settings (e.g., textual structure, parameters) and what combinations of codebook components to include in the prompt, LLMs can effectively support the theory-driven qualitative coding.
Our practice helps to reduce human-LLM coding disagreements caused by inadequate prompts or inaccurate/deviant LLM interpretations, so that remaining disagreements are more likely to arise from potential limitations within the coding scheme and the theories themselves.

We observed that LLMs and humans may interpret certain ambiguous or difficult-to-define constructs differently. 
Analysis of these \textbf{discrepancies} (\textbf{T2}) helped reveal data patterns that warranted deeper qualitative analysis. 
These patterns may indicate that the codebook lacked sufficiently clear guidance for coders to reach a consensus, which allowed the human coders to discover potential imperfections in the theoretical framework and explore opportunities for theory development.
In doing so, the LLMs also helped the human coders recover from some of their oversight and re-evaluate coding decisions by flagging certain details they may have missed.
The disagreement analysis also unveiled semantic, pragmatic, and/or psycholinguistic markers that uniquely challenge LLMs, which helped to understand how to improve LLMs' ability to understand high-level socio-psychological constructs.

Through \textbf{inductive coding} (\textbf{T3}), we further conceptualize human-LLM disagreement to enrich the existing coding scheme. 
The disagreement-resolving process leverages the complementary strengths of humans and LLMs -- humans bring social awareness and empathetic interpretations based on social norms and personal experience, while LLMs excel at consistent, semantically precise coding derived from their vast training corpora.
Though tackling human-human disagreement \cite{disagreement_zade_2018} is undoubtedly important, addressing human-LLM disagreement offers additional benefits beyond adding another human coder -- LLM codes, based on vast and diverse training data, can be produced rapidly, help human coders reconsider concepts from multiple viewpoints, and potentially generate serendipitous discoveries, facilitating a more holistic unpacking of complex social phenomena like mental-illness stigma.
Here, we should emphasize that while the derived new codes and theoretical constructs offer qualitative insights and are partially supported by existing theoretical frameworks \cite{patronization_douglas_2011, paternalism_stull_2013, microaggression_stigma_barber_2020}, our primary contribution lies in proposing a \textbf{methodology} that leverages human-LLM synergy to facilitate such theoretical discovery and in demonstrating the value of human-LLM coding discrepancies in theory-led qualitative analysis.

\subsection{LLMs' Capabilities for Advancing Qualitative Analysis} 
\subsubsection{Prompts for Superior Human-LLM Coding Agreement}

Prior works \cite{label_llm_huang_2023, classify_llm_sun_2023, labeling_llm_fu_2024, label_llm_kuzman_2023} examined the use of simple prompts and/or discussed common techniques such as chain-of-thought prompting and few-shot/zero-shot settings, and showed LLMs' satisfactory performance in various text annotation tasks, including objective \cite{llm_label_method_gilardi_2023, labeling_llm_uymaz_2023, labeling_llm_savelka_2023, labeling_llm_prasad_2024, labeling_llm_azad_2023} and subjective \cite{label_llm_huang_2023, labeling_llm_maceda_2023, chatbot_various_task_amin_2023} tasks. 
Generally, our case study showed similar results to these studies \cite{label_llm_politic_twi_heseltine_2024} -- we found that LLMs performed comparably to human coders in identifying mental-illness stigma, with human-LLM agreement (Cohen's $\kappa$=0.69) approaching human-human agreement (Cohen's $\kappa$=0.71).
More importantly, we went further by systematically exploring prompt design and leveraging human-LLM disagreements to detect potential gaps in codebooks and theories, which differs from existing literature that typically uses simplistic prompts and aims to minimize disagreements for ``accuracy" \cite{llm_deductive_kirsten, prompt_practice_reiss_2023, prompt_practice_dunivin_2024, llm_deductive_hou_2024}.

Our case-study results showed that providing all codebook components for all codes in prompts generally improves LLM-human agreement, with two exceptions: \textbf{pity} and \textbf{coercive segregation}.
For \textbf{pity}, codebook components from other codes can act as noise, with Cohen's $\kappa$ decreasing from 0.50 to 0.46 when comparing the target-code condition to the all-code condition (Table \ref{tab:exp1}). 
This decrease could be attributed to how \textit{pity} is presumably different from other attributions in the stigmatization context and their potential overlap in expression. 
For attributions like \textit{anger} or \textit{fear}, the presence of these emotions indicates stigma, whereas for \textit{pity} it is the opposite -- showing no pity reflects stigmatizing attitudes.
This vagueness becomes more pronounced when the LLM is exposed to codebook components from multiple attributions, as messages that should be coded as \textit{Stigmatizing (pity)} due to lack of sympathy might instead be misclassified as \textit{Stigmatizing (anger)} or \textit{Stigmatizing (fear)} if these emotional responses appear in the messages.

For \textbf{coercive segregation}, we unexpectedly observed from Table \ref{tab:exp1} that providing only code names for all codes yielded better LLM-human agreement (Cohen's $\kappa = 0.58$) compared to including full codebook components (Cohen's $\kappa = 0.54$).
One possible reason is that human coders also evolve their understanding from collaborative codebook creation, discussion, and verbal guidance beyond the written codebook. 
For example, human coders debated what constitutes a forced action and dissected the varying intensities of different phrasings, considering the differing severity implied by terms such as ``\textit{therapy}", ``\textit{psychiatric treatment}", ``\textit{hospitalization}," and ``\textit{see a doctor}."
LLMs, however, were not provided the same historical interactive context to learn and relied solely on written descriptions \cite{prompt_practice_dunivin_2024}, where the codebook didn't explicitly articulate these subtle term distinctions. 
Thus, the added components likely contained only superficial knowledge without teaching LLMs the gist of semantically differentiating coercive or segregative behaviors.

Recognition of LLMs' struggle with such \textbf{potential code overlaps} (as in \textit{pity}) and \textbf{tonal subtleties} (as in \textit{coercive segregation}) further informed prompt design. 
These prompt-related exceptions drew our attention to hardly discernible patterns, such as those in Table \ref{table:lf}, prompted coders to re-examine both the boundaries between multiple emotional responses and the spectrum from strongly coercive, sermonizing statements to softer suggestions, revealed codebook ambiguities, and catalyzed new code definitions and conceptualizations.

It should be noted that our primary goal is not to find optimally engineered prompts for all tasks; rather, we believe that for most of the given dataset and codebook, the proposed exploration could train LLMs to a level of consistency that allows them to learn from human-created codebooks to perform deductive coding and collaborate with human coders.
Our results suggest that it would be beneficial if future research investigates how LLMs perform when allowed to assign \textbf{multiple codes} to a single message, which better reflects the complexity of real-world qualitative-coding scenarios, and examines whether false positives in multi-code predictions reflect code overlap, LLM limitations, or vagueness in the coding scheme that needs clarification. 
Future work should also explore systematic prompting approaches to capture and \textbf{formalize the non-written, tacit, experiential, and consensual knowledge that emerges through human coding practices} into LLM training processes to enhance human-LLM collaboration further.

\subsubsection{Locating LLMs' Limitation in Coding Psychological Constructs}

In addition to the methodological contributions, our \textbf{case study} itself also has practical value.
Recent studies have explored the use of LLMs to automatically analyze human psychological constructs through language \cite{psych_construct_language_hur_2024, psych_construct_language_peters_2024, dc_meng_2025}, the medium through which these constructs are expressed \cite{linguistic_ernala_2017, linguistic_li_2018}. 
Our practice contributes to this body of research by identifying key limitations of LLMs' ability to interpret these constructs.

Through human-LLM disagreement analysis, we found that accurate coding of psychological constructs requires unique human capacities that LLMs currently lack, including rich \textbf{interpersonal experience, societal awareness, and the ability to perceive social cues}. 
For instance, LLMs struggle to recognize ``\textit{distancing language}" \cite{distancing_language_nook_2017}, where the use of passive voice, euphemisms, and avoidance of first-person pronouns signals psychological distance from people with mental illness. 
We therefore encourage future research to address these limitations and also call for greater attention to the psycholinguistic nuances \cite{meng_stigma_corpus_2025} between social-media data, which is commonly used for language-based inference \cite{psych_construct_language_peters_2024}, and conversational texts, such as those sourced from human-chatbot interactions, when training LLMs to code psychological constructs.

\subsection{Collaborative Human-LLM Approach for Theory-driven Qualitative Analysis}

\subsubsection{From Manual Hybrid Deductive/Inductive Coding to Human-LLM Collaboration}
\label{sec:discussion_collaborative}

Our study contributes to research on hybrid, purely human-led qualitative coding \cite{traditional_fereday_2006, hybrid_bowe_2020, hybrid_proudfoot_2023, hybrid_swain_2018, hybrid_xu_2020} through human-LLM collaboration, particularly through the analysis of human-LLM disagreements.

First, incorporating a consistent, semantically driven, and rule-abiding LLM coder into the conceptualization process can help \textbf{uncover ambiguities that might be missed when analyzing human-human disagreements alone, while also increasing the certainty of their identification}. 
From our practice, we observed that even with standardized codebook guidance, different human coders may have slight variations in perceiving the scope of each code during the human deductive-coding phase prior to LLM involvement. 
These differences result from human coders' innate subjective interpretations of the underlying intent of messages, which are shaped by their positionality \cite{positionality_tracy_2024} -- including their social positions, identities, and worldviews.
When encountering messages that appear to conform to an existing code but differ from other messages within that code, human coders may experience hesitation about whether to develop a new code or fit it into the existing theoretical framework, especially when there is potential overlap between how different coders sense the scope of each code.

Adding an LLM coder can presumably contribute to coding decisions when human coders are uncertain about code boundaries, because the LLM's consistent, linguistically focused coding may provide a clearer signal. 
That is, when the LLM indicates that forcing a message into an existing code will lead to disagreement, it gives human coders more \textbf{confidence} to revisit the message and consider developing a new code.
Interestingly, of the 273 messages with human-LLM disagreements, 120 (43.96\%) also showed human-human disagreements.
Of these, 86 (71.67\%) were ultimately defined as new codes. 
These figures show that the analysis of human-LLM disagreements not only helps identify more potentially ambiguous messages (273 vs. 120) but also increases the certainty of coding decisions by transforming initially uncertain instances into better-defined new codes (86 vs. 0).
To illustrate this specifically, during our initial deductive coding, the coders encountered the following message:

\begin{quote}
``\textit{I would give them pointers on how they themselves could do the project, but I would not complete it for them. Hopefully, it would open their eyes that they can be successful despite challenges.}" (P422)
\end{quote}

One coder initially coded it as \textit{Non-stigmatizing}, while another coded it as \textit{Stigmatizing (helping)}. 
When discussing this human-human disagreement, the coders identified the phrase ``\textit{give them pointers}" as a form of \textit{indirect help}. 
According to the codebook, the coders felt that this message could be interpreted either way -- as a form of assistance (\textit{Non-stigmatizing}) or, despite the willingness to help, as reflecting an unequal power dynamic (\textit{Stigmatizing (helping)}). 
The coders chose the latter after discussion, although some uncertainty lingered about this coding decision.
Whichever variant of the prompt was used, the LLM consistently coded it as \textit{Non-stigmatizing}, reminding the human coders of the fact that this type of \textit{help} was far from overtly stigmatizing messages, such as P368's ``\textit{I don't want to help because I don't want to babysit petulant child}". 
This LLM output allowed the coders to revisit the message with confidence and recognize that while it was not purely \textit{Non-stigmatizing}, it did not fit neatly into the existing \textit{Stigmatizing (helping)} code. 
This led the coders to create a new code, \textit{Condescension}, which captured the implicit stigma in the participant's behavior of ``\textit{giving pointers}" and emphasized how this action would ``\textit{open their eyes}", exemplifying how human-LLM disagreement analysis can uncover equivocal messages that might escape detection in human-only analysis.

Second, human-LLM disagreement analysis also helps \textbf{locate human coding errors} that might otherwise go undetected in human-only approaches. 
Specifically, in our case study, human coders initially agreed on 28 messages, but later disagreement analysis with LLMs revealed that these were human coding errors, accounting for 54.9\% of the total 51 human coding errors identified. 
These cases emerged when both human coders, after observing consistently different interpretations by LLMs, were prompted to recognize their initial inaccurate coding decisions and subsequently arrived at potentially more thoughtful and precise codes.
This highlights LLMs' potential to mitigate occasional human errors due to fatigue, carelessness, or inattention. 
Such potential suggests that human coders, while skilled, are not infallible, and LLMs could help to ``\textit{read between the lines}" -- to capture what was not said but was meant -- a capability particularly valuable when working with human coders who are non-native speakers and may fail to pick up on such nuances.

Finally, LLMs can \textbf{simulate different coding outcomes and rationales, rapidly generating diverse interpretations of messages}. 
The results in \textit{Supplementary Materials} show that LLMs can assign at least three valid codes to the same message when given different prompt variants.
Of the 4,153 messages analyzed, 599 (14.42\%) received at least three different codes from LLMs, with the distribution across stigma attributions being: \textit{helping} (127), \textit{coercive segregation} (112), \textit{pity} (124), \textit{anger} (96), \textit{social distance} (66), \textit{fear} (49), and \textit{responsibility} (25).
For instance, P357's statement ``\textit{I would feel comfortable entrusting it to Avery -- but only if she were seeking help to manage her mental health}" was interpreted differently depending on the prompt used. 
It was coded as \textit{Stigmatizing (social distance)} due to the awareness of conditional comfort, as \textit{Stigmatizing (responsibility)} because of the emphasis on Avery's need to seek help independently, or as \textit{Non-stigmatizing} when prioritizing the willingness to entrust tasks. 
This rapid production of diverse codes allows human coders to quickly explore a wider range of potential interpretations, incentivizing reflection on areas where the codebook may be vague or where code may overlap, much like discussing the codes with another set of human coders. 
Fruitful avenues for future research include systematically comparing human-human and human-LLM disagreements, specifically analyzing when and why these disagreements coincide or occur separately, as such analysis may reveal distinct patterns in how humans and LLMs navigate language that falls in the interstices between theoretical constructs.

\subsubsection{Trade-off between Human Involvement and LLM Automation}
\label{sec:dis_discussion}

Our \texttt{CHALET} method's three main tasks carefully balance human expertise with LLM capabilities, reflecting our finding that while LLMs show promise in automating qualitative coding, \textbf{strategic human involvement} remains critical.
First, before involving LLMs in \textbf{T1} (i.e., \textit{LLM deductive coding}), we intentionally had human coders analyze a substantial volume of messages (e.g., 4,200 in our case study) in a deductive way, given that human coders' interpretations, developed through codebook creation and team discussions \cite{prompt_practice_dunivin_2024}, can establish a reliable baseline, directly influence how human-LLM disagreements are identified, and guide subsequent conceptualization.

Second, our \textit{human-LLM disagreement analysis} (\textbf{T2}) extends previous studies \cite{ncoder_zambrano_2023, deductive_labelling_chew_2023} by using LLM-human discrepancies to foster a more efficient, targeted, and less labor-intensive coding process. 
Specifically, it prioritizes and guides human coders toward \textbf{a smaller but potentially more theoretically valuable subset of the data}, rather than requiring them to inductively code large volumes of data that may be homogeneous.
However, we also recognize that this efficiency potentially comes at a cost -- in our case study, for instance, selecting only messages where all LLM-generated codes are inconsistent with human-assigned codes risks overlooking some human-coding errors and codebook ambiguities that exist outside our disagreement analysis.

And third, in \textbf{T3} (i.e., \textit{collaborative inductive coding}), our case study empirically demonstrates that LLMs could independently produce only relatively superficial inductive codes (see Appendix \ref{sec:app_llm_ind}), and combining these codes with \textbf{human-led induction} proved important in broadening and clarifying what and how underlying psychological constructs should be used for coding mental-illness stigma.
With these considerations in mind, we encourage future research to pursue two directions: 1) normative discussions exploring flexible coding strategies in which human involvement adapts to language complexity, particularly for data rich in implicit meaning or contextual nuance, and 2) systematic ways to select which disagreements to analyze and determine appropriate inclusion criteria, for example, relaxing current selection criteria to include messages with a certain small percentages of human-LLM agreement, high human-human disagreement, or high variance in LLM codes across prompts.

\subsection{Epistemological and Ontological Reconfigurations in Human-LLM Qualitative Analysis}

\subsubsection{Redistributing Agency Between Human and LLMs}

Our work initiates a critical discussion about \textbf{how agency is operated and re-allocated} in human-LLM collaborative qualitative analysis generally, given that \textit{agency} has long been central to qualitative research and interpretivist social science \cite{agency_emirbayer_1998}. 
\textbf{Barad's agential realism} \cite{agency_barad_2007} theorizes the ontological inseparability of intra-acting agencies from post-structuralist and post-humanist perspectives.
Specifically, this theory proposes the notion of \textit{intra-action}, challenges conventional notions of pre-figured entities by positing that entities come into being through their mutual interconnections, and captures how agency arises as a relational effect rather than a property possessed by individual actants.

Under Barad's influence, recent scholars have attempted to reframe human-algorithm relationships \cite{agency_modern_baumer_2017}, move beyond assumptions that humans and algorithms exist as separate entities that subsequently interact, and further reveal how algorithms enliven subjectivities through processes of entangled co-constitution \cite{agency_baumer_2024}.
In such human-algorithm assemblages, the knots threaded between computational processes and human sensemaking reveal how agency disperses across relationalities where humans do not merely \textit{use} algorithms, but are \textit{interwoven} with them, insofar as both are actualized and co-evolve into specific types of subjects \cite{agency_baumer_2024}. 
The increasingly agentic nature of LLM-based agents is further changing how agency is construed in human-LLM qualitative analysis, prompting a shift from asking \textit{who} holds agency to examining \textit{how} it is embodied, situated, and hybridized through partnered coding processes.

In response, our method, \texttt{CHALET}, draws from insights related to Barad's agential realism \cite{agency_barad_2007}, wherein human coders, on the one hand, maintain agency by 1) categorizing interpretive disagreements with LLM coders (\textbf{T2}), 2) transforming those discrepancies into potential new theoretical constructs (\textbf{T3}), and 3) making ultimate decisions by curating overall theoretical coherence, with their subjectivities themselves partly shaped by the engagement with LLM-generated coding results. 
Empirically, such interplay was observed in our case study, where human coders were occasionally persuaded by LLM-suggested codes (51 out of 273 disagreements) and/or began to reflect on the fluid boundaries between codes, prompted by the LLM's interpretive alternatives.
Conversely, LLMs also enact agency by 1) surfacing inconsistencies with human coders (\textbf{T2}), 2) rapidly proposing multiple, alternative code suggestions (\textbf{T1, T3}), and 3) modulating their outputs in response to human-written prompts (\textbf{T1}) -- prompt design, in this sense, becomes not mere instruction-giving but a practice through which human coders participate in constituting the LLM's agency.
Together, these agentic movements, we believe, are not unidirectional; rather, in \texttt{CHALET}, human and LLM coders are co-evolving through \textbf{gradual, intertwined interpretations and productive disagreement} to finally yield theoretical insights.
We call for future theoretical work in HCI that articulates how displaying LLMs' coding \textit{confidence} affects agency distribution in human-LLM qualitative coding, building on research showing that when AI reveal their confidence levels, humans feel more in control (i.e., agency) \cite{confidence_vantrepotte_2022} and stay more aware of their own analytical capabilities (i.e., metacognition) \cite{confidence_li_2025}.

\subsubsection{Bridging Computational Methodology with Interpretive Inquiry}

Our methodology empirically contributes to the understanding of \textbf{how computational approaches can productively engage with interpretive inquiry rather than displace it}, with implications that extend across qualitative HCI research.
As discussed by Schroeder et al. \cite{paradigm_schroeder_2024}, a tension exists where, on one hand, qualitative researchers internalize a responsibility to preserve hermeneutic power through immersive, pen-and-paper exploration with data, honoring its multiplicity and thickness; on the other hand, computational tools, particularly LLMs, can generate luminous, seemingly seductive summaries, labels, and/or topics that can possibly draw researchers away from situated, on-the-ground engagement, thus blurring the boundaries between positivist and interpretive paradigms \cite{paradigm_messeri_2024} and risking contentious shifts toward scale, abstraction, and delegation that might homogenize the rich diversity of innate human elements \cite{paradigm_soden_2024, paradigm_schroeder_2024}.

Accordingly, we envision that the \texttt{CHALET} approach could offer a potential way to navigate this tension. 
Rather than having humans code a subset and then delegating the remainder to LLMs as labor-saving tools (i.e., \textit{sequencing}), we propose that humans and LLMs code overlapping data and analyze \textit{disagreements} (i.e., \textit{parallelism}), a methodology that transforms LLMs from being used as shortcuts into agentic collaborators, maintains qualitative researchers' deep immersion in data, and openly harnesses LLMs' broad knowledge to absorb diverse coding rationales (\textbf{T2}). 
In particular, \texttt{CHALET} creates a middle ground that helps preserve the thickness valued in qualitative research (\textbf{human deductive coding} and \textbf{T3}) on one side, and on the other side leverages capabilities of LLMs beyond mere automation, such as drawing on vast training corpora and rapidly simulating different analytical voices (\textbf{T1}).
Moving forward, we hope our paper will spark cross-disciplinary conversations that help HCI researchers with an interpretive orientation better harness LLMs by recognizing the boundaries of their capabilities \cite{qualitative_research_llm_bano_2023, qualitative_research_llm_torii_2024}, and support those with computing backgrounds to ground their work in social theories, ultimately nurturing emerging fields like computational social science and strengthening HCI's theoretical foundations for more thoughtful socio-technical systems.

\subsection{Practical Guidance for Applying \texttt{CHALET} to Human-LLM Qualitative Analysis}

\subsubsection{Scope and Generalizability of \texttt{CHALET}}
\label{sec:generalizability}

We view \texttt{CHALET} as a conceptual-methodological framework that operates independently of specific data sources (e.g., human-chatbot conversations), domains (e.g., mental-illness stigma), and theories (e.g., attribution model \cite{attribution_model_corrigan_2003}), with its three-task structure designed at a general level to highlight the value of \textbf{human-AI collaboration} in theory-driven qualitative analysis.
Our approach shows particular promise in domains with existing social theories, where researchers can build upon these theories to better conceptualize complex sociological and psychological constructs \cite{attribution_model_other_graham_1997, attribution_model_other_martinko_2007} from new empirical data -- for example, applying \texttt{CHALET} to better understand human decision-making processes that emerge from interview data using the \textit{Theory of Planned Behavior} \cite{tpb_ajzen_1985} as a starting point. 
However, \texttt{CHALET} may be less appropriate for 1) \textit{objective classification tasks} (e.g., topic categorization, document classification \cite{llm_label_method_he_2024}), where human-LLM disagreements represent errors rather than opportunities for theoretical discovery, 2) \textit{tasks that yield limited theoretical insights} (e.g., categorizing emotions as simply positive or negative \cite{labeling_llm_maceda_2023}), where there is little need to generate new codes/(sub-)themes, and 3) \textit{purely exploratory data-driven studies} without existing conceptual foundations (e.g., interviews for system usability testing \cite{ux_interview_jo_2022}), where the absence of theory prevents structured data collection and grounding for human deductive coding.

To help researchers apply these \textbf{boundary conditions} to their own domains, here is a concrete example: HCI researchers studying ``\textit{how users experience harm from AI systems}'' could begin with an established theory-informed coding scheme (e.g., representational, allocative, and interpersonal harms \cite{aiharm_shelby_2023}) to code interview transcripts with affected users, have LLMs code the same transcripts using the same scheme (\textbf{T1}), then analyze human-LLM disagreements (\textbf{T2}) and collaborate with LLMs through inductive coding (\textbf{T3}) to discover new harm \textit{sub-themes/codes}, such as \textit{relational transgressions} \cite{aiharm_zhang_2025} within the broader \textit{theme} of \textit{interpersonal harm}. 
We hope these clarifications provide practical future directions for HCI researchers to 1) test constructs derived through \texttt{CHALET} in quantitative studies to assess its methodological rigor, and 2) apply \texttt{CHALET} to qualitative research across different theoretical paradigms and cultural contexts, building more systematic evidence for its \textbf{transferability}.

\subsubsection{Cross-cultural Considerations: Decolonizing Human-LLM Collaboration}
\label{sec:crosscultural}

We remind readers in any HCI qualitative research context to remain mindful that \textbf{socio-cultural factors} fundamentally shape how \texttt{CHALET} applies across disciplines and cultural settings.
Sociohistorically, \textit{colonialism} reshaped socio-cultural structures by amplifying and institutionalizing divisions within local communities as a means of control, a logic that persists as coloniality and continues to be studied by \textit{postcolonialism} \cite{colonial_chatterjee_2020}.
Recent \textit{postcolonial computing} scholars \cite{colonial_irani_2010} contend that socio-technical systems, particularly LLMs, operate through \textit{colonial impulse} \cite{colonial_das_2024}, design choices that reanimate universalist assumptions, encode \textit{reductionist} representations, and reinforce colonial hierarchies \cite{colonial_dourish_2012}, producing cultural hegemony where certain identities become normative through practices that marginalize subcultural groups as deviant and/or non-normative \cite{colonial_crenshaw_2013, colonial_collins_2022}.
This colonial impulse leaves an imprint through mechanisms such as cultural assimilation \cite{colonial_fanon_2023}, a Western gaze that imposes Western interpretations onto non-Western contexts \cite{colonial_kotliar_2020}, tokenistic representations that flatten difference into stereotype \cite{colonial_said_1977}, and broader ideological structures (e.g., imperialism and masculinist) \cite{colonial_ghosh_2024} that result in the essentialization of diverse identities \cite{colonial_hanna_2020}.

When applied to our human-LLM collaboration scenario, these concerns become salient, as our integration of LLMs into theory-driven qualitative coding necessitates \textit{socio-cultural text processing} in which performativity \cite{colonial_butler_1990}, the practice by which language constructs and enacts identities through vernacular, sociolects, and culturally embedded norms, is central, yet the algorithmic translation of such dynamic, context-dependent performances into computational representations intelligible to LLMs remains opaque, potentially reductive, and vulnerable to perturbations that may reinscribe colonial logics \cite{colonial_dourish_2012}. 
Furthermore, the agential nature of LLMs may actually intensify this challenge, as their lack of reflexivity, ambiguous grounding, and potential to embody Westernized perspectives can complicate interpretive processes in disagreement analysis -- a setting in which coding outcomes are jointly shaped by \textit{who the human coders are} (e.g., their cultural positionalities), \textit{which LLMs are used} (e.g., their training data and institutional origins), and \textit{what is being coded} (e.g., topics and theories with high socio-cultural sensitivity, such as mental-illness stigma) \cite{positionality_tracy_2024}.

As such, we present \textbf{actionable recommendations} for practitioners adopting \texttt{CHALET} in cross-cultural contexts. 
First, akin to human-only qualitative coding requires reflexive awareness of who the human coders are and how their backgrounds shape interpretation \cite{positionality_tracy_2024}, agential human-LLM collaboration likewise demands practitioners' careful attention to \textbf{which proprietary LLMs are used, the stereotypes embedded in their training data}, and how these biases intersect with the selected theories (e.g., attribution model \cite{attribution_model_corrigan_2003}) and the data being coded (e.g., interview about mental illness). 
Second, practitioners should \textbf{actively involve local community members in the development and validation of codebooks}, creating opportunities for the codes produced by Western-trained AI systems to be questioned or reshaped by indigenous knowledge systems. 
And third, practitioners must \textbf{remain attentive to the power dynamics in human-LLM disagreement analysis}, especially whether LLMs can offer diverse or counterfactual explanations that challenge dominant narratives, and how such disagreements may either reinforce Western interpretive frames or create space for alternative perspectives.

\subsection{Broader Implications}

\subsubsection{Navigating AI Aversion and Dependence through Negotiation}


Our \texttt{CHALET} methodology, which treats disagreement as a site of productive dialogue and can be viewed as a form of \textbf{structured human-LLM negotiation}, has potential, in human-AI collaboration broadly, to help temper the common vacillation between AI aversion and AI dependence \cite{agency_mahmud_2022, agency_burton_2020}, resisting both the dismissal of AI contributions, which may stem from a sense of \textit{agency threat}, and AI overtrust, which can contribute to \textit{agency loss}.


Specifically, rather than reacting to disagreements with \textit{distrust} and automatically attributing them to AI errors, regardless of whether they are driven by intentional non-use \cite{agency_baumer_2014} or concerns over AI accessibility and explainability \cite{agency_mahmud_2022}, \texttt{CHALET} invites human coders to openly approach such divergence as an opportunity for epistemic inquiry, asking if the misalignment signals human oversight, AI/LLM limitations, or previously unarticulated concepts (\textbf{T2}), and to give LLM-produced codes a ``\textit{fair hearing}".
Analogously, instead of defaulting to LLM outputs as authoritative, \texttt{CHALET} requires human coders to actively assess, interpret, and decide on these suggestions through reflexive, dialogic engagement (\textbf{T3}) that preserves the critical agency of both human and LLM coders (i.e., actants).
A greater focus on how algorithmic factors (e.g., LLM sycophancy \cite{sycophancy_sharma_2023}, anthropomorphism) and individual elements (e.g., personality, task familiarity) \cite{agency_mahmud_2022} correlate with human coders' responses to disagreements with LLMs could produce implications that account more for the design of human-LLM partnerships in qualitative analysis.

\subsubsection{Moving beyond Human-centered Design in Qualitative Analysis}

Our \texttt{CHALET} approach may spark general discussions pertaining to HCI design principles that can be applied to systems supporting qualitative analysis.
Among these, \textbf{human-centered design (HCD)} \cite{hcd_jones_2016} stands out as particularly important to examine.
HCD is a widely championed and adopted theoretical framework in HCI, with its pronounced influence expanding into human-centered data science \cite{hcd_aragon_2022}, algorithm design \cite{hcd_baumer_2017}, computing \cite{hcd_bannon_2011}, and AI systems \cite{hcd_capel_2023}, and has guided how we regulate and build artifacts, technologies, and AI agents that collaborate with human coders for decades \cite{content_analysis_coding_toolgpt_gao_2023, nlp_qc_marathe_2018}.
Recent theoretical work, such as \textit{post-userism} \cite{userism_baumer_2017}, however, invites frame reflection on the assumptions underlying HCD, including the idea that humans exist as pre-given, stable subjects who can articulate clear needs, and that our role as designers is to build AI systems that faithfully serve those requirements, arguing instead that humans and technologies co-constitute one another through ongoing intra-actions, with design processes actively shaping human/user identities per se \cite{agency_baumer_2024}.

Our \texttt{CHALET} methodology partly follows, instantiates, and adds value to the concept of ``\textbf{uncentered design}" \cite{agency_baumer_2024} by emphasizing human–LLM disagreement and partnership. 
Rather than positioning human coders as privileged \textit{overseers} who direct AI \textit{tools} to fulfill predetermined intentions, it creates dynamics in which humans and LLMs mutually define and adjust to each other through productive disagreement and negotiation -- shifting from optimizing predefined user experiences toward co-creating knowledge.
Albeit promising, considerably more work will need to be done to explore the impacts and potential risks of treating LLMs as independent qualitative coders on equal footing -- particularly given their unclear positionality and unknown provenance \cite{paradigm_schroeder_2024}, in contrast to human coders with explicit social locations and lived experiences.
We touched on these crucial questions preliminarily in Section \ref{sec:crosscultural}.

\subsubsection{Informing the Design of LLM-integrated Qualitative-coding Tools}

For systems implementing \textit{theory-driven} frameworks like \texttt{CHALET}, our paper could offer actionable design considerations that extend beyond existing platforms such as NVivo, ATLAS.ti, MaxQDA, and AI-assisted coding tools like CollabCoder \cite{content_analysis_coding_toolgpt_gao_2023}, which primarily support \textit{data-driven} qualitative coding.
Specifically, future UI/UX research may explore: 1) \textbf{theory-integration interfaces} that allow users to import, manage, and update theoretical frameworks with visual indicators that distinguish theory-grounded codes from data-derived ones; 2) \textbf{disagreement-analysis panels} that automatically flag human-LLM coding discrepancies and provide categorization menus for users to classify disagreements as coding errors or potential new theoretical constructs (\textbf{T2}); and 3) \textbf{interactive review dashboards} for users to quickly review LLM-assigned codes (\textbf{T1}) and evaluate LLM-suggested coding schemes (\textbf{T3}) through intuitive side-by-side comparison views.

To provide a potential use scenario, consider the example in Section \ref{sec:generalizability} of HCI researchers studying "\textit{how people experience harm from AI systems}": users could 1) upload participant messages into the \textbf{theory-integration interface} and import their initial coding scheme (e.g., representational, allocative, and interpersonal harms with codes like "\textit{verbal abuse}" and "\textit{opportunity loss}" tagged in \textcolor{darkred}{red} as theory-grounded codes), then view LLM-generated deductive codes side-by-side with human codes where discrepancies are highlighted with \colorbox{lightgray}{gray} backgrounds; 2) use the \textbf{disagreement-analysis panel} to classify these flagged messages through drop-down menus as "\textit{human coding error}," "\textit{LLM coding error}," or "\textit{new code}," with messages marked as "\textit{new code}" automatically routed to the next page; and 3) refine emergent codes in the \textbf{interactive review dashboard} where LLMs generate data-derived codes (tagged in \textcolor{darkgreen}{green}) like "\textit{relational transgression}," which users can revise through prompts or manual editing, producing an expanded coding scheme that visually distinguishes theory-grounded codes (e.g., "\textit{opportunity loss}" under allocative harm) from data-derived codes (e.g., "\textit{relational transgression}" under interpersonal harm), with the entire coding process logged for transparency.

\subsubsection{Facilitating Psychological-dataset Creation}

The \texttt{CHALET} methodology may inspire future research in creating well-curated, real-world, domain-valuable datasets for training LLMs across various domains. 
Our case study applies most directly to the psychology and social science fields.
As highlighted by previous research \cite{llm_psycho_demszky_2023}, investing in the creation of such datasets is a top priority for the effective use of generative AI tools in HCI research. 
By using LLM-powered chatbots for data collection, we showed how they can facilitate the acquisition of rich, cross-cultural conversational data for training and testing LLMs in qualitative coding. 
The subsequent LLM-assisted coding can help generate high-quality codes that meet the stringent standards for expert evaluation \cite{llm_psycho_demszky_2023}. 
These dataset-creation steps also serve as a foundation for researchers to adopt our approach to theory-guided qualitative analysis, paving the way for more robust and influential long-term theoretical contributions to HCI studies.

\subsection{Limitations and Future Research}

Several limitations warrant discussion. 
First, \textbf{our demonstration of \texttt{CHALET} through a single case study using one LLM (i.e., GPT-4) may limit claims about its broader generalizability}, as the case-study findings are relatively descriptive and exploratory, and recent research indicates that different LLMs can exhibit different predispositions and patterns in their responses \cite{replace_participant_dillion_2023}.
A natural extension of our work would be to test \texttt{CHALET} with other state-of-the-art LLMs like Claude, across various research domains and theories, and through more systematic evaluation methods, helping establish best practices for human-LLM qualitative analysis across platforms and model versions.

Second, \textbf{socio-cultural factors present concerns at both the technological and empirical levels}. 
Given the nascent nature of LLM technology, \textbf{LLMs may harbor and inadvertently reinforce societal biases} (e.g., historical, representational, and measurement biases \cite{suresh_lifecycle_2021}), cultural stereotypes, and Western-centric perspectives \cite{llm_stereotype_ma_2023}.
These biases can manifest throughout the machine-learning lifecycle, from data preparation to model building and implementation \cite{suresh_lifecycle_2021}. 
This is of particular concern when performing qualitative coding of socially and culturally sensitive topics like stigma-related content, as LLMs might be trained on data that reflect predominantly Western, liberal notions of mental health and other culturally bounded interpretations \cite{colonial_das_2024}. 
A critical next step is to systematically benchmark whether, and to what extent, LLMs reproduce biases like mental-illness stigma in their generated codes, particularly in cultural contexts beyond the West.
Besides, it is also worth noting that \textbf{the participant pool of our case study, while diverse, was not representative of any specific population}.
Because psychological constructs are deeply influenced by socio-cultural factors \cite{culture_difference_krendl_2020}, our case-study findings may not have captured the full spectrum of culture-specific nuances.
Future research could explore \texttt{CHALET} with more culturally diverse participants to better understand how socio-cultural factors influence stigma-related expression and human-LLM disagreement patterns.

And third, there are important potential \textbf{differences in the dynamics of human-human and human-chatbot interactions}. 
While qualitative data like interviews and focus groups allow researchers to observe crucial paralinguistic and kinesic features \cite{nonverbal_denham_2013}, including prosodic variation, proxemics, and gestural behavior, which can provide additional layers of meaning in understanding stigmatizing attitudes, the absence of these non-verbal communicative elements in human-chatbot interactions may limit our ability to fully capture the covert form of stigma. 
In addition, unlike human interviews that capture immediate, spontaneous responses and natural speech patterns, chatbot interactions introduce a temporal buffer that may affect how participants formulate and express their thoughts.
Although studies suggest that chatbots can facilitate meaningful self-disclosure \cite{disclosure_lee_2022, disclosure_li_2020}, more research is needed to uncover differences in the expression of beliefs, attitudes, and behavioral intentions between these interaction modes. 
A promising direction is to explore chatbot designs incorporating audio and avatar features \cite{interviewer_suen_2023} to enhance tangibility and immediacy and to better approximate the rich multimodal nature of human interviews for richer qualitative-data collection.

Finally, the message-level analysis of our approach may have overlooked important dynamics that emerged and evolved in complex ways \textbf{over the course of entire conversations}. 
While analyzing every individual message provides valuable qualitative insights, it may overlook how stigmatizing attitudes evolve through turn-taking, how participants refine their views over the course of the interaction, and how conversational elements, such as chatbot responses, previous exchanges, and the overall discussion climate, shape the expression of stigma. 
Moving forward, researchers would benefit from expanding the analysis to include aspects such as trajectories, temporal patterns, intersubjectivity, and narrative development across full conversations \cite{conversation_level_paakki_2024}.

\section*{Acknowledgments}

Our research has been supported by the Ministry of Education,
Singapore (A-8002610) and the NUS Artificial Intelligence Institute (NAII) (A-8003879). 
We are grateful to all the reviewers for their valuable comments and suggestions that helped improve this paper, and to all the participants whose time and efforts made this research possible.

\appendix


\clearpage
\section{Full Results of LLM Deductive Coding}
\label{app:full}

Tables \ref{tab:exp1} and \ref{tab:exp2} present the full results of the LLM deductive coding (Section \ref{sec:casestudy_llmdeduc}).
Table \ref{tab:exp1} shows how varying components of the codebook affected the agreement between LLMs and human coders. 
Building on these results, Table \ref{tab:exp2} explores the extent to which additional components not covered by the codebook influenced coding agreement, with each value representing the change in agreement compared to our baseline condition (highlighted in red in Table \ref{tab:exp1}).

\begin{table}[h]
\small
    \centering
    \caption{Cohen's $\kappa$ between LLM and human codes for each attribution across different prompt configurations, with the \textit{total} row showing aggregated agreement across all attributions. 
    Prompts are structured in two main settings: \textit{All-code comp.} (providing components for all codes) and \textit{Target-code comp.} (providing components only for the target attribution, with others coded as \textit{Stigmatizing (others)}). 
    These components are cumulative from left to right, with each column adding to the previous one: \textit{name} (code name only), \textit{+vig} (vignette), \textit{+rule} (coding rules), \textit{+keyword} (keywords), and \textit{+exp} (example). 
    The \textbf{\textcolor{darkred}{red}} column contains the same codebook components as the human-written codebook, with only format differences, and \textbf{bold} values show the highest human-LLM agreement for each attribution. Codebook components were systematically varied to assess their individual contributions to human-LLM agreement. See Section \ref{sec:casestudy_llmdeduc} for analysis and interpretation.}
    \label{tab:exp1}
    \begin{adjustbox}{width=\textwidth,center}
    \begin{tabular}{@{}m{2cm}|*{5}{c}|*{5}{c}@{}}
        \toprule
         & \multicolumn{5}{c|}{All-code comp.} & \multicolumn{5}{c}{Target-code comp.} \\
         \cmidrule{2-11}
         & name & $+$vig & $+$rule & $+$keyword& \textcolor{darkred}{\textbf{$+$exp}} & name & $+$vig & $+$rule & $+$keyword & $+$exp \\
         \midrule
         \raggedright Responsibility & \gradientcell{0.68} & \gradientcell{0.67} & \gradientcell{0.63} & \gradientcell{0.64} & \gradientcellbold{0.69} & \gradientcell{0.61} & \gradientcell{0.63} & \gradientcell{0.62} & \gradientcell{0.63} & \gradientcell{0.64} \\
         \midrule
         \raggedright Anger & \gradientcell{0.57} & \gradientcell{0.48} & \gradientcell{0.6} & \gradientcell{0.62} & \gradientcellbold{0.65} & \gradientcell{0.3} & \gradientcell{0.31} & \gradientcell{0.51} & \gradientcell{0.52} & \gradientcell{0.54} \\
         \raggedright Pity & \gradientcell{-0.04} & \gradientcell{-0.03} & \gradientcell{0.42} & \gradientcell{0.45} & \gradientcell{0.46} & \gradientcell{0.13} & \gradientcell{0.12} & \gradientcell{0.44} & \gradientcell{0.48} & \gradientcellbold{0.5} \\
         \raggedright Fear & \gradientcell{0.69} & \gradientcell{0.66} & \gradientcell{0.63} & \gradientcell{0.67} & \gradientcellbold{0.71} & \gradientcell{0.58} & \gradientcell{0.61} & \gradientcell{0.68} & \gradientcell{0.69} & \gradientcell{0.69} \\
         \midrule
         \raggedright Helping & \gradientcell{0.15} & \gradientcell{0.22} & \gradientcell{0.68} & \gradientcellbold{0.69} & \gradientcellbold{0.69} & \gradientcell{-0.03} & \gradientcell{0.01} & \gradientcell{0.58} & \gradientcell{0.62} & \gradientcell{0.64} \\
        \raggedright Coercive Segregation & \gradientcellbold{0.58} & \gradientcell{0.53} & \gradientcell{0.49} & \gradientcell{0.51} & \gradientcell{0.54} & \gradientcell{0.26} & \gradientcell{0.24} & \gradientcell{0.35} & \gradientcell{0.38} & \gradientcell{0.35} \\
        \raggedright Social Distance & \gradientcell{0.59} & \gradientcell{0.67} & \gradientcell{0.73} & \gradientcell{0.74} & \gradientcellbold{0.76} & \gradientcell{0.25} & \gradientcell{0.34} & \gradientcell{0.59} & \gradientcell{0.54} & \gradientcell{0.61} \\
         \midrule
        \raggedright Total & \gradientcell{0.37} & \gradientcell{0.39} & \gradientcell{0.64} & \gradientcell{0.67} & \gradientcellbold{0.69} & \gradientcell{0.28} & \gradientcell{0.3} & \gradientcell{0.58} & \gradientcell{0.59} & \gradientcell{0.6} \\
        \bottomrule
    \end{tabular}
    \end{adjustbox}
\end{table}

\begin{table}[t]
  \centering
  \setlength{\tabcolsep}{3pt}
  \caption{Changes in Cohen's $\kappa$ when comparing prompts with additional components outside the codebook to prompts with exactly the same components as the human-written codebook. 
  The notations used include \textit{CoT}/\textit{NoCoT} for chain-of-thought presence/absence, \textit{S}/\textit{NS}/\textit{O} for \textit{Stigmatizing}/\textit{Non-stigmatizing}/\textit{Stigmatizing (others)}, and \textit{S\_NS\_O} indicates presenting three examples in the prompt with \textit{Stigmatizing} first, followed by \textit{Non-stigmatizing} and \textit{Stigmatizing (others)}. 
  The other notations similarly indicate other possible permutations.
  \textcolor{purple}{\textbf{Purple}} cells show where adding components improved agreement compared to using exact codebook components, while \textcolor{darkgreen}{\textbf{green}} cells show where additions decreased agreement, with darker shades indicating larger differences. These comparisons assess whether components beyond the standard codebook affect human-LLM agreement. See Section \ref{sec:casestudy_llmdeduc} for analysis and interpretation.}
  \label{tab:exp2}
  \begin{adjustbox}{width=\textwidth,center}
  \begin{tabular}{@{}m{2cm}|*{1}{c|}*{12}{c}@{}}
    \toprule
     & \multirow{-1}{*}{+ CoT} & \multicolumn{12}{c}{+ CoT \& + More Examples} \\
     \cmidrule(lr){3-14} 
     & & \multicolumn{2}{c|}{S\_NS\_O} & \multicolumn{2}{c|}{S\_O\_NS} & \multicolumn{2}{c|}{NS\_S\_O} & \multicolumn{2}{c|}{NS\_O\_S} & \multicolumn{2}{c|}{O\_S\_NS} & \multicolumn{2}{c}{O\_NS\_S} \\
     \cmidrule(lr){3-14} 
     & & CoT & NoCoT & CoT & NoCoT & CoT & NoCoT & CoT & NoCoT & CoT & NoCoT & CoT & NoCoT \\
     \midrule
    \raggedright Responsibility & \g{0.03} & \g{0.06} & \g{0.03} & \g{0.04} & \g{0.04} & \g{-0.02} & \g{-0.02} & \g{-0.03} & \g{0} & \g{0.02} & \g{0.02} & \g{-0.01} & \g{0.02} \\
    \midrule
    \raggedright Anger & \g{0.03} & \g{0.11} & \g{0.09} & \g{0.13} & \g{0.1} & \g{0.08} & \g{0.08} & \g{0.08} & \g{0.09} & \g{0.1} & \g{0.1} & \g{0.07} & \g{0.08} \\
    \raggedright Pity & \g{0.05} & \g{0.01} & \g{-0.02} & \g{-0.01} & \g{-0.07} & \g{0} & \g{-0.08} & \g{0.03} & \g{-0.06} & \g{-0.05} & \g{-0.1} & \g{-0.01} & \g{-0.08} \\
    \raggedright Fear & \g{0.02} & \g{0.02} & \g{0.01} & \g{0.02} & \g{0.03} & \g{0.02} & \g{0.03} & \g{0.02} & \g{0.01} & \g{0.02} & \g{0.01} & \g{0.02} & \g{0.02} \\
    \midrule
    \raggedright Helping & \g{0.02} & \g{-0.12} & \g{-0.09} & \g{-0.18} & \g{-0.12} & \g{-0.15} & \g{-0.13} & \g{-0.14} & \g{-0.09} & \g{-0.15} & \g{-0.09} & \g{-0.16} & \g{-0.09} \\
    \raggedright Coercive Segregation & \g{-0.04} & \g{0.05} & \g{0.04} & \g{0.06} & \g{0.08} & \g{0.02} & \g{0} & \g{0} & \g{0} & \g{0.03} & \g{0.02} & \g{0.01} & \g{0.05} \\
    \raggedright Social Distance & \g{0.03} & \g{0.07} & \g{0.06} & \g{0.03} & \g{0.06} & \g{0.03} & \g{0.05} & \g{0.04} & \g{0.05} & \g{0.02} & \g{0.07} & \g{0.07} & \g{0} \\
    \midrule
    \raggedright Total & \g{0.02} & \g{0.03} & \g{0.02} & \g{0.01} & \g{0.02} & \g{0} & \g{-0.01} & \g{0} & \g{0} & \g{0} & \g{0.01} & \g{0} & \g{0} \\
    \bottomrule
  \end{tabular}
  \end{adjustbox}
\end{table}

\section{Prompt Templates} 
\label{app:prompt}

\begin{figure}
    \centering
    \includegraphics[width=\textwidth]{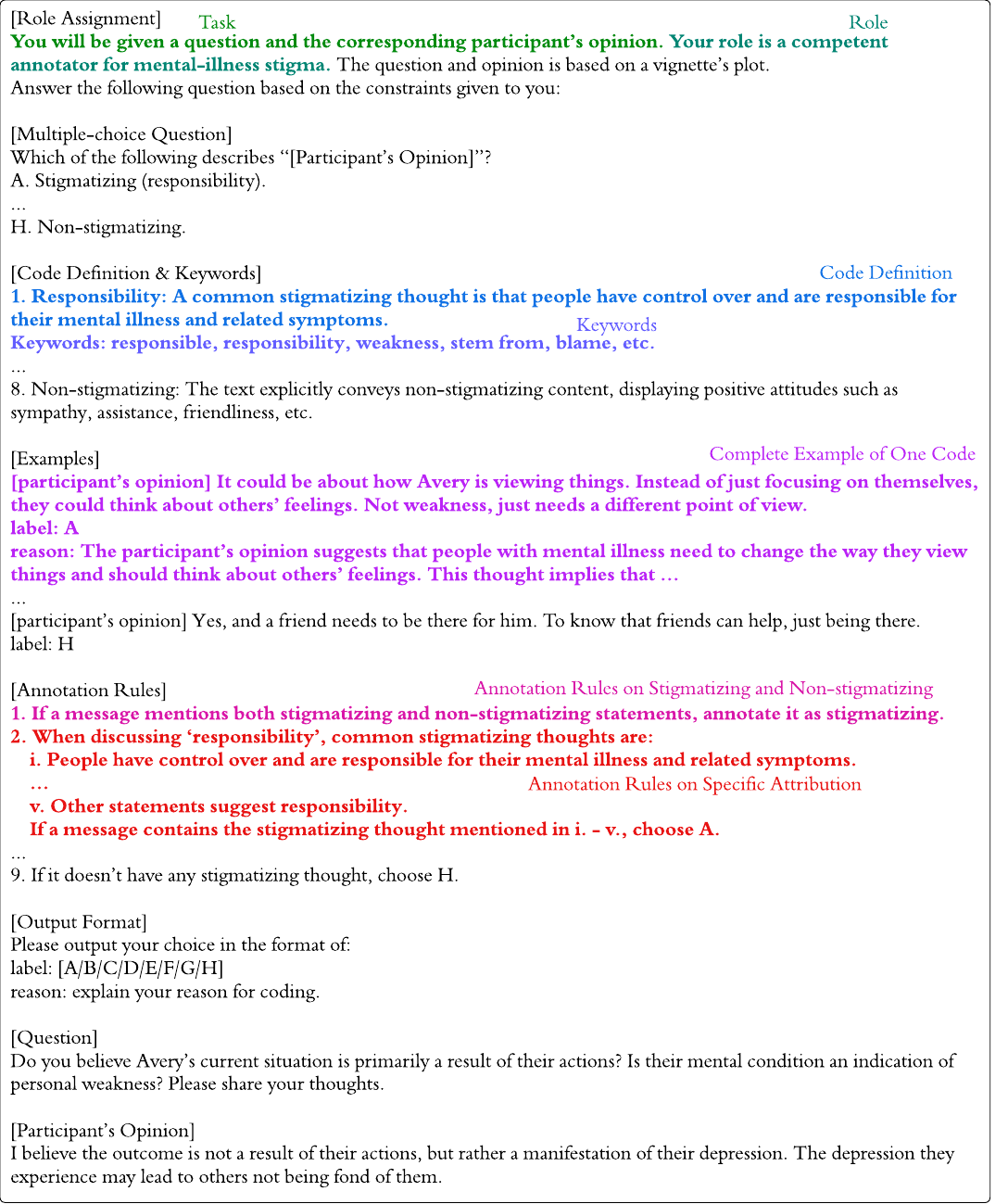}
    \caption{An example prompt template for LLM deductive coding, illustrating the all-code condition where each code's components align exactly with the human-written codebook.}
    \Description{This figure..}
    \label{fig:prompt_structure}
\end{figure}

Figure \ref{fig:prompt_structure} demonstrates an example of our prompt template. 
This specific example shows the all-code condition, where we include all eight codes with their complete codebook components. 
In practice, we modify this template by varying the components included for each code, as discussed in Section \ref{sec:casestudy_llmdeduc}.

\section{Autonomous LLM Inductive Coding Results}
\label{sec:app_llm_ind}

Table \ref{tab:llm_ind} presents the codebook from autonomous LLM inductive coding (Section \ref{sec:induct_res}).

{\small
\begin{longtable}{>{\raggedright\arraybackslash}p{0.2\linewidth} 
                   >{\raggedright\arraybackslash}p{0.75\linewidth}}

\caption{Results of the autonomous LLM inductive coding for mental-illness stigma. Themes and codes were generated through autonomous LLM coding of 181 messages identified as potential new codes from human-LLM disagreements. The coding process and implementation are detailed in Section \ref{sec:induct_res}.} \label{tab:llm_ind} \\

\toprule
\textbf{Themes} & \textbf{Descriptions \& Codes} \\
\midrule
\endfirsthead

\toprule
\textbf{Themes} & \textbf{Descriptions \& Codes} \\
\midrule
\endhead

\midrule
\endfoot

\bottomrule
\endlastfoot

\begin{minipage}[t]{\linewidth} \setlength{\baselineskip}{10pt} \textbf{Theme 1: Health Management}\end{minipage} & 
\begin{minipage}[t]{\linewidth} 
\setlength{\baselineskip}{10pt} 
Topics related to managing health, whether through professional intervention, self-care, or community support. \\
Code 1.1: Hospitalization Debate \\
Code 1.2: Home Treatment Preference \\
Code 1.3: Professional Assessment Advocacy \\
Code 1.4: Anger Management 
\end{minipage} \\

\begin{minipage}[t]{\linewidth} \setlength{\baselineskip}{10pt} \textbf{Theme 2: Support Systems}\end{minipage} & 
\begin{minipage}[t]{\linewidth} 
\setlength{\baselineskip}{10pt} 
Topics focus on the various forms of support available or needed for individuals, including workplace, community, and emotional support. \\
Code 2.1: Community Support Importance \\
Code 2.2: Workplace Support \\
Code 2.3: Sympathy and Concern
\end{minipage} \\

\begin{minipage}[t]{\linewidth} \setlength{\baselineskip}{10pt} \textbf{Theme 3: Mental Health Focus}\end{minipage} & 
\begin{minipage}[t]{\linewidth} 
\setlength{\baselineskip}{10pt} 
Topics that specifically relate to mental health issues, including management, misconceptions, and the importance of professional assessments. \\
Code 3.1: Mental State Misinterpretation \\
Code 3.2: Anger Management \\
Code 3.3: Professional Assessment Advocacy
\end{minipage} \\

\begin{minipage}[t]{\linewidth} \setlength{\baselineskip}{10pt} \textbf{Theme 4: Social Acceptance and Integration}\end{minipage} & 
\begin{minipage}[t]{\linewidth} 
\setlength{\baselineskip}{10pt} 
Topics that deal with the acceptance and integration of individuals into various social settings, including housing and the workplace. \\
Code 4.1: Tenant Acceptance \\
Code 4.2: Workplace Support
\end{minipage} \\

\begin{minipage}[t]{\linewidth} \setlength{\baselineskip}{10pt} \textbf{Theme 5: Emotional and Psychological Well-being}\end{minipage} & 
\begin{minipage}[t]{\linewidth} 
\setlength{\baselineskip}{10pt} 
Topics that touch on the emotional and psychological aspects of well-being, including the management of emotions and the importance of empathy. \\
Code 5.1: Sympathy and Concern \\
Code 5.2: Anger Management \\
Code 5.3: Mental State Misinterpretation
\end{minipage} \\

\begin{minipage}[t]{\linewidth} \setlength{\baselineskip}{10pt} \textbf{Theme 6: Advocacy and Debate}\end{minipage} & 
\begin{minipage}[t]{\linewidth} 
\setlength{\baselineskip}{10pt} 
Topics that involve discussions, debates, or advocacy for certain approaches or understandings related to health and social integration. \\
Code 6.1: Hospitalization Debate \\
Code 6.2: Professional Assessment Advocacy
\end{minipage} \\

\begin{minipage}[t]{\linewidth} \setlength{\baselineskip}{10pt} \textbf{Theme 7: Safety and Concerns}\end{minipage} & 
\begin{minipage}[t]{\linewidth} 
\setlength{\baselineskip}{10pt} 
Topics that are centered around concerns for safety, whether during travel or in the context of mental health. \\
Code 7.1: Traveling Concerns \\
Code 7.2: Mental State Misinterpretation
\end{minipage} \\

\begin{minipage}[t]{\linewidth} \setlength{\baselineskip}{10pt} \textbf{Theme 8: Preference and Choice in Care}\end{minipage} & 
\begin{minipage}[t]{\linewidth} 
\setlength{\baselineskip}{10pt} 
Topics that highlight the preference for where and how care and treatment should be received. \\
Code 8.1: Hospitalization Debate \\
Code 8.2: Home Treatment Preference
\end{minipage} \\

\begin{minipage}[t]{\linewidth} \setlength{\baselineskip}{10pt} \textbf{Theme 9: Community and Social Support}\end{minipage} & 
\begin{minipage}[t]{\linewidth} 
\setlength{\baselineskip}{10pt} 
Topics emphasizing the importance of community and social support in various aspects of life and well-being. \\
Code 9.1: Community Support Importance \\
Code 9.2: Workplace Support
\end{minipage} \\

\begin{minipage}[t]{\linewidth} \setlength{\baselineskip}{10pt} \textbf{Theme 10: Conflict and Management}\end{minipage} & 
\begin{minipage}[t]{\linewidth} 
\setlength{\baselineskip}{10pt} 
Topics related to managing conflicts, whether internal (such as anger) or external (such as tenant acceptance). \\
Code 10.1: Anger Management \\
Code 10.2: Tenant Acceptance
\end{minipage} \\

\begin{minipage}[t]{\linewidth} \setlength{\baselineskip}{10pt} \textbf{Theme 11: Mobility and Accessibility}\end{minipage} & 
\begin{minipage}[t]{\linewidth} 
\setlength{\baselineskip}{10pt} 
Topics that deal with concerns related to mobility, including traveling concerns and the broader implications for individuals needing support. \\
Code 11.1: Traveling Concerns
\end{minipage} \\

\end{longtable}
}

\end{document}